\documentclass{aa}

\usepackage{graphicx}
\usepackage[usenames,dvipsnames]{xcolor}
\usepackage{txfonts}
\usepackage{hyperref}
\usepackage{lscape}
\usepackage{natbib}
\bibpunct{(}{)}{;}{a}{}{,} 

\begin{document} 
	

	\title{Measuring chemical abundances in AGN from infrared nebular lines: \textsc{HII-CHI-Mistry-IR} for AGN}

	\author{Borja Pérez-Díaz\inst{\ref{inst1}}
		\and
		Enrique Pérez-Montero\inst{\ref{inst1}} \and Juan A. Fernández-Ontiveros\inst{\ref{inst2}} \and José M. Vílchez\inst{\ref{inst1}}}
	
	\institute{Instituto de Astrofísica de Andalucía (IAA-CSIC), Glorieta de la Astronomía s/n, 18008 Granada, Spain\label{inst1} \\\email{bperez@iaa.es}\and Centro de Estudios de Física del Cosmos de Aragón (CEFCA), Unidad Asociada al CSIC, Plaza San Juan 1, E--44001 Teruel, Spain\label{inst2}}
	
	\date{Received MONTH DAY, YEAR; accepted MONTH DAY, YEAR}



\abstract
{Future and on-going infrared and radio observatories such as JWST, METIS or ALMA will increase the amount of rest-frame IR spectroscopic data for galaxies by several orders of magnitude. While studies of the chemical composition of the ISM based on optical observations have been widely spread over decades for SFG and, more recently, for AGN, similar studies need to be performed using IR data. This regime can be especially useful in the case of AGN given that it is  less affected by temperature and dust extinction, traces higher ionic species and can also provide robust estimations of the chemical abundance ratio N/O.}
{We present a new tool based on a bayesian-like methodology (\textsc{HII-CHI-Mistry-IR}) to estimate chemical abundances from IR emission lines in AGN. We use a sample of 58 AGN with IR spectroscopic data retrieved from the literature, composed by 43 Seyferts, 8 ULIRGs, 4 LIRGs and 3 LINERs, to probe the validity of our method. The estimations of the chemical abundances based on IR lines in our sample  are later compared with the corresponding abundances derived from the optical emission lines in the same objects.}
{\textsc{HII-CHI-Mistry-IR} takes advantage of photoionization models,  characterized by the chemical abundance ratios O/H and N/O and the ionization parameter $U$, to compare their predicted emission-line fluxes with a set of observed values. Instead of matching single emission lines, the code uses some specific emission-line ratios sensitive to the above free parameters.}
{We report mainly solar and also subsolar abundances for O/H in the nuclear region for our sample of AGN, whereas N/O clusters around solar values. We find a discrepancy between the chemical abundances derived from IR and optical emission lines, being the latter higher than the former. This discrepancy, also reported by previous studies of the composition of the ISM in AGN from IR observations, is independent from the gas density or the incident radiation field to the gas, and it is likely associated with dust obscuration and/or temperature stratification within the gas nebula.}
{}


\maketitle

\defcitealias{Perez-Montero_2019}{PM19}
\defcitealias{Fernandez-Ontiveros_2021}{FO21}
\defcitealias{Thornley_2000}{Thornley et al. 2000}
\defcitealias{Yeh_2012}{Yeh \& Matzner 2012}
\defcitealias{Kewley_2019}{Kewley et al. 2019}
\defcitealias{Nagao_2011}{Nagao et al. 2011}


\section{Introduction}
\label{sec1}

Active Galactic Nuclei (AGN) are among the most luminous objects in the Universe and can hence be studied up to very high redshift. The interstellar medium (ISM) surrounding these nuclei is ionized by very energetic photons radiated from the accretion disc and jets around the supermassive black hole (SMBH). This ionization is partially re-emitted in the form of strong and prominent emission lines which can provide information on the physical and chemical properties of the region where they were originated.

Since the nebular line properties depend on the chemical composition of the ISM gas, their relative fluxes can be used to quantify the abundances of elements heavier than hydrogen and helium, known as metals (see \citealt{Maiolino_2019} for a thorough review). While the primordial Big Bang Nucleosynthesis explains the observed abundances of hydrogen or  deuterium, as well as a significant fraction of helium and a small fraction of lithium \citep{Cyburt_2016}, nearly all other elements are produced by stellar nucleosynthesis in the cores of stars, driven to their surfaces by convective flows and, in the late stages of their lives, are finally ejected into the ISM by stellar winds and supernovae (see review from \citealt{Nomoto_2013}). Thus, the analysis of chemical abundances at different redshifts can provide key information on galactic evolution throughout different cosmological epochs.

The oxygen abundance (usually represented as 12+log(O/H)) is widely used as a proxy of the metal content in the ISM of galaxies, since O is the most abundant metal in mass and its presence can be easily detected through strong emission lines in the ultraviolet (UV), optical and infrared (IR) range \citep{Osterbrock_book}. Another quantity relevant to analyze the past chemical evolution  of the ISM in galaxies is the nitrogen-to-oxygen abundance ratio, represented as log(N/O). This relative abundance provides essential information on the build-up of heavy elements from stellar \citep{Chiappini_2005} to galactic \citep{Vincenzo_2018} scales because it involves a primary metal, O, and another one, N, that may have a secondary origin. In the low-metallicity regime ({\em i.e.} 12+log(O/H) $\lesssim$ 8.0), N is expected to be mainly primary produced by massive stars, thus N/O shows basically a constant value. However, in the high-metallicity regime, N has a significant contribution from a secondary production channel, as it is formed via the CNO cycle in intermediate-mass stars, and therefore N/O tends to increase with O/H \citep[e.g.][]{Perez-Montero_2009}. This correlation between O/H and N/O has been determined in studies of chemical abundances in star-forming galaxies (SFG) and {\sc Hii} regions in galaxies using optical observations from nearby and distant galaxies \citep[e.g.][]{Vila-Costas_1993, Pilyugin_2004, Andrews_2013, Masters_2016, Hayden-Pawson_2022}, although it has also been reported that some groups of galaxies deviate from that behavior \citep{Amorin_2010, Guseva_2020, Perez-Montero_2021}. Thus, the N/O determination does not only provide key information on the metal production in the ISM, but is also a necessary step in the determination of oxygen abundances when nitrogen lines are involved.

For decades, many studies have been devoted to analyze the chemical composition of the gas-phase in SFG using optical emission lines \citep[e.g.][]{McClure_1968, Lequeux_1979, Garnett_1987, Thuan_1995, Pilyugin_2004}. Several techniques have been developed for that purpose: \textit{i)} the T$_{e}$-method (also known as direct method), that is measuring the line ratios of specific collisional emission lines (CELs), sensitive to the electronic temperature and density, to directly derive the abundances of the main ionic species \citep[e.g.][]{Aller_book, Osterbrock_book}; \textit{ii)} by means of photoionization models to reproduce the observed CELs and then constrain chemical and physical properties of the region, using several codes such as \textsc{Cloudy} \citep{Ferland_2017}, \textsc{Mappings} \citep{Sutherland_2017} or \textsc{Suma} \citep{Contini_2001}; and \textit{iii)} the use of empirical or semi-empirical calibrations between accurate chemical abundances and the relative fluxes of strong emission lines \citep[e.g.][]{Perez-Montero_2009, Pilyugin_2016, Curti_2017}. Furthermore, new approaches take advantage of more than one of the above techniques at the same time, such as \textsc{HII-CHI-Mistry} (hereinafter \textsc{HCm}, \citealt{Perez-Montero_2014}), which uses sensitive ratios to chemical abundances to search for the best fit among a grid of photoionization models.

In recent years, the analysis of chemical abundances in the gas-phase of {\sc Hii} regions has been extended to the Narrow Line Region (NLR) in AGN \citep[e.g.][]{Storchi_1998, Contini_2001, Dors_2015, Perez-Montero_2019, Thomas_2019, Flury_2020, Perez-Diaz_2021}. This region of the ISM, located between $\sim$10$^{2}$ pc and few kpc \citep{Bennert_2006, Bennert_2006b}, is characterized by an electronic density n$_{e}$ typically in the 10$^{2}$-10$^{4}$ cm$^{-3}$ range, and an electronic temperature of T$_{e} \sim 10^{4}$ K \citep{Vaona_2012, Netzer_2015}. Although these physical conditions may not depart significantly from to those of the ISM in {\sc Hii} regions \citep{Osterbrock_book}, the source and the shape of the ionizing continuum are completely different in both cases: the accretion disk in AGN, which produces a power-law like continuum extending to high energies; and a thermal-like continuum from massive O and B-type stars in {\sc Hii} regions. This difference has profound effects in the emission-line spectrum, as some highly-ionized species are not found in {\sc Hii} regions, while their contribution is not negligible in AGN due to the harder radiation fields involved \citep{Kewley_2019, Flury_2020}. Thus, the techniques developed for the metal content study in SFG must take these differences into account when applied to the AGN case \citep[e.g.][]{Dors_2015, Perez-Montero_2019, Carvalho_2020, Flury_2020, Perez-Diaz_2021}.

Chemical abundances can also be derived using emission lines in the UV range. This is the case of galaxies at redshift z$\gtrsim $1-2, where UV lines can be measured by optical telescopes, allowing the determination of chemical abundances in SFG \citep[e.g.][]{Erb_2010, Dors_2014, Berg_2016, Perez-Montero_2017} and AGN \citep{Dors_2019}. In these cases, besides the oxygen abundance 12+log(O/H), it is also important to constrain carbon-to-oxygen ratio log(C/O), since C emits strong UV emission lines which are easily detected and, as N, it is also a metal with both primary and secondary origins.

Nevertheless, the determination of chemical abundances using optical and, above all, UV emission lines, can be seriously affected by reddening.
In particular, deeply dust-embedded regions may go unnoticed by optical and UV tracers, which therefore will not be able to probe their content of heavy elements. In addition, the optical and UV CELs present a non-negligible dependence with some physical properties of the ISM such as the electronic temperature (T$_{e}$) or the electronic density (n$_{e}$), difficult to be considered either in empirical calibrations or in models. These problems do not arise when chemical abundances are derived from IR emission lines. The relative insensitivity of IR lines to interstellar reddening allows us to peer through the dusty regions in galaxies \citep{Nagao_2011, Pereira-Santaella_2017, Fernandez-Ontiveros_2021}. In addition, the negligible dependence of the IR line emissivity on T$_{e}$ \citep[see fig.\,1 in][]{Fernandez-Ontiveros_2021} avoids the large uncertainties in the temperature determination affecting the abundances based on optical lines. For instance, \citet{Dors_2013} suggest that extinction effects and temperature fluctuations might be an explanation for the discrepancy between optical and infrared estimations of the neon abundances. Temperature fluctuations have been also reported in previous works, {\em e.g.} \citet{Croxall_2013} analyzing a sample of {\sc Hii} regions in NGC628 used fine-structure IR and also optical emission lines. An example of the effect of dust obscuration can be found in \citet{Fernandez-Ontiveros_2021}, where they estimate twice the metallicity of NGC 3198 from IR emission lines when comparing with their optical estimations. 

Over the past decades, several IR spectroscopic telescopes, such as the \textit{Infrared Space Observatory} (\textit{ISO}, covering the 2.4-197 $\mu$m range, \citealt{Kessler_1996}), the \textit{Spitzer Space Observatory} (5-39 $\mu $m, \citealt{Werner_2004}), the \textit{Herschel Space Observatory} (51-671 $\mu $m, \citealt{Pilbratt_2010}) or the \textit{Stratospheric Observatory for Infrared Astronomy} (\textit{SOFIA}, covering the 50-205 $\mu $m range, \citealt{Fischer_2018}), have provided essential information of these emission lines for a considerable amount of sources. Moreover, upcoming missions such as the \textit{James Webb Space Telescope} (\textit{JWST}, which will cover the 4.9-28.9$\mu$m range with the Mid-InfraRed Instrument MIRI, \citealt{Rieke_2015, Wright_2015}) or the Mid-infrared ELT Imager and Spectograph (METIS, covering the N-band centered at 10$\mu$m, \citealt{Brandl_2021}) will increase the amount of information from IR observations of local galaxies, significantly improving recent studies of chemical abundances based on IR emission lines \citep{Peng_2021, Fernandez-Ontiveros_2021, Spinoglio_2021} and setting the ground to extend the analysis to higher redshifts ($z > 4$) with ALMA \citep{Wootten_2009}.

In this work we present an IR version of the method \textsc{HCm} developed by \citep{Perez-Montero_2014} to derive chemical abundances from optical emission lines in SFG, and later extended to the NLR region of AGN by \citet{Perez-Montero_2019} (hereinafter noted as \citetalias{Perez-Montero_2019}). This version of \textsc{HII-CHI-Mistry-IR} or \textsc{HCm-IR} complements the work done by \citet{Fernandez-Ontiveros_2021} (hereinafter noted as \citetalias{Fernandez-Ontiveros_2021}) for SFG. By taking advantage of a grid of photoionization models covering a wide range in 12+log(O/H), log(N/O) and log($U$), our method computes these three parameters by fitting emission-line ratios sensitive to those quantities.

The work is organized as follows. In Sec. \ref{sec2} we describe a sample of galaxies with available spectroscopic IR data used to check the method. This sample is composed by Seyferts, ULIRGs, LIRGs and LINERs, and showing Ne$^{4+}$ emission lines characteristic of the AGN activity \citep{Genzel_1998, Perez-Torres_2021}. In Sec. \ref{sec3} we describe the methodology underlying \textsc{HCm-IR}, including the emission-line ratios used to estimate chemical abundances and the differences with those proposed by \citetalias{Fernandez-Ontiveros_2021} when applied to the AGN case. In Sec. \ref{sec4} we present the main results from \textsc{HCm-IR} for our sample of galaxies, also comparing them with estimations from optical observations. In Sec. \ref{sec5} we present a full discussion on these results and we summarize in Sec. \ref{sec6} the main conclusions from this work.

\section{Sample}
\label{sec2}
To probe the validity of the diagnostics detailed in Sec. \ref{sec3}, we compiled a sample of 58 AGN with spectroscopic observations in the mid- and far-IR ranges from \textit{Spitzer}/IRS \citep{Werner_2004,Houck_2004} and \textit{Herschel}/PACS \citep{Pilbratt_2010,Poglitsch_2010}, respectively. Most of the galaxies (48) have been drawn from the IR spectroscopic atlas in \citet{Fernandez-Ontiveros_2016}, corresponding to those objects with a detection of a hydrogen recombination line in the IR range, namely Brackett-$\alpha$ at $4.05\, \rm{\mu m}$ (Br$\alpha$), Pfund-$\alpha$ at $7.46\, \rm{\mu m}$ (Pf$\alpha$) or Humphreys-$\alpha$ at $12.4\, \rm{\mu m}$ (Hu$\alpha$). The sample was completed with 9 galaxies with available  additional \textit{SOFIA}/FIFI-LS observations \citep{Temi_2014,Fischer_2018} of the [NIII]$57\, \rm{\mu m}$ and/or the [OIII]$52,88\, \rm{\mu m}$ lines from \citet{Spinoglio_2021}. Thus, the sample selection maximizes the number of AGN galaxies with detections of these lines, which allows us to obtain N/O abundance ratios that are independently derived from the oxygen abundance.

The Br$\alpha$ and Pf$\alpha$ line fluxes were collected from the literature, while new measurements of the Hu$\alpha$ line for 11 galaxies are presented in this work (see Tab. \ref{TabA1}). The latter were obtained from the calibrated and extracted \textit{Spitzer}/IRS high-resolution spectra ($R = 600$) in the CASSIS database \citep{Lebouteiller_2015}. The line flux was measured by direct integration of the spectrum at the rest-frame wavelength of the line, subtracting the continuum level derived from a linear polynomial fit to the adjacent continuum at both sides of the line.

The final sample thus consists of  17 Seyfert 1 nuclei (Sy1), 14 Seyfert nuclei with hidden broad lines in the polarized spectrum (Sy1h), 12 Seyfert 2 nuclei (Sy2), 3 Low-Ionization Nuclear Emission-line Regions (LINERs), and 12 luminous and ultraluminous IR galaxies (LIRGs and ULIRGs, respectively).

\begin{figure}
	\centering
	\includegraphics[width=\hsize]{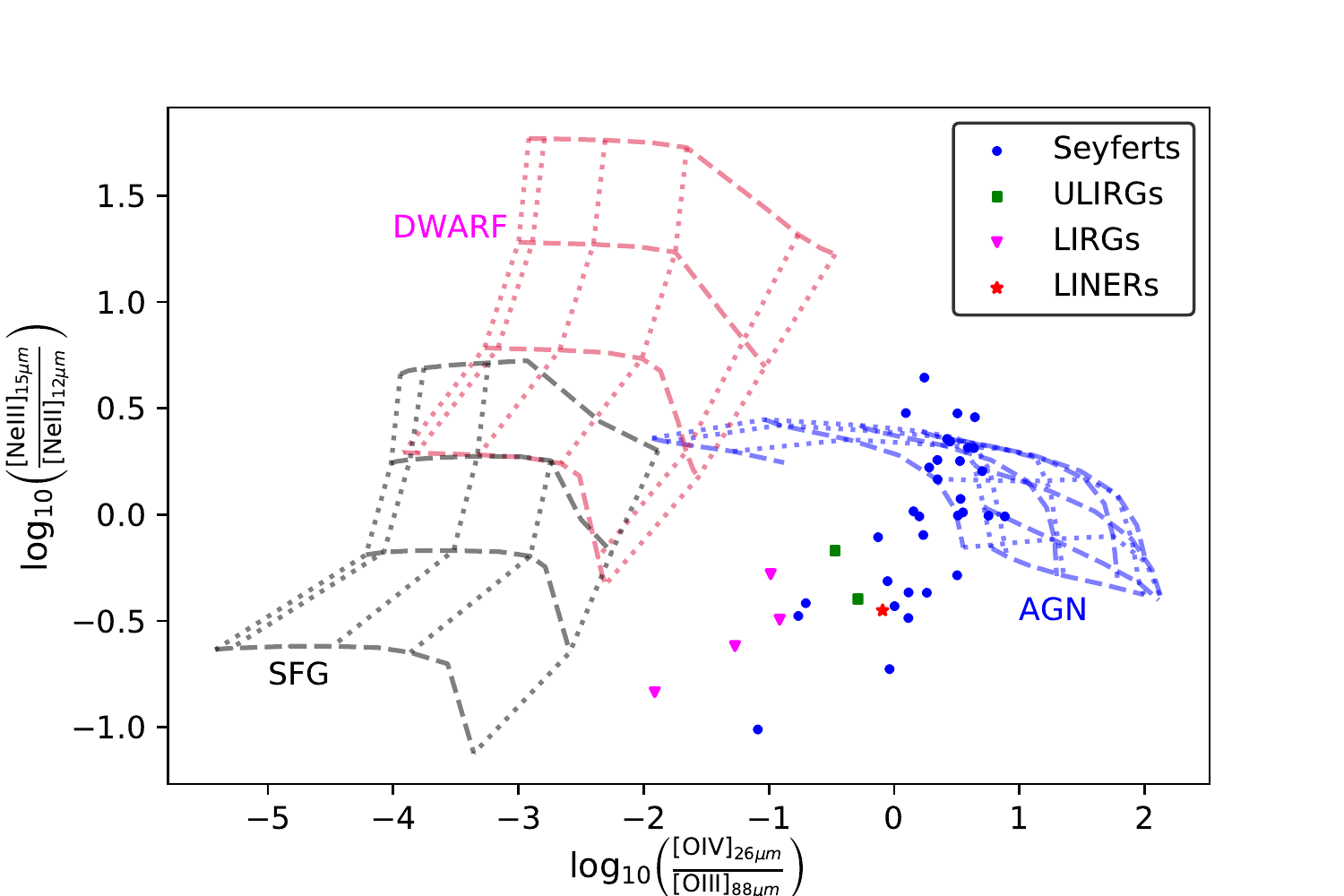} 
	\caption{BPT or diagnostic diagram in the IR range, proposed to distinguish between spectral types \citep{Fernandez-Ontiveros_2016}. Models computed with \textsc{Cloudy} v17.02 \citep{Ferland_2017} are presented as lines: black for SFG models computed from library STARBUST99 \citep{Gonzalez-Delgado_1999} with 12+log(O/H) = 8.69 and log(N/O) = -0.86; magenta for Dwarf models computed from a starburst of $\sim$ 10$^{6}$ yr with 12+log(O/H) = 8.0 and log(N/O) = -0.86; blue for AGN models computed from the same SED used in this work ($\alpha_{OX} = -0.8$ and $\alpha_{UV} = -1.0$, see Sec. \ref{sec3}) with 12+log(O/H) = 8.69 and log(N/O) = -1.0. For all models, dotted lines trace models with the same electronic density (ranging 10-10$^{7}$ cm$^{-3}$) while dashed lines represent fixed values of the ionization parameter $U$.}
	\label{Fig1aux}
\end{figure}

We present in Fig. \ref{Fig1aux} a classification of our sample of AGN based on the so-called \textit{BPT-IR} diagram \citep{Fernandez-Ontiveros_2016}. In contrast with the optical diagnostic diagrams \citep{Baldwin_1981, Kauffmann_2003, Kewley_2006}, the axis on the BPT-IR diagram represent ratios of the different ionized states for the same element, {\em i.e.}, they do not show any dependence on the chemical abundances. Although pure AGN models (blue) do not cover the region where ULIRGs, LIRGs and LINERs fall, a significant fraction of our Seyfert sample are in agreement with AGN dominated models. We also represent the same models from \citet{Fernandez-Ontiveros_2016} for Dwarf Galaxies and SFG, and we obtain that they do not cover the region where our sample of AGN is located, implying that part of our sample shows both star-formation and AGN activity. Nevertheless, the detection Ne$^{4+}$ IR lines in these galaxies supports that AGN activity dominates our sample \citep{Genzel_1998, Armus_2007, Izotov_2012, Perez-Torres_2021}.


\section{Photoionization models and abundance estimations}
\label{sec3}
Chemical abundances (O/H and N/O) and ionization parameter ($U$) were derived  using an updated version of the Python code \textsc{HII-CHI-Mistry-IR}\footnote{All versions of the {\sc Hii-Chi-mistry} code are publicly available at: \url{http://www.iaa.csic.es/~epm/HII-CHI-mistry.html}.} \citepalias{Fernandez-Ontiveros_2021} from IR emission lines, adapting it to work for AGNs, by using the same models 
used for the optical version of the code, and described in \citetalias{Perez-Montero_2019}. We denote the optical version as \textsc{HCm}, while its infrared version presented here will be denoted as \textsc{HCm-IR}. We follow the methodology used by \citetalias{Fernandez-Ontiveros_2021}, taken into account some differences that arise when considering AGN instead of SFG models.

\subsection{Grids of AGN photoionization models}
\label{subsec31}

\textsc{HCm-IR} estimates chemical abundances (O/H and N/O) and $U$ using a Bayesian-like calculation, by  comparing certain observed  IR emission-line flux ratios with the corresponding values as predicted by large grids of photoionization models. The models were computed using \textsc{Cloudy} v17 \citep{Ferland_2017} and they are the same employed by \citetalias{Perez-Montero_2019}. In these models, the gas-phase is characterized by an  electronic density of $n_{e} = 500 $ cm$^{-3}$. 
All models assume a standard dust-to-gas mass ratio and the gas-phase abundances were scaled in each model to O following the solar 
photospheric proportions reported by \citet{Asplund_2009}, with the exception of N, which is considered as a free parameter in the grids for an independent estimation of the N/O ratio. 
The source of ionization is an AGN SED composed by a Big Blue Bump peaking at 1 Ryd and a power law for X-ray non-thermal emission characterized by $\alpha_{X} = -1.0$. The continuum between UV and X-ray ranges is represented by a power law with an index of $\alpha_{OX} = -0.8$ or $\alpha_{OX} = -1.2$. The filling factor was set to 0.1 while two stopping criteria were considered: a fraction of free electrons of 2$\%$ or 98$\% $. Thus, a total of 4 grids of photoionization models ({\em i.e.} assuming two different values for $\alpha_{OX}$ and two different stopping criteria) can be considered and selected by the user in an iterative process while running \textsc{HCm-IR}. Hereafter, we discuss the grid of models corresponding to $\alpha_{OX} = -0.8$ and a stopping criteria of 2$\%$ of free electrons. The effects of using different grids will be discussed in Sec. \ref{subsec35}.

The grids cover a range from 12+log(O/H) = 6.9 to 9.1 in steps of 0.1 dex, a range from log(N/O) = -2.0 to 0.0 in steps of 0.125 dex, and a range from log($U$) = -4.0 to -0.5 in steps of 0.25 dex. The behavior of some of the emission-line ratios used by the code shows a bivaluation in the [-4.0, -2.5] and [-2.5, -0.5] ranges. This behavior for some optical emission-line ratios was also discussed in \citetalias{Perez-Montero_2019} and reported by  \citet{Perez-Diaz_2021}. For this reason, the grids are constrained to certain $U$ ranges by considering two branches: the low-ionization branch which covers the [-4.0, -2.5) range and the high-ionization branch covering [-2.5, -0.5]. 

The code takes as input the following IR emission lines: \ion{H}{i}$\lambda $4.05$\mu$m, \ion{H}{i}$\lambda $7.46$\mu$m, [\ion{S}{iv}]$\lambda $10.5$\mu $m, \ion{H}{i}$\lambda $12.4$\mu$m, [\ion{Ne}{ii}]$\lambda $12.8$\mu$m, [\ion{Ne}{v}]$\lambda $14.3$\mu$m, [\ion{Ne}{iii}]$\lambda $15.6$\mu$m, [\ion{S}{iii}]$\lambda $18$\mu$m, [\ion{Ne}{v}]$\lambda $24$\mu$m, [\ion{O}{iv}]$\lambda $26$\mu$m,
[\ion{S}{iii}]$\lambda $33$\mu$m, [\ion{O}{iii}]$\lambda $52$\mu$m, [\ion{N}{ii}]$\lambda $57$\mu$m, [\ion{O}{iii}]$\lambda $88$\mu$m, [\ion{N}{ii}]$\lambda $122$\mu$m and [\ion{N}{ii}]$\lambda $205$\mu$m, which can be introduced in arbitrary units and not necessarily reddening-corrected. Since AGN models considered a higher electronic density than SFG models used by \citetalias{Fernandez-Ontiveros_2021} (AGN models assumed $n_{e} = 500$ cm$^{-3}$ while SFG $n_{e} = 100$ cm$^{-3}$), special attention must be paid to the critical density of the emission lines, which are much lower than those characterizing optical emission lines \citep[e.g][]{Osterbrock_book}. Considering different electronic temperatures T$_{e}$ characteristic of the ionic species in our models, we summarized in see Tab. \ref{Critical_densities} critical densities for all emission lines used as inputs with \textsc{Pyneb} (\citealt{Luridiana_2015}). Therefore, from the set of lines used by \textsc{HCm-IR} for SFG \citepalias{Fernandez-Ontiveros_2021}, we omitted three of them due to their relatively low critical densities when compared to the electronic density adopted for the models: [\ion{N}{ii}]$\lambda $205$\mu$m  [\ion{N}{ii}]$\lambda $122$\mu$m and [\ion{O}{iii}]$\lambda $88$\mu$m. The code only takes into account [\ion{O}{iii}]$\lambda $88$\mu$m when [\ion{O}{iii}]$\lambda $52$\mu$m is missing, although we warn that observed emission lines may deviate from predictions due to contributions of diffuse ionized gas (DIG), leading to uncertainties in the estimated chemical abundances.

\begin{table*}
	\caption{Computed critical densities (cm$^{-3}$) for fine structure IR lines from \textsc{Pyneb} for different electronic temperatures.}
	\label{Critical_densities}     
	\centering          
	\begin{tabular}{lllllll}
		\textbf{Emission line} & {\boldmath $T_{e} = 5000$ K} & {\boldmath $T_{e} = 10000$ K} & {\boldmath $T_{e} = 15000$ K} & {\boldmath $T_{e} = 20000$ K} & {\boldmath $T_{e} = 25000$ K} & {\boldmath $T_{e} = 30000$ K}  \\ \hline 
		{[}\ion{S}{iv}{]}$\lambda $10.5$\mu$m & 4.0$\cdot$10$^{4}$ & 5.6$\cdot$10$^{4}$ & 6.9$\cdot$10$^{4}$ & 7.9$\cdot$10$^{4}$ & 8.8$\cdot$10$^{4}$ & 9.6$\cdot$10$^{4}$ \\
		{[}\ion{Ne}{ii}{]}$\lambda $12.8$\mu$m & 4.6$\cdot$10$^{5}$ & 6.3$\cdot$10$^{5}$ & 7.4$\cdot$10$^{5}$ & 8.4$\cdot$10$^{5}$ & 9.2$\cdot$10$^{5}$ & 1.0$\cdot$10$^{6}$\\
		{[}\ion{Ne}{v}{]}$\lambda $14.3$\mu$m & 1.8$\cdot$10$^{4}$ & 3.2$\cdot$10$^{4}$ & 4.7$\cdot$10$^{4}$ & 6.3$\cdot$10$^{4}$ & 7.8$\cdot$10$^{4}$ & 9.3$\cdot$10$^{4}$\\
		{[}\ion{Ne}{iii}{]}$\lambda $15.6$\mu$m & 1.5$\cdot$10$^{5}$ & 2.1$\cdot$10$^{5}$ & 2.5$\cdot$10$^{5}$ & 2.8$\cdot$10$^{5}$ & 3.0$\cdot$10$^{5}$ & 2.9$\cdot$10$^{5}$\\
		{[}\ion{S}{iii}{]}$\lambda $18$\mu$m & 8173 & 1.2$\cdot$10$^{4}$ & 1.5$\cdot$10$^{4}$ & 1.7$\cdot$10$^{4}$ & 1.8$\cdot$10$^{4}$ & 2.0$\cdot$10$^{4}$\\
		{[}\ion{Ne}{v}{]}$\lambda $24$\mu$m & 3579 & 5952 & 8426 & 1.1$\cdot$10$^{4}$ & 1.3$\cdot$10$^{4}$ & 1.6$\cdot$10$^{4}$\\
		{[}\ion{O}{iv}{]}$\lambda $26$\mu$m & 8387 & 9905 & 1.2$\cdot$10$^{4}$ & 1.3$\cdot$10$^{4}$ & 1.4$\cdot$10$^{4}$ & 1.5$\cdot$10$^{4}$\\
		{[}\ion{S}{iii}{]}$\lambda $33$\mu$m & 980 & 1417 & 1801 & 2140 & 2425 & 2686 \\
		{[}\ion{O}{iii}{]}$\lambda $52$\mu$m & 2753 & 3530 & 3879 & 4082 & 4233 & 4418 \\
		{[}\ion{N}{iii}{]}$\lambda $57$\mu$m & 1180 & 1519 & 1723 & 1859 & 1938 & 1971 \\
		{[}\ion{O}{iii}{]}$\lambda $88$\mu$m & 388 & 501 & 569 & 620 & 662 & 704 \\
		{[}\ion{N}{ii}{]}$\lambda $122$\mu$m & 198 & 238 & 256 & 267 & 276 & 283 \\
		{[}\ion{N}{ii}{]}$\lambda $205$\mu$m & 30 & 38 & 44 & 47 & 51 & 54\\
	\end{tabular}      
\end{table*}

In addition, we omitted the emission line [\ion{S}{iii}]$\lambda $33$\mu$m, which is an input for the SFG version of the code. Despite being stronger than [\ion{S}{iii}]$\lambda $18$\mu$m, the introduction of this emission line in the calculations of the code leads to wrong estimations of both O/H and $U$ (see Sec. \ref{subsec34} for more details). Also in contrast with the input used for SFG, we consider [\ion{Ne}{v}] emission lines which are preferentially detected in infrared observations of galaxies hosting AGN \citep{Genzel_1998, Armus_2007, Izotov_2012, Perez-Torres_2021} and the [\ion{O}{iv}] emission line, characteristic of the AGN activity \citep{Melendez_2008, Rigby_2009, LaMassa_2010} since it can only be marginally produced in extremely high-ionized (logU $\gtrsim$  -1.5) star-forming regions.

Instead of matching single emission lines to predicted observations, \textsc{HCm-IR} uses particular emission-line ratios (listed below) to match observations and predictions. Then, the abundances O/H,  N/O, and $U$ are calculated following a $\chi^{2}$-methodology, being the mean of all input values of the models weighted by the quadratic sum of the differences between observations and predictions, the same methodology described in \citetalias{Perez-Montero_2019}. After the first iteration, N/O is fixed and the grid of models is constrained. Then, O/H and $U$ values are calculated in later iterations considering the already constrained model grids. When errors on the emission-line fluxes are provided, the code also takes them into account in  the final uncertainty of the estimations with a Monte Carlo simulation by perturbing the nominal input emission-line fluxes in the range delimited by their corresponding error.

\subsection{N/O estimation}
\label{subsec32}

\begin{figure*}
	\begin{tabular}{cccc}
		\begin{minipage}{0.05\hsize}\begin{flushright}\textbf{(a)} \end{flushright}\end{minipage}  &  \begin{minipage}{0.43\hsize}\centering{\includegraphics[width=1\textwidth]{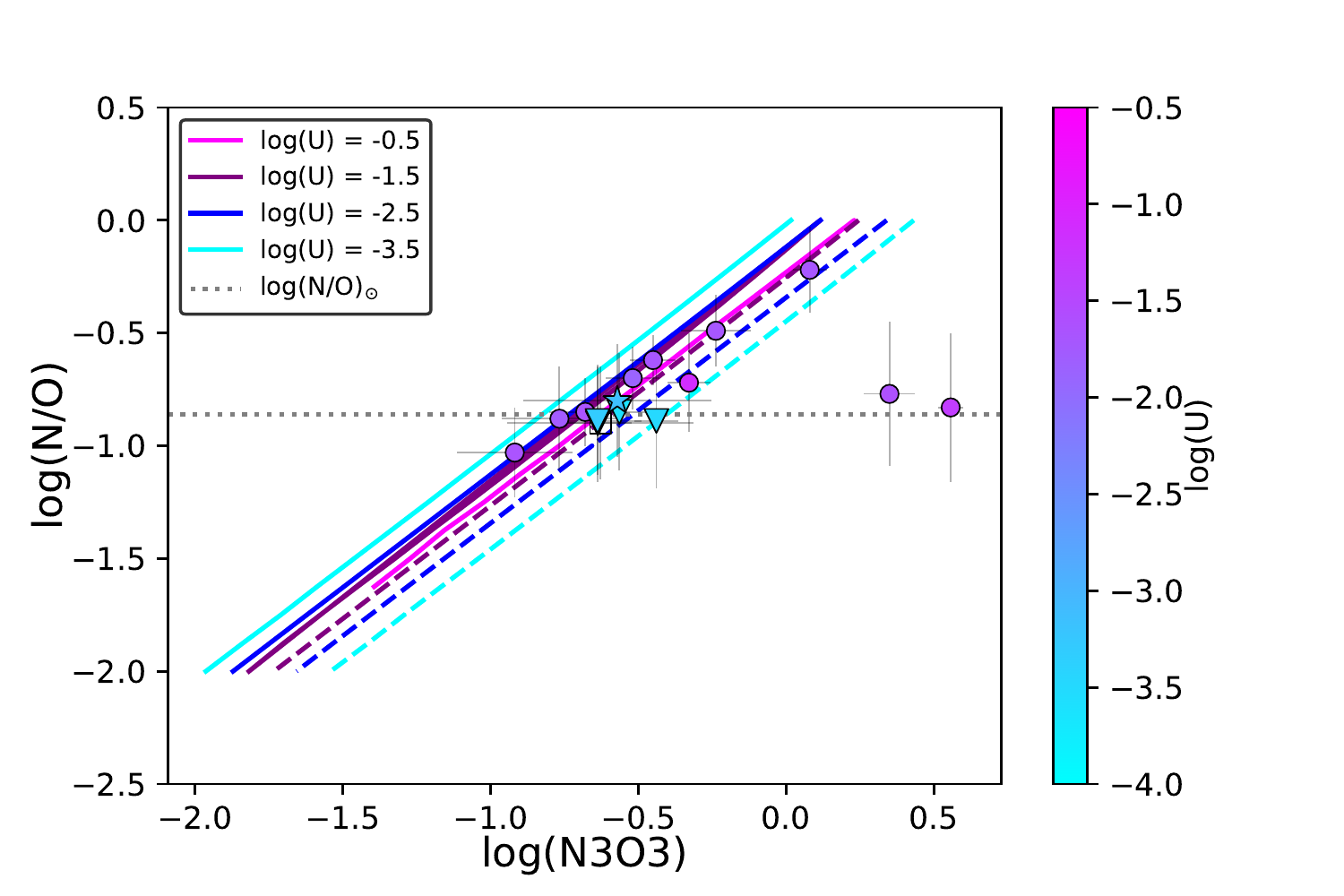}} \vspace{-0.2in} \end{minipage} & \begin{minipage}{0.05\hsize}\begin{flushright}\textbf{(b)} \end{flushright}\end{minipage}  &  \begin{minipage}{0.43\hsize}\centering{\includegraphics[width=1\textwidth]{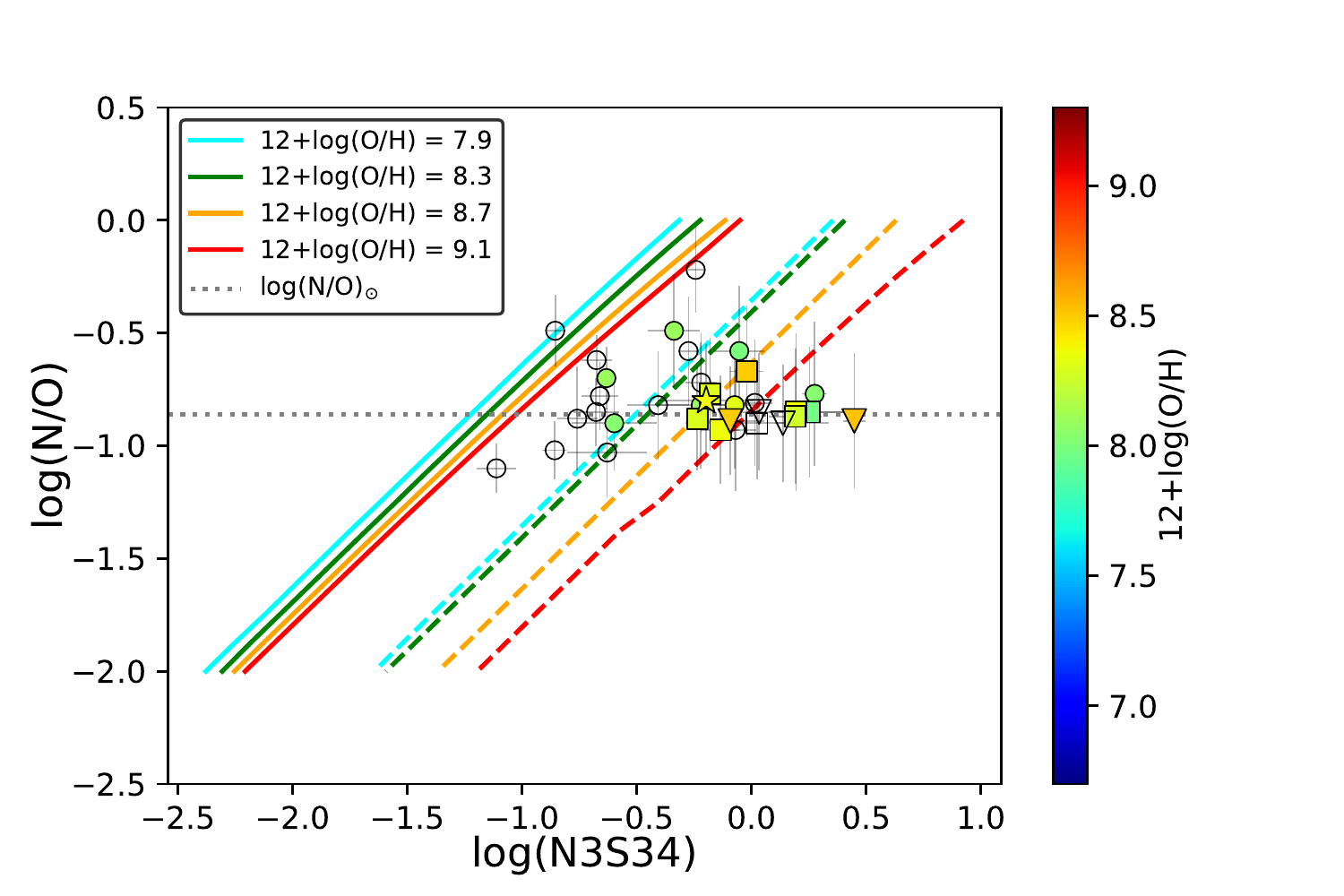}} \vspace{-0.2in} \end{minipage} 
	\end{tabular}
	\caption{Relations between different IR emission-line ratios with N/O . (a) Relation with N3O3 in our sample. Colorbar shows estimations of log$U$ . AGN models for a fixed value of 12+log(O/H) = 8.6 are presented as continuous lines while SFG models correspond to dashed lines for the same fixed value. (b) Relation with the N3S34 estimator in our sample. Colorbar shows estimations of 12+log(O/H). AGN models for a fixed value of log($U$) = -2.0 are presented as continuous lines while SFG models correspond to dashed lines for the same fixed value. For both plots: blank points indicate that no estimation can be provided of the colored quantity. The following spectral types are represented:  Seyferts 2 as circles; ULIRGs as squares; LIRGs as triangles; LINERs as stars.}
	\label{Fig1}
\end{figure*}

To estimate N/O, \textsc{HCm-IR} considers two emission-line ratios. The first one is N3O3, proposed by several authors \citep[e.g.][]{Nagao_2011, Pereira-Santaella_2017, Peng_2021, Fernandez-Ontiveros_2021}, an infrared analogue of the estimator N2O2 usually used in the optical range due to its effectiveness for both AGN and SFG \citep{Perez-Montero_2019} to derive the same abundance ratio \citep[e.g.][]{Perez-Montero_2009}.
The N3O3 is defined as:
\begin{equation}
\label{N3O3} \log \left( \mathrm{N3O3} \right) = \log \left( \frac{ \mathrm{I} \left(  \left[ \mathrm{\ion{N}{iii}} \right] _{57\mu m} \right) }{  \mathrm{I} \left( \left[ \mathrm{\ion{O}{iii}} \right] _{52\mu m} \right) }   \right)
\end{equation}
Our definition of N3O3 only takes into account [\ion{O}{iii}]$\lambda $52$\mu$m because the other [\ion{O}{iii}] IR emission line presents a critical density close to the electronic density of our models, in contrast to the same estimator defined for star-forming regions as described by \citetalias{Fernandez-Ontiveros_2021} which also takes into account [\ion{O}{iii}]$\lambda $88$\mu$m. When [\ion{O}{iii}]$\lambda $52$\mu$m is not provided, the code calculates N3O3 using [O\textsc{iii}]$\lambda $88$\mu$m. This modification is only taken into account when the code is applied for AGN, but not in the case of SFG since the grid of models for that case is calculated for $n_{e} \sim 100$ cm$^{-3}$ \citepalias{Fernandez-Ontiveros_2021}. As for its optical analogue, this estimator has little dependence on $U$ as shown in Fig. \ref{Fig1} (a). Moreover, our grid of AGN models and the grid of SFG show that this parameter is almost independent of the spectral type, {\em i.e.}, there is little distinction between AGN and SFG models.

The second emission-line ratio used by  the code to derive N/O is N3S34, which takes advantage on the primary origin of sulfur (as in the case of oxygen), defined as:
\begin{equation}
\label{N3S34} \log \left( \mathrm{N3S34} \right) = \log \left( \frac{ \mathrm{I} \left( \left[ \mathrm{\ion{N}{iii}} \right] _{57\mu m} \right) }{ \mathrm{I} \left( \left[ \mathrm{\ion{S}{iii}} \right] _{18\mu m} \right) + \mathrm{I} \left( \left[  \mathrm{\ion{S}{iv}} \right] _{10\mu m} \right) } \right) 
\end{equation}
This ratio also correlates with N/O,  presenting little dependence with O/H, as shown in Fig. \ref{Fig1} (b). Moreover, this tracer also presents a tight correlation with N/O when SFG models are considered. Thus, we have also added this estimator in the calculations of \textsc{HCm-IR} for SFG, with the particular difference that emission line [\ion{S}{iii}]$\lambda $33$\mu$m is also considered in SFG.

Considering the behavior of the AGN models in both Fig. \ref{Fig1} (a) and (b), we obtain the following linear calibrations using data from the photoionization models:

\begin{equation}
\label{fit_N3O3} \log \left( \mathrm{N/O} \right) = \left( 0.9839 \pm 0.0016 \right) \log \left( \mathrm{N3O3} \right) + \left( -0.1389 \pm 0.0015 \right)
\end{equation}
\begin{equation}
\label{fit_N3S34} \log \left( \mathrm{N/O} \right) = \left( 0.727 \pm 0.007 \right) \log \left( \mathrm{N3S34} \right) + \left( 0.326 \pm 0.013 \right)
\end{equation}
which can be used as alternative to the code to directly estimate N/O.

\begin{figure*}
	\begin{tabular}{cccc}
		\begin{minipage}{0.05\hsize}\begin{flushright}\textbf{(a)} \end{flushright}\end{minipage}  &  \begin{minipage}{0.43\hsize}\centering{\includegraphics[width=1\textwidth]{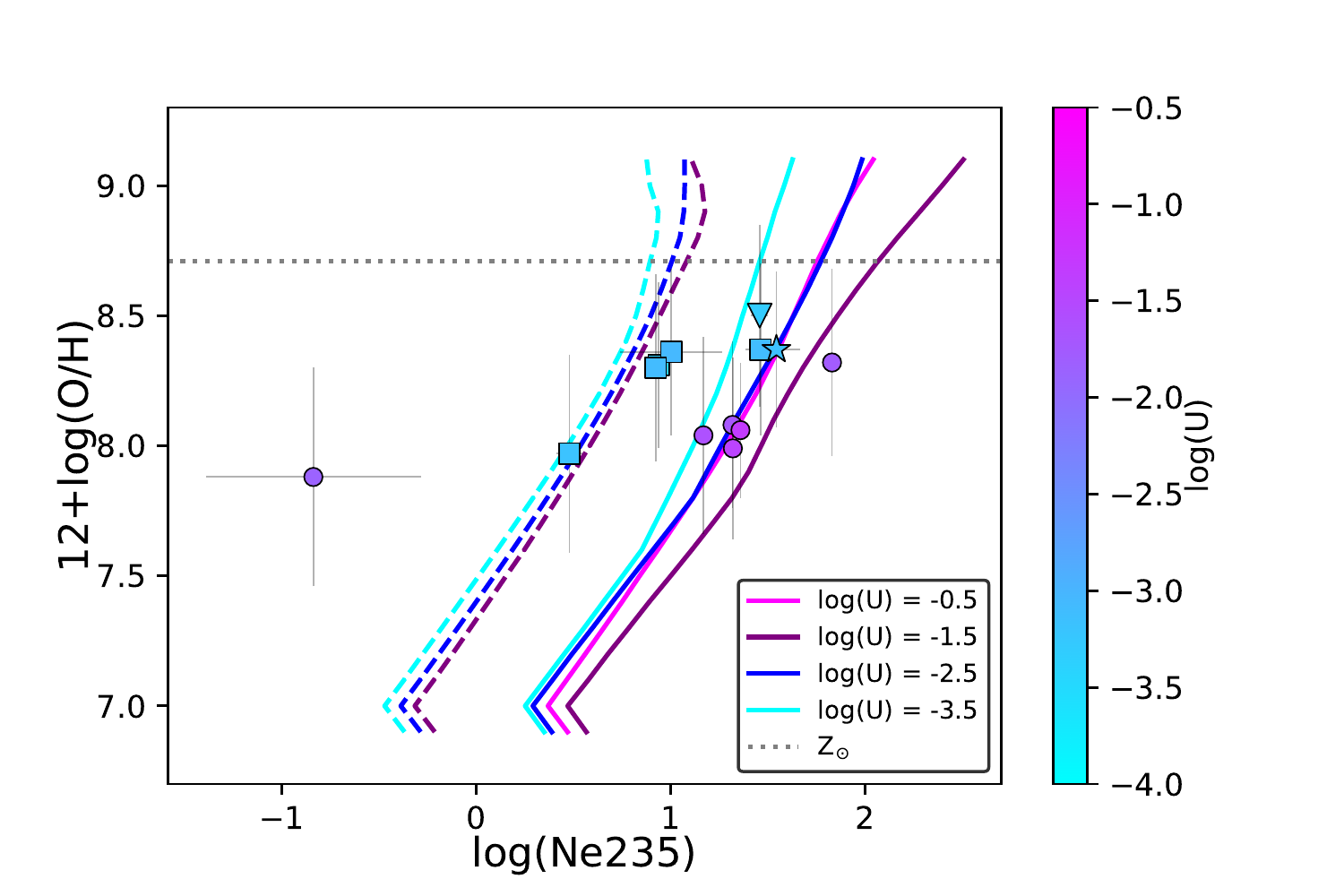}} \vspace{-0.2in} \end{minipage} & \begin{minipage}{0.05\hsize}\begin{flushright}\textbf{(b)} \end{flushright}\end{minipage}  &  \begin{minipage}{0.43\hsize}\centering{\includegraphics[width=1\textwidth]{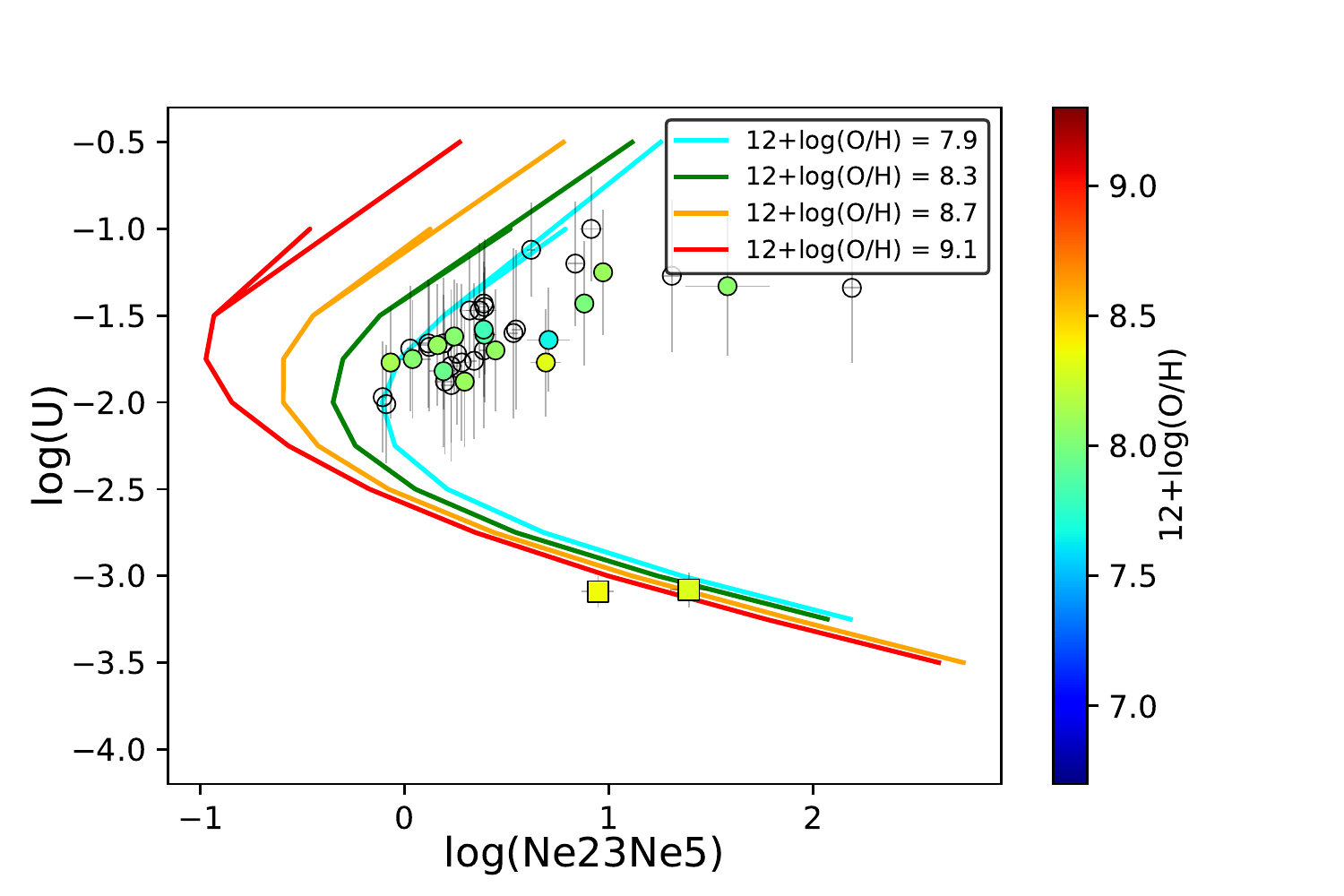}} \vspace{-0.2in} \end{minipage} 
	\end{tabular}
	\caption{Relations of different ratios involving Ne IR emission-lines with O/H and log($U$). (a) Relation between the  Ne235 estimator, using H\textsc{i} 4$\mu $m, and 12+ log(O/H) in our sample. Colorbar shows estimations of log($U$). (b) Relation between estimator Ne23Ne5 and log($U$) in our sample. Colorbar shows estimations of 12+log(O/H). For both plots: blank points indicate that no estimation can be provided of the colored quantity. AGN models for a fixed value of log(N/O) = -1.0 are presented as continuous lines while SFG models correspond to dashed lines for the same fixed value. The following spectral types are represented:  Seyferts 2 as circles; ULIRGs as squares; LIRGs as triangles; LINERs as stars.}
	\label{Fig2}
\end{figure*}

\begin{figure*}
	\begin{tabular}{cccc}
		\begin{minipage}{0.05\hsize}\begin{flushright}\textbf{(a)} \end{flushright}\end{minipage}  &  \begin{minipage}{0.43\hsize}\centering{\includegraphics[width=1\textwidth]{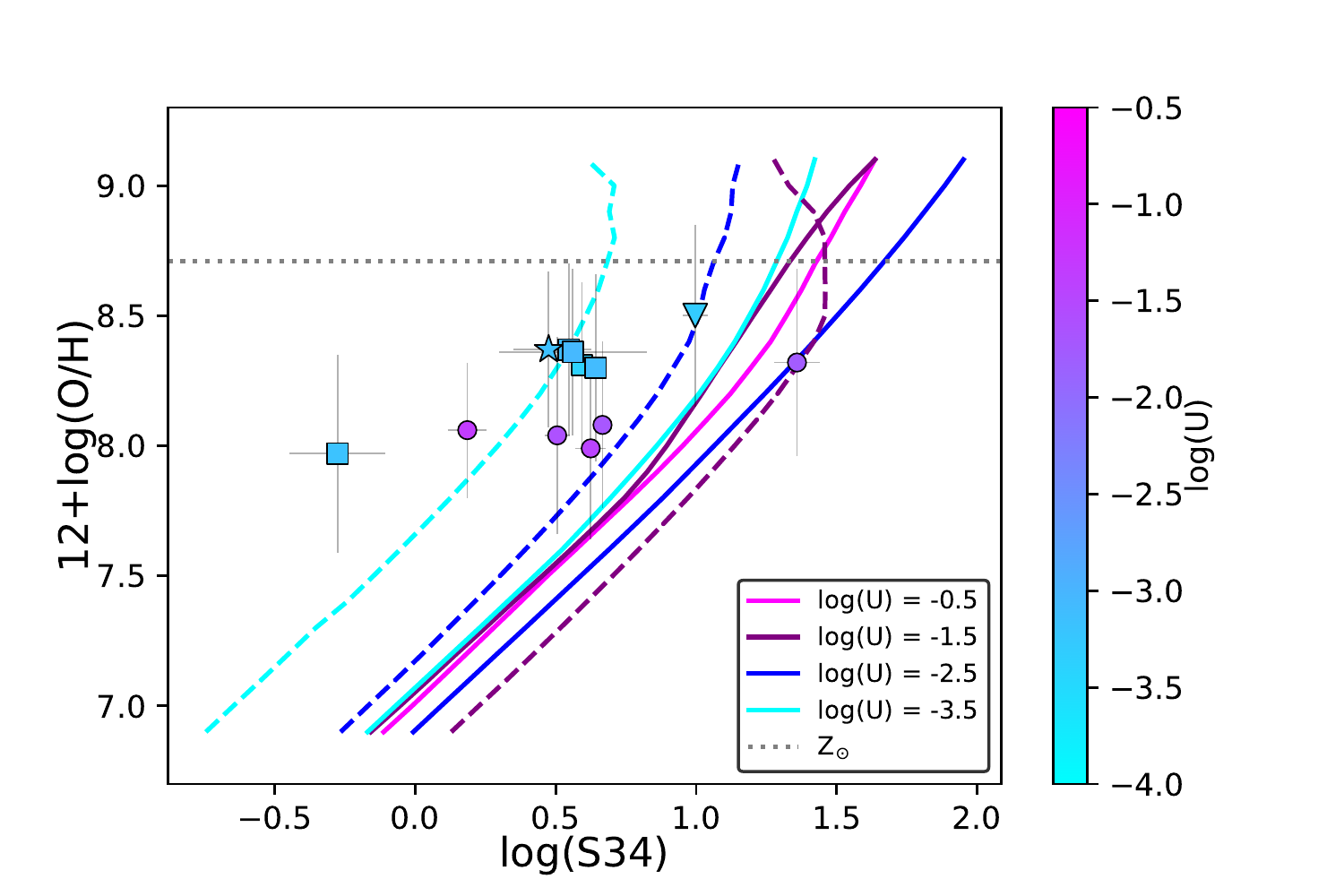}} \vspace{-0.2in} \end{minipage} & \begin{minipage}{0.05\hsize}\begin{flushright}\textbf{(b)} \end{flushright}\end{minipage}  &  \begin{minipage}{0.43\hsize}\centering{\includegraphics[width=1\textwidth]{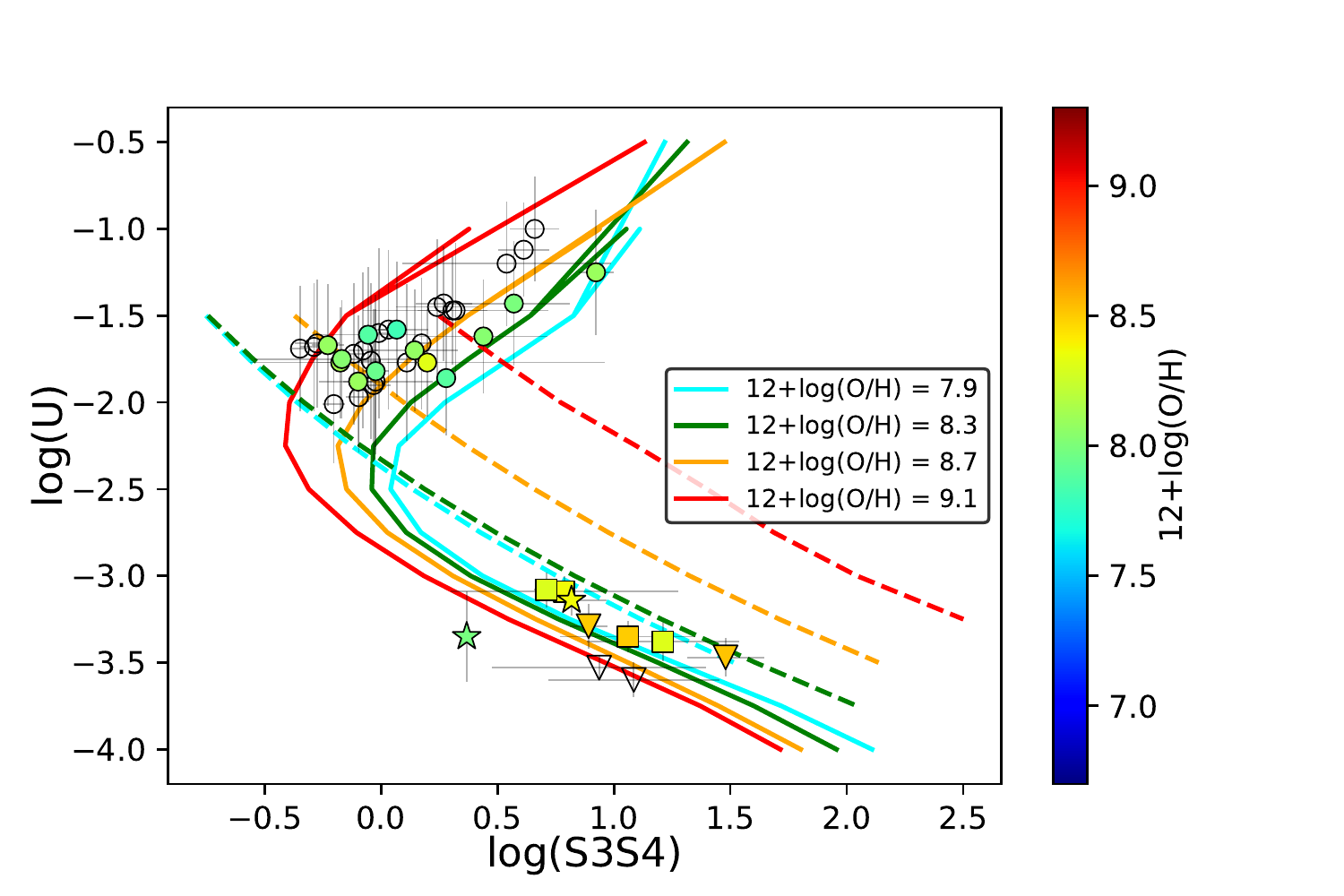}} \vspace{-0.2in} \end{minipage} 
	\end{tabular}
	\caption{Same as Fig. \ref{Fig2} but for S34 and S3S4.}
	\label{Fig3}
\end{figure*}

\subsection{O/H and $U$ estimations}
\label{subsec33}
Once determined  N/O in a first iteration, as this involves  emission-line ratios with little dependence on other input parameters, the code estimates in a second iteration both O/H and $U$ from a subgrid of models compatible with the previous estimation of N/O. Therefore, this guarantees that no previous assumption between O/H and N/O is introduced . In case N/O cannot be constrained due to the lack of key emission lines, a relation between O/H and N/O is assumed by the code. By default, this relation is that obtained by \citet{Perez-Montero_2014} for star-forming regions using chemical abundances based on  optical emission-lines. However, this relation can be modified by any user of the code adopting alternative laws within the corresponding libraries.

To estimate O/H and $U$, the code uses multiple emission-line ratios sensitive to the above quantities. One of them, is based on neon lines and it comes from a modification of the estimator Ne23 proposed by \citet{Kewley_2019} and \citetalias{Fernandez-Ontiveros_2021} to account for [\ion{Ne}{v}] lines that are more prominent in AGNs than in SFG. Then, accordingly, the estimator Ne235 for O/H can be defined as:
\begin{equation}
\label{Ne235} \begin{aligned} \log \left( \mathrm{Ne235} \right) &  = & \log \left( \frac{ \mathrm{I} \left( \left[ \mathrm{\ion{Ne}{ii}} \right] _{12\mu m} \right) +  \mathrm{I} \left( \left[ \mathrm{Ne\textsc{iii}} \right] _{15\mu m} \right) }{ \mathrm{I} \left( \mathrm{\ion{H}{i}}  _{i} \right) } + \right. \\ & & + \left. \frac{ \left[ \mathrm{I} \left( \mathrm{\ion{Ne}{v}} \right] _{14\mu m} \right) + \mathrm{I} \left( \left[ \mathrm{\ion{Ne}{v}} \right] _{24\mu m}  \right) }{  \mathrm{I} \left( \mathrm{\ion{H}{i}}  _{i} \right) }   \right)
\end{aligned}
\end{equation}
being $\mathrm{\ion{H}{i}}  _{i}$ one of the hydrogen lines that the code can take as input. In case that more than one of the hydrogen lines are introduced as input, \textsc{HCm-IR} calculates Ne235 for each hydrogen line, taking all considered ratios in the resulting weighted-distribution. In addition, these same neon lines are used to estimate log($U$) from the ratio Ne23Ne5 defined as:

\begin{equation}
\label{Ne23Ne5} \log \left( \mathrm{Ne23Ne5} \right) = \log \left(  \frac{ \mathrm{I} \left( \left[ \mathrm{\ion{Ne}{ii}} \right] _{12\mu m} \right) + \mathrm{I} \left( \left[ \mathrm{\ion{Ne}{iii}} \right] _{15\mu m} \right)   }{  \mathrm{I} \left( \left[ \mathrm{\ion{Ne}{v}} \right] _{14\mu m} \right) + \left[ \mathrm{I} \left( \mathrm{\ion{Ne}{v}} \right] _{24\mu m} \right) }  \right)
\end{equation}
which is a modification of the estimator Ne2Ne3 proposed by several authors \citepalias{Thornley_2000, Yeh_2012, Kewley_2019, Fernandez-Ontiveros_2021} to account for high ionic species of Ne which are found in AGN. Fig. \ref{Fig2} (a) shows the behavior of Ne235 with 12+log(O/H). There is little dependence with $U$ and there is clear separation between SFG and AGN models, which is explained with the little capacity of SFG models to produce [Ne\textsc{v}]. Fig. \ref{Fig2} (b) shows the relation between Ne23Ne5 and log($U$). The behavior of the models, well reproduced by the estimations in our sample, clearly shows the bi-valuation that forces the code to distinguish between low and high ionization AGN. For the former ones, little dependence is found with O/H, while  it has a more significant impact for the upper branch. The lack of SFG models in this figure justifies the omission of [Ne\textsc{v}] lines in SFG estimators.

Another set of IR emission lines that \textsc{HCm-IR} uses to estimate O/H and $U$ are the sulfur lines. To calculate both quantities, the code uses the estimators S34 \citepalias{Fernandez-Ontiveros_2021} and S3S4 \citepalias{Yeh_2012, Fernandez-Ontiveros_2021} respectively, defined as:

\begin{equation}
\label{S34} \log \left( \mathrm{S34} \right) = \log \left( \frac{ \mathrm{I} \left( \left[ \mathrm{\ion{S}{iii}} \right] _{18\mu m} \right) + \mathrm{I} \left( \left[ \mathrm{\ion{S}{iv}} \right] _{10\mu m} \right) }{  \mathrm{I} \left( \mathrm{\ion{H}{i}}_{i} \right) }  \right)
\end{equation}
\begin{equation}
\label{S3S4} \log \left( \mathrm{S3S4} \right) = \log \left( \frac{ \mathrm{I} \left( \left[ \mathrm{\ion{S}{iii}} \right] _{18\mu m} \right) }{  \mathrm{I} \left( \left[ \mathrm{\ion{S}{iv}} \right] _{10\mu m} \right) }  \right)
\end{equation}
Our definitions of these two estimators differ from those used in  \citetalias{Fernandez-Ontiveros_2021} since we omit [\ion{S}{iii}]$\lambda $33$\mu$m in our calculations. Fig. \ref{Fig3} (a) shows that S34 correlates with 12+log(O/H) and has also little dependence with $U$, although in this case there is no clear separation between SFG and AGN models. Fig. \ref{Fig3} (b) reinforces our preliminary statement in the need of distinguishing between low and high ionization AGNs due the bi-valuation of log($U$) with S3S4. We also observed that for low ionization parameters (log($U$) $<$ -2.5), the behavior of AGN and SFG models is similar, although they cover different regions of the diagram.

\begin{figure*}
	\begin{tabular}{cccc}
		\begin{minipage}{0.05\hsize}\begin{flushright}\textbf{(a)} \end{flushright}\end{minipage}  &  \begin{minipage}{0.43\hsize}\centering{\includegraphics[width=1\textwidth]{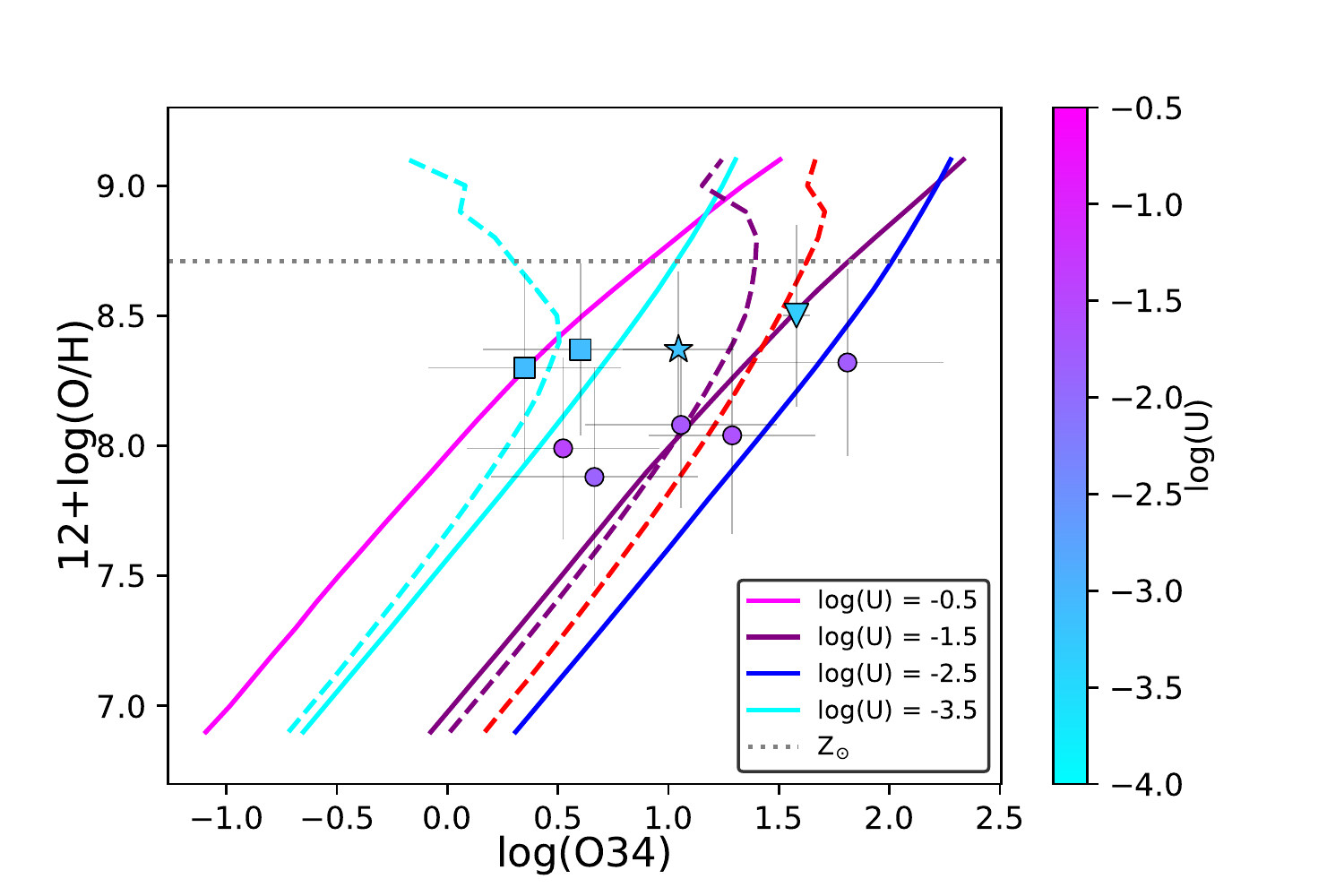}} \vspace{-0.2in} \end{minipage} & \begin{minipage}{0.05\hsize}\begin{flushright}\textbf{(b)} \end{flushright}\end{minipage}  &  \begin{minipage}{0.43\hsize}\centering{\includegraphics[width=1\textwidth]{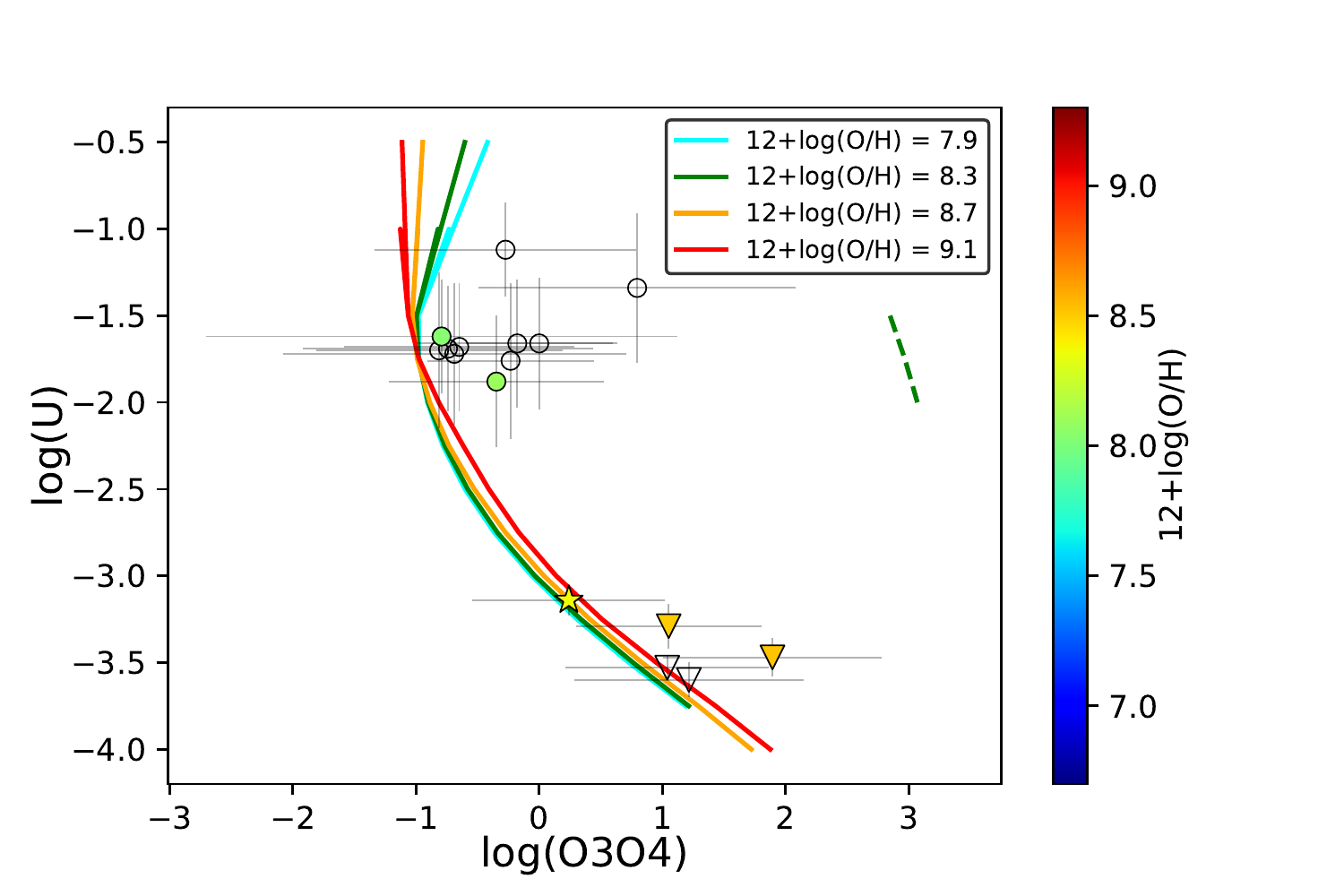}} \vspace{-0.2in} \end{minipage} 
	\end{tabular}
	\caption{Same as Fig. \ref{Fig2} but for O34 and O3O4.}
	\label{Fig4}
\end{figure*}

In the same fashion as we proceed with the sulfur lines, the code takes into account estimators based on IR oxygen lines. We define O34 and O3O4 to estimate 12+log(O/H) and log($U$) respectively as:

\begin{equation}
\label{O34} \log \left( \mathrm{O34} \right) = \log \left( \frac{ \mathrm{I} \left( \left[ \mathrm{\ion{O}{iii}} \right] _{52\mu m} \right) +  \mathrm{I} \left( \left[ \mathrm{\ion{O}{iv}} \right] _{26\mu m}  \right)  }{ \mathrm{I} \left( \mathrm{\ion{H}{i}}_{i} \right) }  \right)
\end{equation}
\begin{equation}
\label{O3O4} \log \left( \mathrm{O3O4} \right) = \log \left( \frac{ \mathrm{I} \left( \left[ \mathrm{\ion{O}{iii}} \right] _{52\mu m} \right) }{ \mathrm{I} \left( \left[ \mathrm{\ion{O}{iv}} \right] _{26\mu m} \right) }  \right)
\end{equation}
Here again we have omitted the use of [\ion{O}{iii}]$\lambda $88$\mu$m due to its very low critical density. However, if [\ion{O}{iii}]$\lambda $52$\mu$m is not provided, the code calculates both estimators O34 and O3O4 with [\ion{O}{iii}]$\lambda $88$\mu$m. Fig. \ref{Fig4} (a) shows that O34 has a strong dependence with $U$. However, O3O4 does not show any dependence with O/H (see Fig. \ref{Fig4} (b)), so it can be used to constrain $U$ and estimate 12+log(O/H) with O34. In addition, Fig. \ref{Fig4} (b) shows that this estimator might be employed in SFG but few models are able to produce [\ion{O}{iv}]$\lambda $26$\mu$m. Nevertheless, these ratios have been also implemented to the SFG version of the code \citepalias{Fernandez-Ontiveros_2021} as they can further constrained SFG models to account for the presence of [\ion{O}{iv}]$\lambda $26$\mu$m, only found for a very reduced small number of models characterized by hard radiation fields (log($U$) > -1.5). Analogous results are obtained if  [\textsc{Oiii}]$\lambda $88$\mu$m emission line is considered.  

\begin{figure}
	\centering
	\includegraphics[width=0.86\hsize]{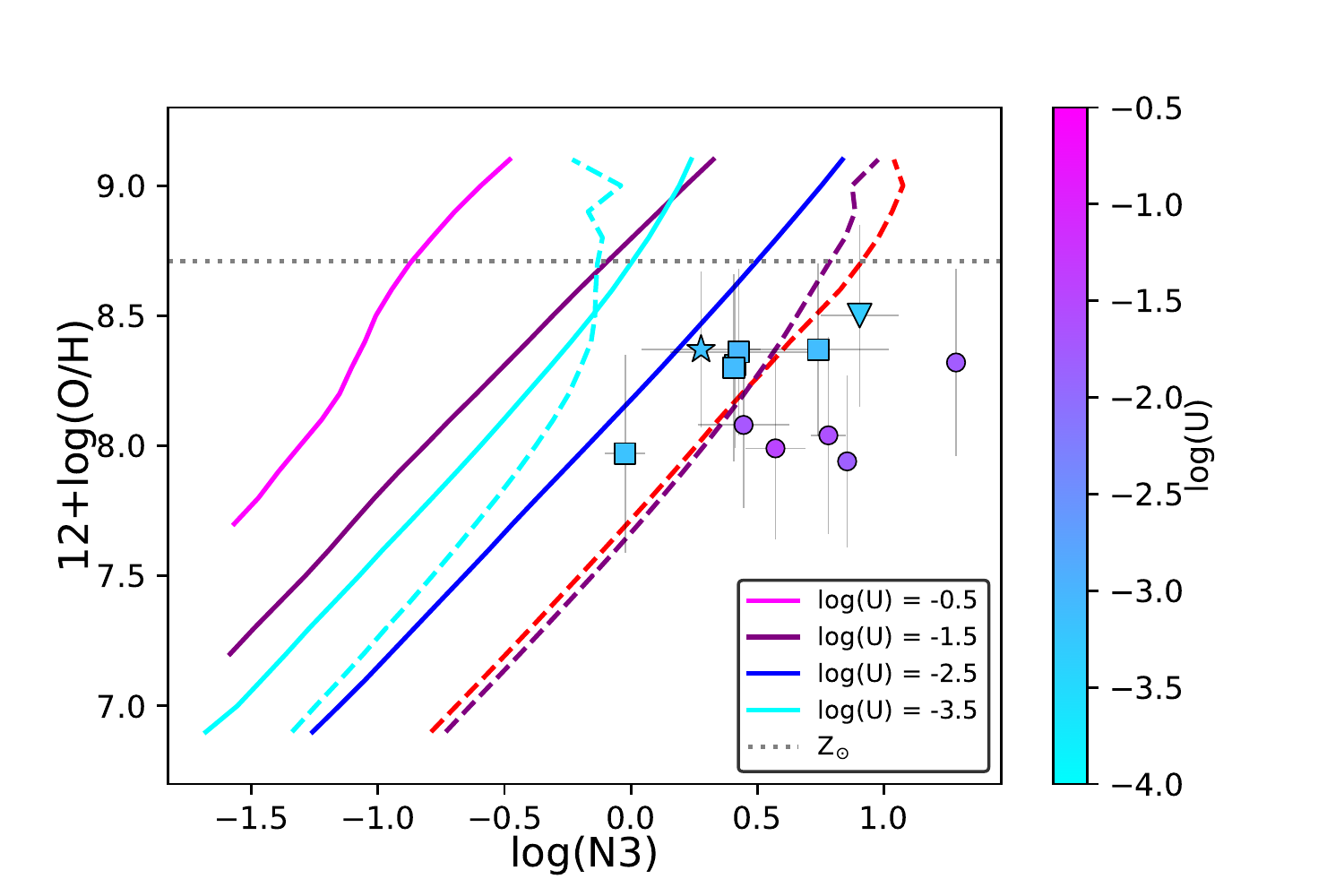} 
	\caption{Same as Fig. \ref{Fig2} (a) but for N3.}
	\label{Fig5}
\end{figure}

Although \citetalias{Fernandez-Ontiveros_2021} considered IR N lines to estimate both O/H and $U$ in SFGs,  both estimators (N23 and N2N3, \citetalias{Nagao_2011, Kewley_2019, Fernandez-Ontiveros_2021}) imply the use of [\ion{N}{ii}], whose critical density is below the electronic density of our models. Therefore, we define the estimator N3 based only on the [\ion{N}{iii}]57$\mu $m line as:
\begin{equation}
\label{N3} \log \left( \mathrm{N3} \right) = \log \left( \frac{  \mathrm{I} \left( \left[ \mathrm{\ion{N}{iii}} \right] _{57\mu m} \right)   }{ \mathrm{I} \left( \mathrm{\ion{H}{i}}  _{i} \right) }  \right)
\end{equation}
which can be used since N/O has already been constrained in a first iteration. However, as in the case of O34, Fig. \ref{Fig5} shows that this estimator also depends on $U$, although for a fixed value of log($U$) N3 shows a tight correlation with 12+log(O/H).

It is important to notice that we cannot consider neither the estimator O3N2 defined by \citetalias{Fernandez-Ontiveros_2021}  as the ratio between [\ion{O}{iii}] and [\ion{N}{ii}] IR lines due to their very low critical densities. Therefore, there is no possible estimation of 12+log(O/H) if none of the three IR H lines (\ion{H}{i}$\lambda$4.05$\mu$m, \ion{H}{i}$\lambda$7.46$\mu$m or \ion{H}{i}$\lambda$12.4$\mu$m) is provided.

Although we have defined estimators based on the most common observed IR emission lines from \textit{Spitzer}, \textit{Herschel} or ISO, additional estimators and modifications will be introduced to the code to account for more spectral lines as new and better resolved spectroscopic data will be released from the upcoming IR missions such as JWST. For instance, emission line [\ion{Ne}{vi}]$\lambda $7.7$\mu $m, which is now unresolved due to low resolution in the near-IR, or fainter emission lines as [\ion{Ar}{ii}]$\lambda $7$\mu$m or [\ion{Ar}{iii}]$\lambda $9$\mu $m will be accessible from JWST. Nevertheless, with the current available IR data is not possible to check the validity of their use, so this will be discussed when new data will be released.

\subsection{Subsets of emission lines}
\label{subsec34}
Although  \textsc{HCm-IR}, as described in the previous section, can take as input a set of IR emission lines in order to estimate chemical abundances and ionization, through the appropriate emission-line ratios (Eq. \ref{N3O3}-\ref{N3}), the code can also reach to a solution   with a small subset of the input emission lines. Nevertheless, the capability of the code to find an accurate solution will depend on the available emission lines used as an input. For instance, if no measurement of [\ion{N}{iii}]$\lambda $57$\mu$m is provided, then the code will be unable to calculate N/O since both estimators (N3O3 and N3S34) involve this emission line. In the case of estimating 12+log(O/H), it is necessary to provide one of the three IR hydrogen recombination lines.

In this section we explore the results from \textsc{HCm-IR} when different sets of emission lines are introduced as inputs. To compare the estimations from the code with reliable results, we use as input emission lines from the models, whose chemical abundances and $U$ are known, by randomly perturbing at 10$\%$ the flux of the lines, simulating observational uncertainties.

\begin{table*}
	\caption{Median offsets and RSME of the residuals between theoretical abundances and log($U$) values (AGN model inputs) and estimations from \textsc{HCm-IR} using different sets of emission lines.}
	\label{Offsets_theoretical}     
	\centering          
	\begin{tabular}{lllllll}
		\textbf{Set of lines} & \boldmath$\Delta_{OH} $ & \boldmath$\mathrm{RSME}_{OH}$ & \boldmath$\Delta_{NO} $ & \boldmath$\mathrm{RSME}_{NO}$ & \boldmath$\Delta_{U} $ & \boldmath$\mathrm{RSME}_{U}$\\ \hline 
		All lines & -0.01 & 0.24 & 0.04 & 0.13 & -0.05 & 0.24 \\ \hline
		\ion{H}{i}$\lambda $4.05$\mu$m, \ion{H}{i}$\lambda $7.46$\mu$m, \ion{H}{i}$\lambda $12.4$\mu$m, & & & & & & \\ 
		{[}\ion{Ne}{ii}{]}$\lambda $12.8$\mu$m, [\ion{Ne}{v}]$\lambda $14.3$\mu$m, & -0.01 & 0.17 &     - &     - & -0.16 & 0.36\\
		{[}\ion{Ne}{iii}{]}$\lambda $15.6$\mu$m, [\ion{Ne}{v}]$\lambda $24$\mu$m & & & & & & \\ \hline
		\ion{H}{i}$\lambda $4.05$\mu$m, \ion{H}{i}$\lambda $7.46$\mu$m,  & & & & & & \\ \ion{H}{i}$\lambda $12.4$\mu$m,  [\ion{O}{iv}]$\lambda $26$\mu$m, & 0.07 & 0.29 & 0.03 & 0.08 & -0.10 & 0.40\\ {[}\ion{O}{iii}{]}$\lambda $52$\mu$m, [\ion{N}{iii}]$\lambda $57$\mu$m  & & & & & & \\ \hline
		\ion{H}{i}$\lambda $4.05$\mu$m, \ion{H}{i}$\lambda $7.46$\mu$m, & & & & & & \\ \ion{H}{i}$\lambda $12.4$\mu$m, [\ion{S}{iv}]$\lambda $10$\mu$m, & -0.03 & 0.21 & 0.02 & 0.30 & -0.07 & 0.27\\ {[}\ion{S}{iii}{]}$\lambda $18$\mu$m, [\ion{N}{iii}]$\lambda $57$\mu$m & & & & & & \\ \hline
		{[}\ion{Ne}{ii}{]}$\lambda $12.8$\mu$m, [\ion{Ne}{v}]$\lambda $14.3$\mu$m, & & & & & & \\ {[}\ion{Ne}{iii}{]}$\lambda $15.6$\mu$m, [\ion{Ne}{v}]$\lambda $24$\mu$m, [\ion{S}{iv}]$\lambda $10$\mu$m, &     - &     - & 0.02 & 0.31 & -0.05 & 0.20\\  {[}\ion{S}{iii}{]}$\lambda $18$\mu$m, [\ion{N}{iii}]$\lambda $57$\mu$m  & & & & & & \\ \hline
		\ion{H}{i}$\lambda $4.05$\mu$m, \ion{H}{i}$\lambda $7.46$\mu$m, H\textsc{i}$\lambda $12.4$\mu$m, & & & & & & \\		
		{[}\ion{S}{iv}{]}$\lambda $10$\mu$m, {[}\ion{Ne}{v}{]}$\lambda $14$\mu$m, {[}\ion{Ne}{v}{]}$\lambda $24$\mu$m, & 0.24 & 0.41 & -0.44 & 0.56 & -0.09 & 0.53\\
		{[}\ion{O}{iv}{]}$\lambda $26$\mu$m, [\ion{N}{iii}]$\lambda $57$\mu $m & & & & & & \\ \hline
		All lines + [\ion{S}{iii}]$\lambda $33$\mu$m & 0.14 & 0.48 & 0.06 & 0.15 & -0.12 & 0.58\\ 
		
	\end{tabular}      
\end{table*}

In Tab. \ref{Offsets_theoretical} we present the statistics of the residuals between the input values  used in the models and the corresponding predictions from \textsc{HCm-IR}. When all the possible emission lines of a set are used as input, we
obtain low median offsets for 12+log(O/H), log(N/O) and log($U$).
Since we introduce a $10\% $ of uncertainty in the emission-line fluxes, considering the error propagation in the involved line ratios, the values of RSME for each quantity are compatible with the uncertainty carried in the estimation. Moreover, considering the steps $\sigma $ of the grid (0.1 dex for O/H; 0.125 dex for N/O; 0.25 dex for log($U$), the RSME are in either case below 3$\sigma $.

If only high ionized emission lines ([\ion{S}{iv}], [\ion{O}{iv}], [\ion{Ne}{v}]) are introduced as inputs, systematic offsets appear for all three quantities. Although low ionized emission lines are key to estimate $U$, we obtain a significant offset even for N/O estimation ($\Delta $N/O $\sim $ -0.44 dex), since the code is assuming a relation between N/O and O/H as there is no independent estimation of both quantities. Overall, we conclude that the best estimations for O/H and $U$ are obtained when neon or sulfur lines are used, similar to what is obtained for SFG \citepalias{Fernandez-Ontiveros_2021}. 
Analyzing N/O, best estimations involved oxygen emission line [O\textsc{iii}]$\lambda $52$\mu $m ({\em i.e.} estimator N3O3), since using only sulfur lines lead to higher dispersion (RSME $\sim $ 0.3 dex).

\begin{figure}
	\begin{tabular}{cc}
		\begin{minipage}{0.05\hsize}\begin{flushright}\textbf{(a)} \end{flushright}\end{minipage}  &  \begin{minipage}{0.9\hsize}\centering{\includegraphics[width=1\textwidth]{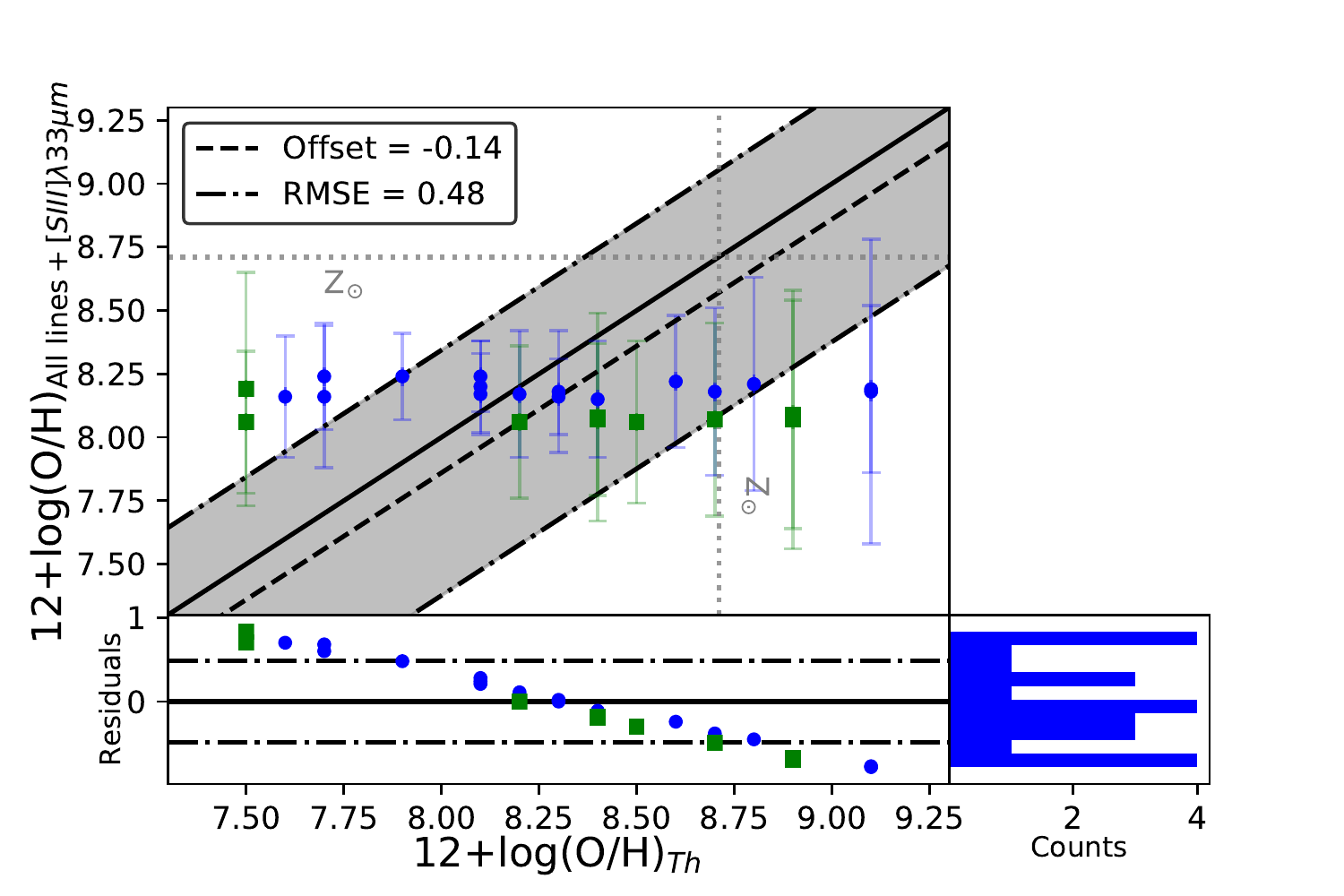}} \vspace{-0.2in} \end{minipage} \\ \begin{minipage}{0.05\hsize}\begin{flushright}\textbf{(b)} \end{flushright}\end{minipage}  &  \begin{minipage}{0.9\hsize}\centering{\includegraphics[width=1\textwidth]{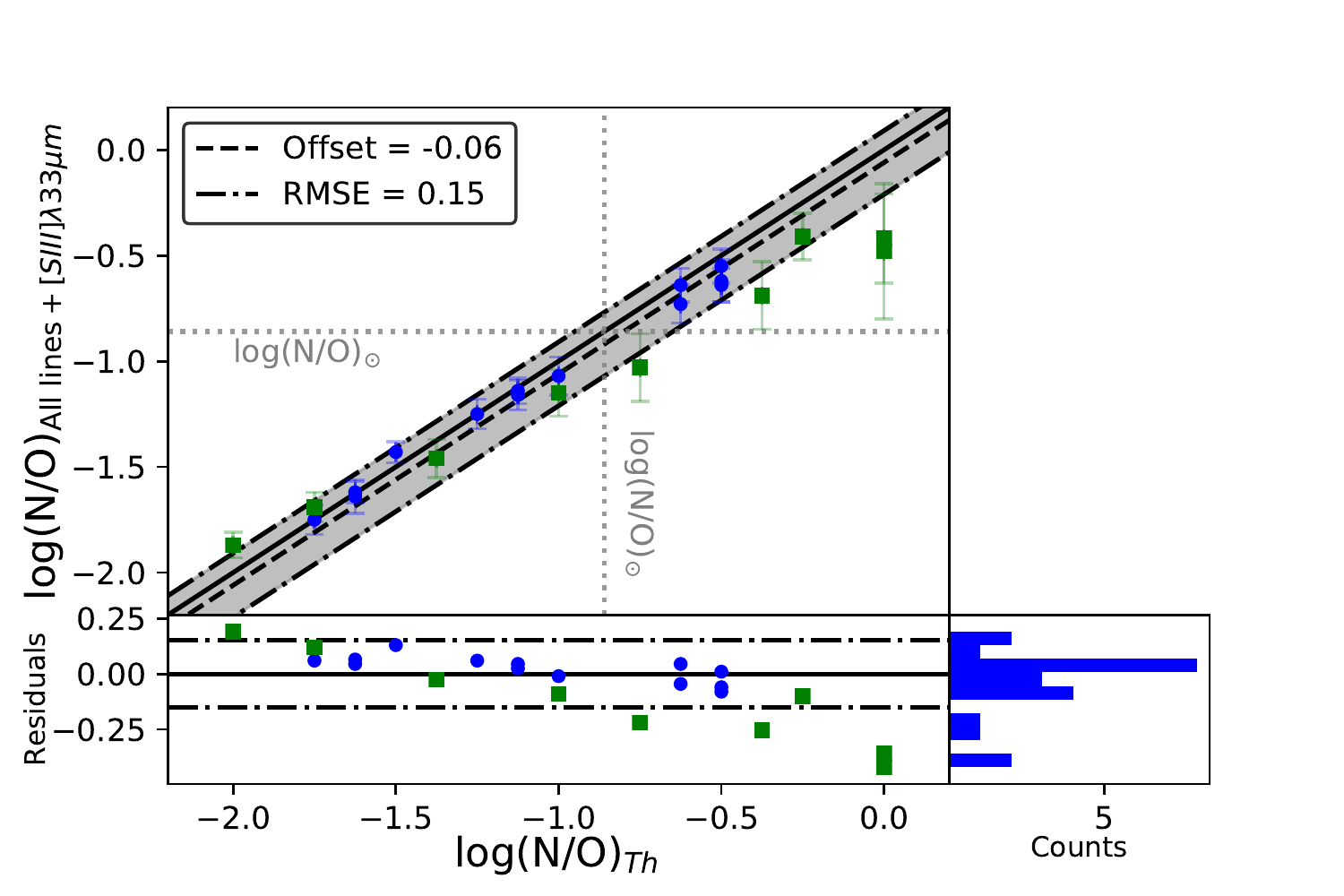}} \vspace{-0.2in} \end{minipage} \\
		\begin{minipage}{0.05\hsize}\begin{flushright}\textbf{(c)} \end{flushright}\end{minipage}  &  \begin{minipage}{0.9\hsize}\begin{flushleft}\includegraphics[width=1\textwidth]{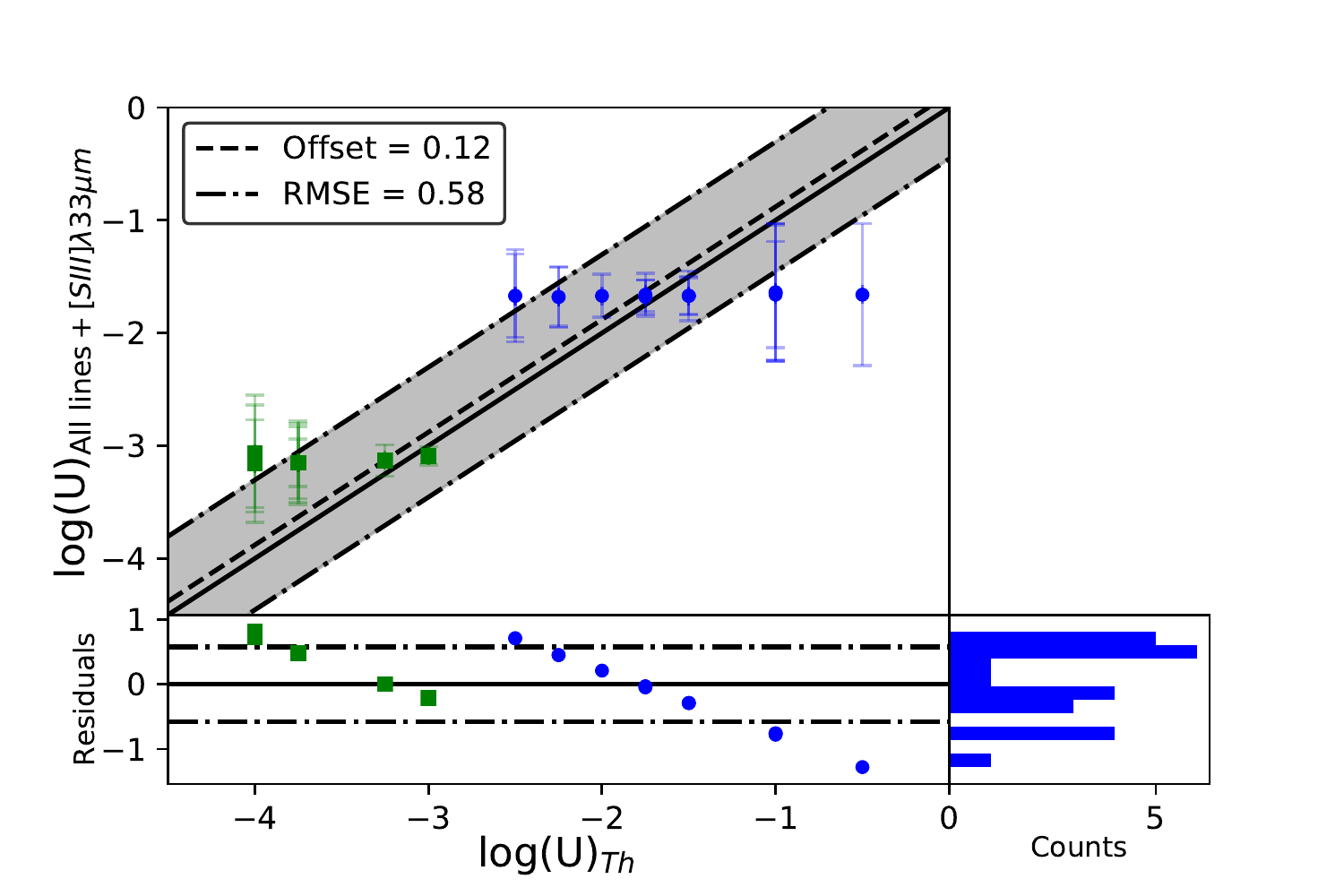}\end{flushleft} \vspace{-0.2in} \end{minipage} \\		 
	\end{tabular}
	\caption{Comparison between the  chemical abundances and $U$ values introduced as inputs for the models (x-axis) with the  estimations from \textsc{HCm-IR} when all lines plus [\ion{S}{iii}]$\lambda $33$\mu $m is used (y-axis). For all plots we present: Seyferts as blue circles and LINERs as green squares. The offsets are given using the median value (dashed line) and RMSE (dot-dashed lines). Bottom plots show the residuals from the offset and their distribution in a histogram (bottom-right plot).}
	\label{Fig6}
\end{figure}

Tab. \ref{Offsets_theoretical} also shows the reason why emission line [\ion{S}{iii}]$\lambda $33$\mu $m was omitted from the calculations: the offsets and RSME for O/H and $U$ increase even if we consider all set of emission lines, leading to wrong estimations. Moreover, Fig. \ref{Fig6} clearly shows that while predictions of N/O fit well with inputs for models, the second iteration of the code to estimate O/H and $U$ shows an almost constant behavior: values of O/H cluster around 8.2 while values of U cluster around -1.7 for high-ionization AGNs and -3.1 for low-ionization AGNs. Thus, this emission line was omitted in the code.

\subsection{Selection of the grid}
\label{subsec35}
Although we used the grid of AGN models computed from a SED characterized by $\alpha_{OX} = -0.8$ and selecting an stopping criteria of 2$\% $ fraction of free electrons, \textsc{HCm-IR} provides more default grids where $\alpha_{OX}$ can change from -0.8 to -1.2 and the fraction of free electrons from 2$\% $ to 98$\% $. Moreover, we included in the last update a new feature for the users to introduce any grid of models. In this section, we explore the effects of changing the default grid of models to estimate chemical abundances from IR emission lines.

\begin{figure*}
	\begin{tabular}{cccc}
		\begin{minipage}{0.05\hsize}\begin{flushright}\textbf{(a)} \end{flushright}\end{minipage}  &  \begin{minipage}{0.43\hsize}\centering{\includegraphics[width=1\textwidth]{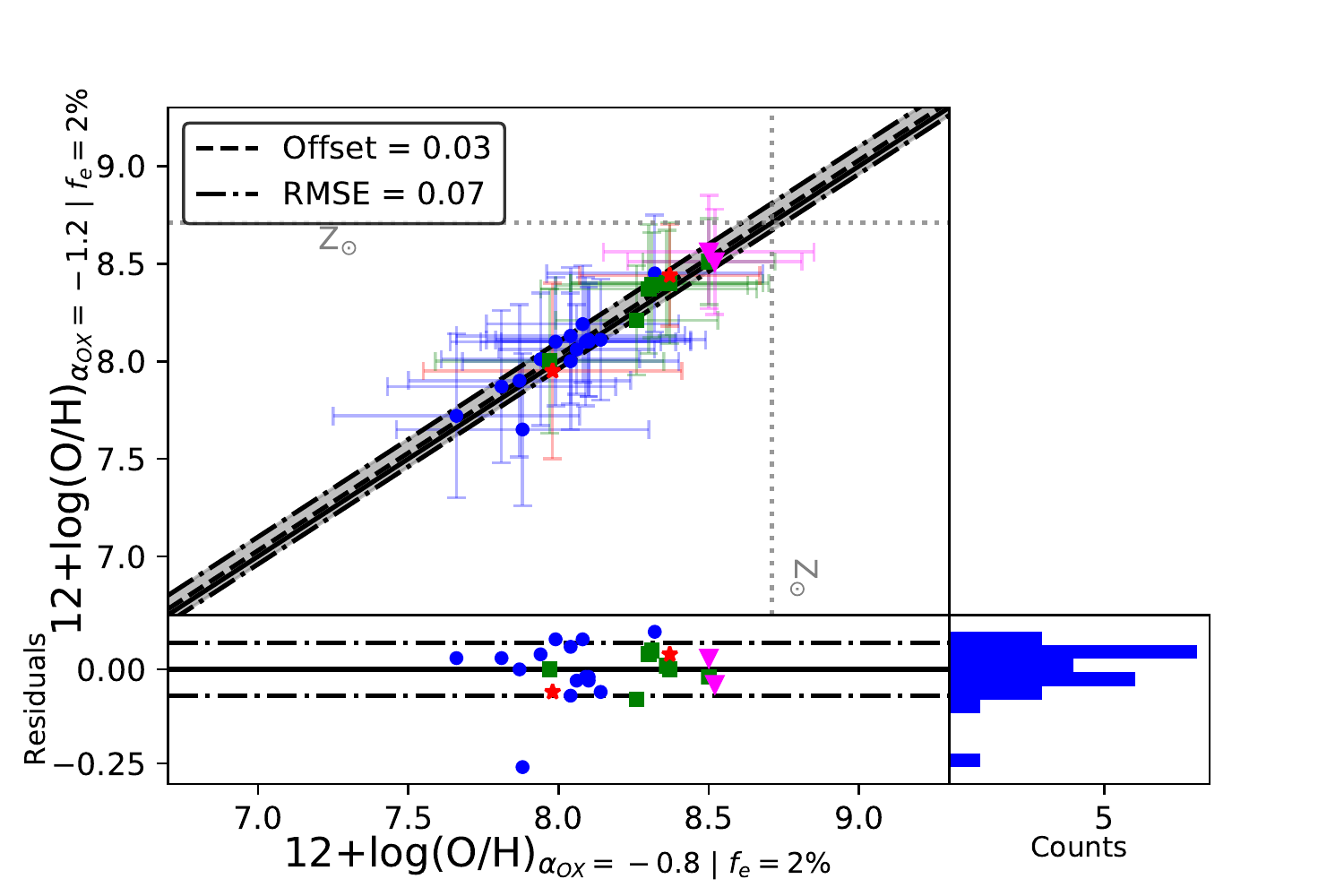}} \vspace{-0.2in} \end{minipage} & \begin{minipage}{0.05\hsize}\begin{flushright}\textbf{(b)} \end{flushright}\end{minipage}  &  \begin{minipage}{0.43\hsize}\centering{\includegraphics[width=1\textwidth]{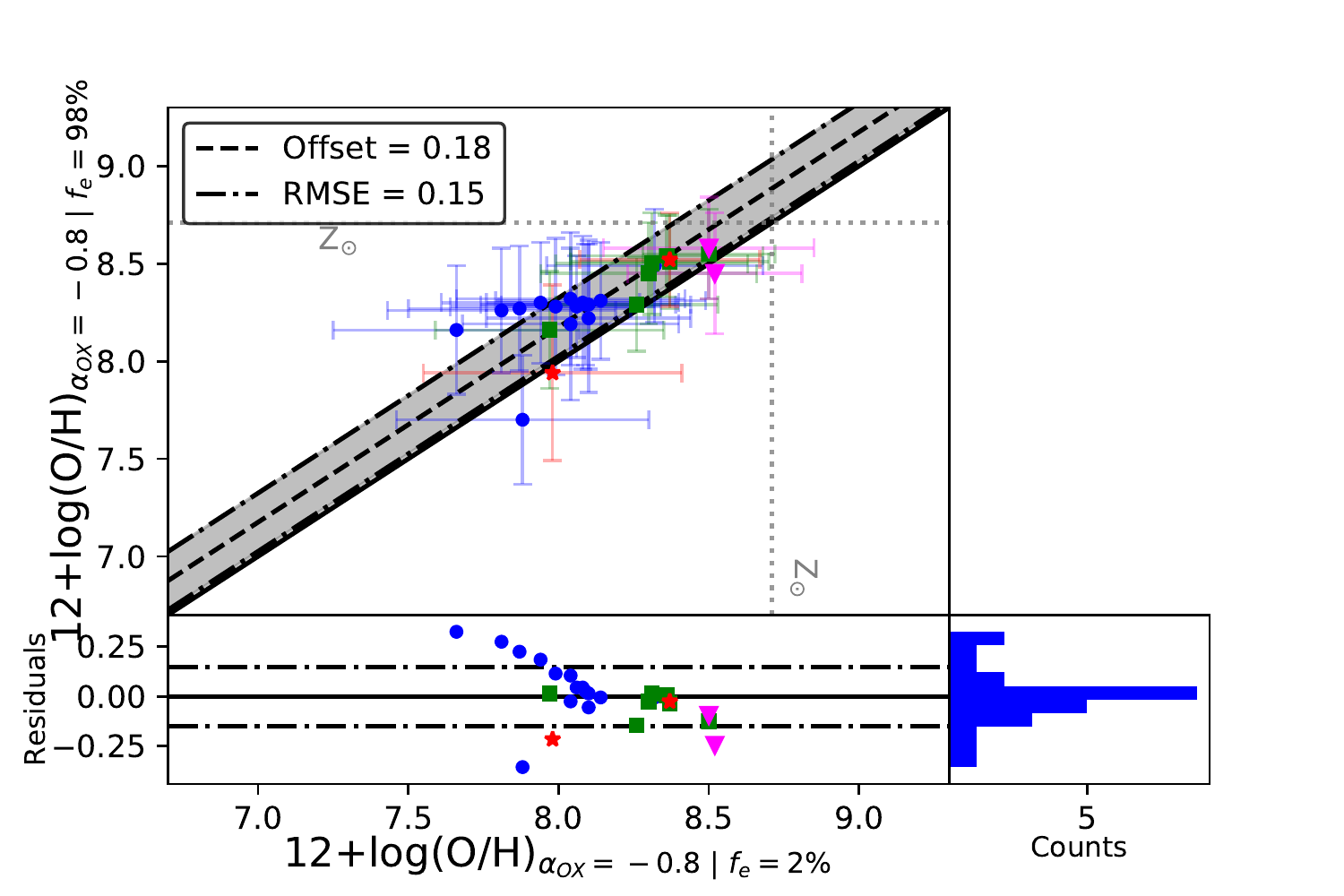}} \vspace{-0.2in} \end{minipage} \\
		\begin{minipage}{0.05\hsize}\begin{flushright}\textbf{(c)} \end{flushright}\end{minipage}  &  \begin{minipage}{0.43\hsize}\centering{\includegraphics[width=1\textwidth]{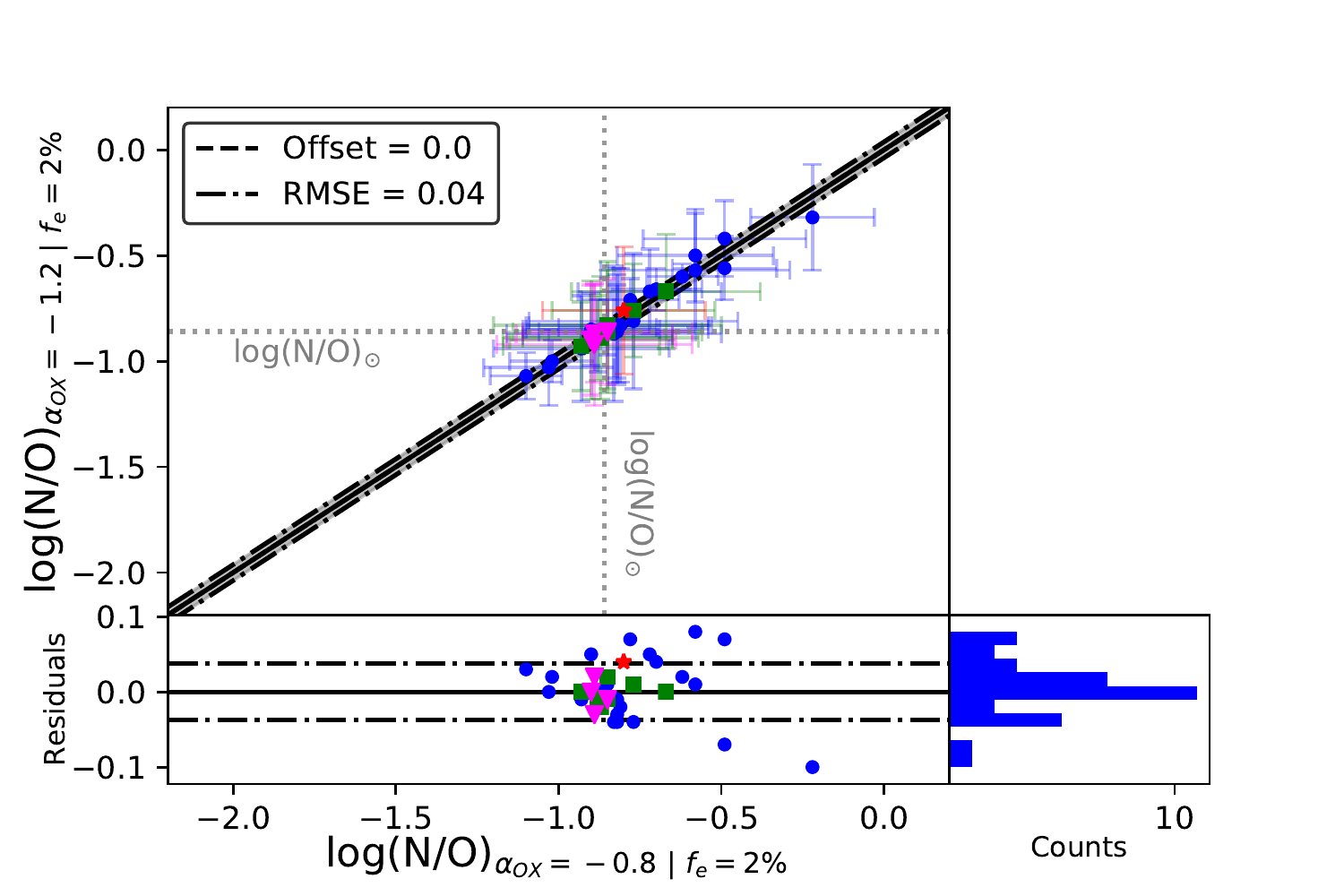}} \vspace{-0.2in} \end{minipage} & \begin{minipage}{0.05\hsize}\begin{flushright}\textbf{(d)} \end{flushright}\end{minipage}  &  \begin{minipage}{0.43\hsize}\centering{\includegraphics[width=1\textwidth]{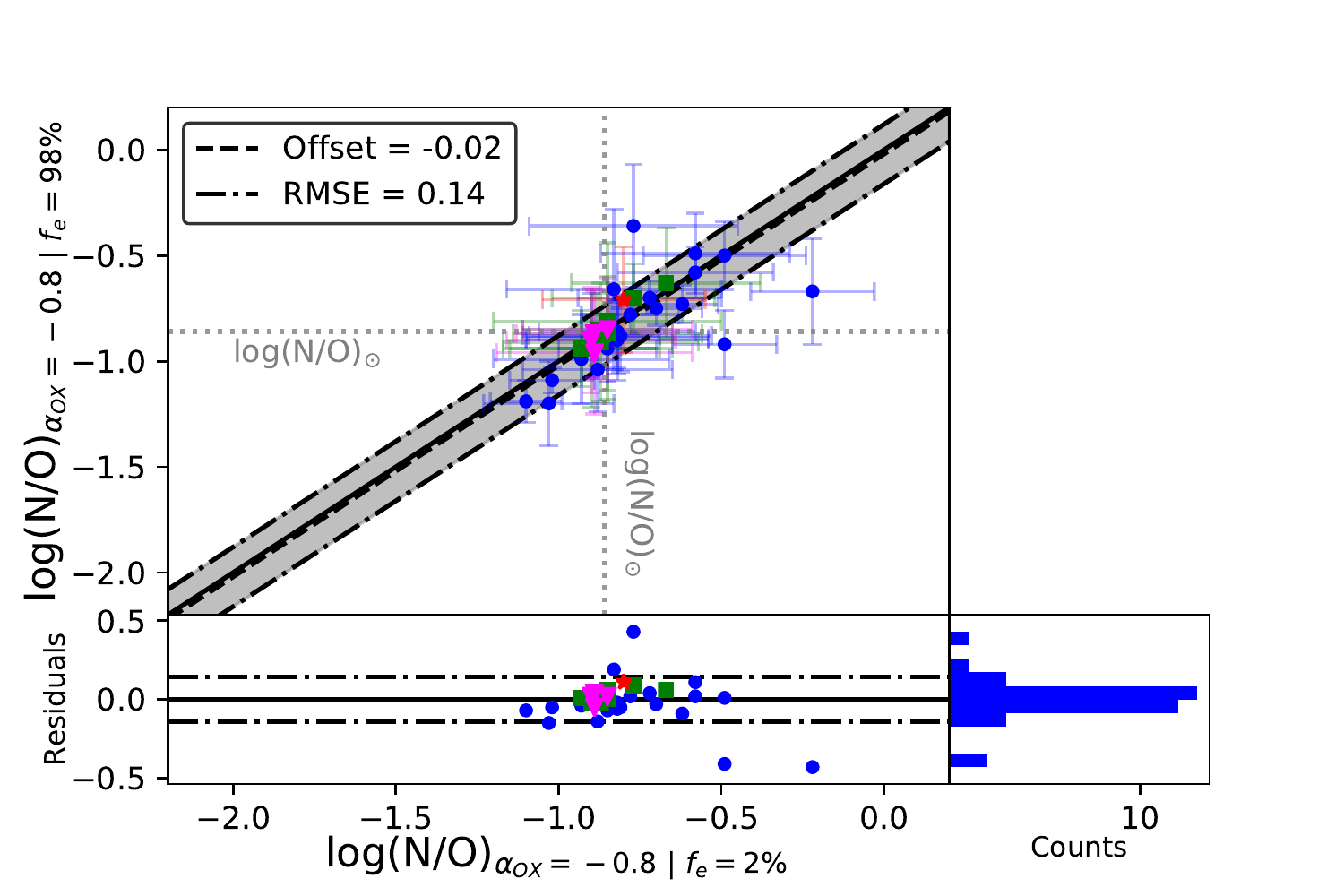}} \vspace{-0.2in} \end{minipage} \\
		\begin{minipage}{0.05\hsize}\begin{flushright}\textbf{(e)} \end{flushright}\end{minipage}  &  \begin{minipage}{0.43\hsize}\centering{\includegraphics[width=1\textwidth]{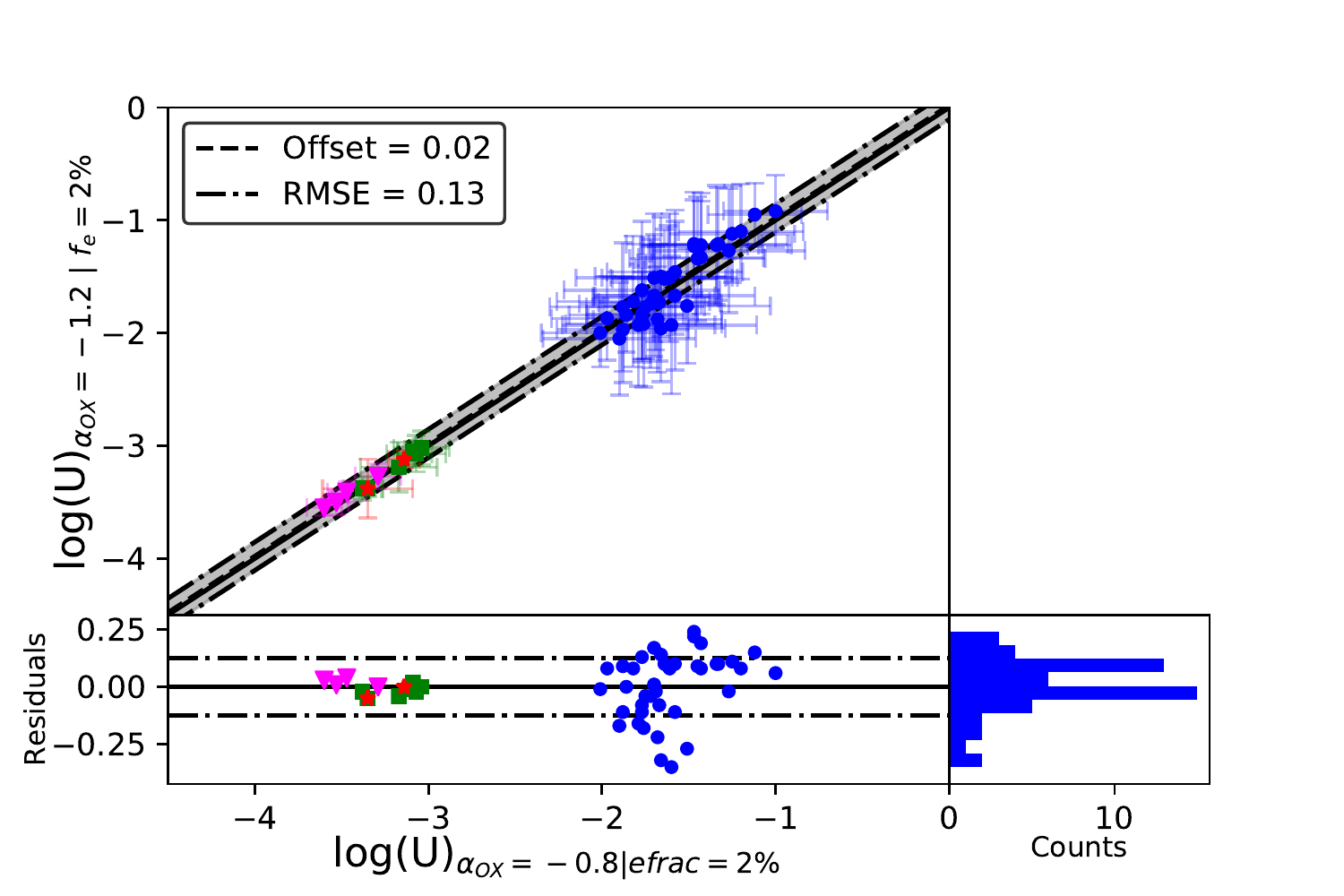}} \vspace{-0.2in} \end{minipage} & \begin{minipage}{0.05\hsize}\begin{flushright}\textbf{(f)} \end{flushright}\end{minipage}  &  \begin{minipage}{0.43\hsize}\centering{\includegraphics[width=1\textwidth]{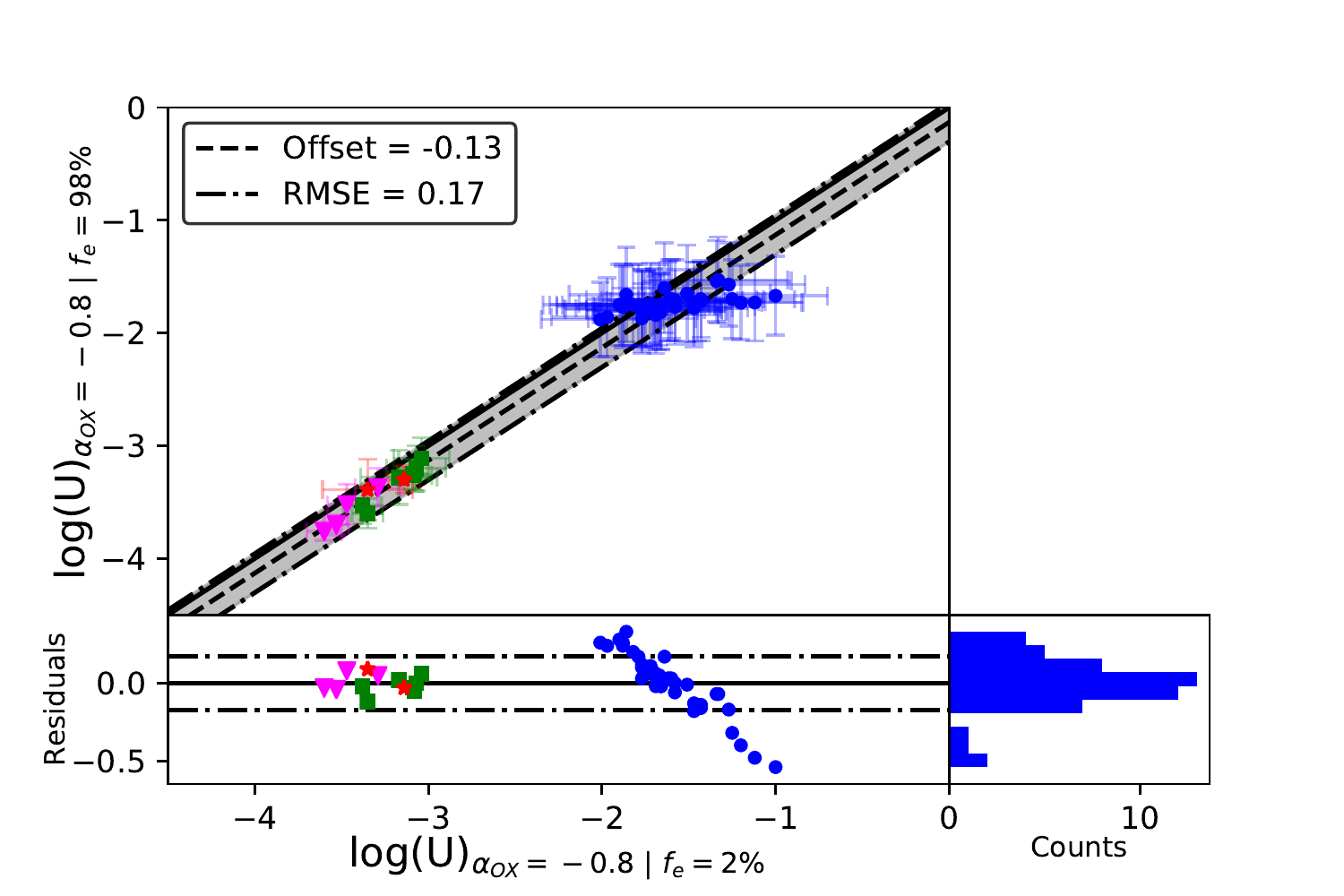}} \vspace{-0.2in} \end{minipage} 
	\end{tabular}
	\caption{Comparison between the  chemical abundances and $U$ values introduced obtained from the grid characterized by $\alpha_{OX} = -0.8$ and $2\%$ fraction of free electrons (x-axis) with models with $\alpha_{OX} = -1.2$ ((a), (c) and (e)) and models with $98\%$ as stopping criteria ((b), (d) and (f)). For all plots we present: Seyferts as blue circles, ULIRGs as green squares, LIRGs as magenta triangles and LINERs as red stars. The offsets are given using the median value (dashed line) and RMSE (dot-dashed lines). Bottom plots show the residuals from the offset and their distribution in a histogram (bottom-right plot).}
	\label{grids}
\end{figure*}

From Fig. \ref{grids} (a), (c) and (e), we conclude that no significant change is introduced in the chemical abundances or ionization parameters derived when the grid of models is computed assuming a SED characterized by $\alpha_{OX} = -1.2$: the offsets are below 0.05 dex and the RSME is always below the step considered in the grid for the given quantity.

The effects of selecting a different stopping criteria are more notorious in the determination of the ionization parameter: high-ionization ($U$ > -2.5) AGN present a higher scatter, with values clustering around log($U$) $\sim$ -1.8. This effect is mainly caused due to changes in the emission lines of high-ionizing species such as Ne$^{4+}$ or O$^{3+}$. In the case of O/H, there seems to be an slight overestimation when models with stopping criteria of 98$\%$ free electrons are considered. N/O does not show any significant change.

\section{Results}
\label{sec4}

We present in this section the chemical abundances and ionization parameters estimated for our sample of AGN using \textsc{HCm-IR}. Due to the lack of alternatives to estimate these parameters from IR observations of AGN, we use optical spectroscopic information of the same sample in order to compare results from both sets of information.

\begin{table*}
	\caption{Statistics of the chemical abundances and log($U$) values derived from \textsc{HCm-IR} for our sample of galaxies.}
	\label{ir_results}     
	\centering          
	\begin{tabular}{ll|lll|lll|lll}
		\multicolumn{2}{l}{} &
		\multicolumn{3}{|l}{\boldmath$12+\log_{10} \left( O/H \right) $} & \multicolumn{3}{|l}{\boldmath$\log_{10} \left( N/O \right) $} & \multicolumn{3}{|l}{\boldmath$\log_{10} \left( U \right) $}  \\ \hline \textbf{Sample} & \textbf{N\boldmath$_{tot}^{\circ}$} & \textbf{N\boldmath$^{\circ}$} & \textbf{Median} & \textbf{Std. Dev.} & \textbf{N\boldmath$^{\circ}$} & \textbf{Median} & \textbf{Std. Dev.} & \textbf{N\boldmath$^{\circ}$} & \textbf{Median} & \textbf{Std. Dev.} \\
		All galaxies & 58 & 26 &  8.05 &  0.24 & 35 & -0.83 &  0.17 & 52 & -1.73 &  0.80\\ 
		Seyferts & 43 & 15 &  7.99 &  0.16 & 22 & -0.81 &  0.20 & 39 & -1.67 &  0.33\\ 
		ULIRGs & 8  & 7  &  8.32 &  0.20 & 8  & -0.86 &  0.07 & 7  & -3.08 &  0.15\\ 
		LIRGs & 4  & 2  &  8.495 &  0.005 & 4  & -0.885 &  0.019 & 4  & -3.58 &  0.14\\ 
		LINERs & 3  & 2  &  8.17 &  0.18 & 1  & -0.80 &  - & 2  & -3.24 &  0.10\\  
		
	\end{tabular}      
\end{table*}

\subsection{Infrared estimations}
\label{subsec41}
We summarize in Tab. \ref{ir_results} the statistics of the  estimations of the chemical abundances and log($U$) values obtained with \textsc{HCm-IR} from the IR emission lines in our sample distinguishing between different types of galaxies. Tab \ref{TabA2} shows these results in detail for each galaxy in our sample.

As expected from our preliminary distinction of AGNs, we have two main subgroups based on $U$ results consistent with our prior distinction: Seyferts belong to high-ionization AGN category as they usually present log($U$) $>$ -2.5 \citep{Ho_1993, Villar-Martin_2008, Zhuang_2019} while LINERs fall in the category of low-ionization AGNs with log($U$) $<$ -2.5 \citep{Ferland_1983, Halpern_1983, Binette_1985, Kewley_2006}. In the case of ULIRGs and LIRGs, the study by \citet{Pereira-Santaella_2017} showed that low (log($U$) $<$ -2.5) ionization parameters are needed to reproduce observations from photoionization models, thus we assume they fall in the category of low-ionization AGN.  

Although we have relatively low statistics, the three spectral types considered as low-ionization AGN differ in their median ionization: ULIRGs show the highest value ($U \sim -3$), followed by LINERs ($U \sim -3.25$) and then by LIRGs ($U \sim -3.6$). Despite being these differences higher than their dispersions, they are still close to the step for $U$ used in the grid (0.25 dex) of models, thus they must be revisited in larger sample of galaxies.

Analyzing chemical abundances, we obtained median subsolar values for all types of galaxies, being Seyferts in average metal-poorer than the other three spectral types. However, since the estimation of 12+log(O/H) is only available in the few galaxies with detected hydrogen recombination lines, this result must be revisited in larger samples of galaxies. N/O shows a similar median value for all types of galaxies, clustering around the solar value of log(N/O)$_{\odot }$ = -0.86 \citep{Asplund_2009}.

\begin{figure}
	\centering
	\includegraphics[width=\hsize]{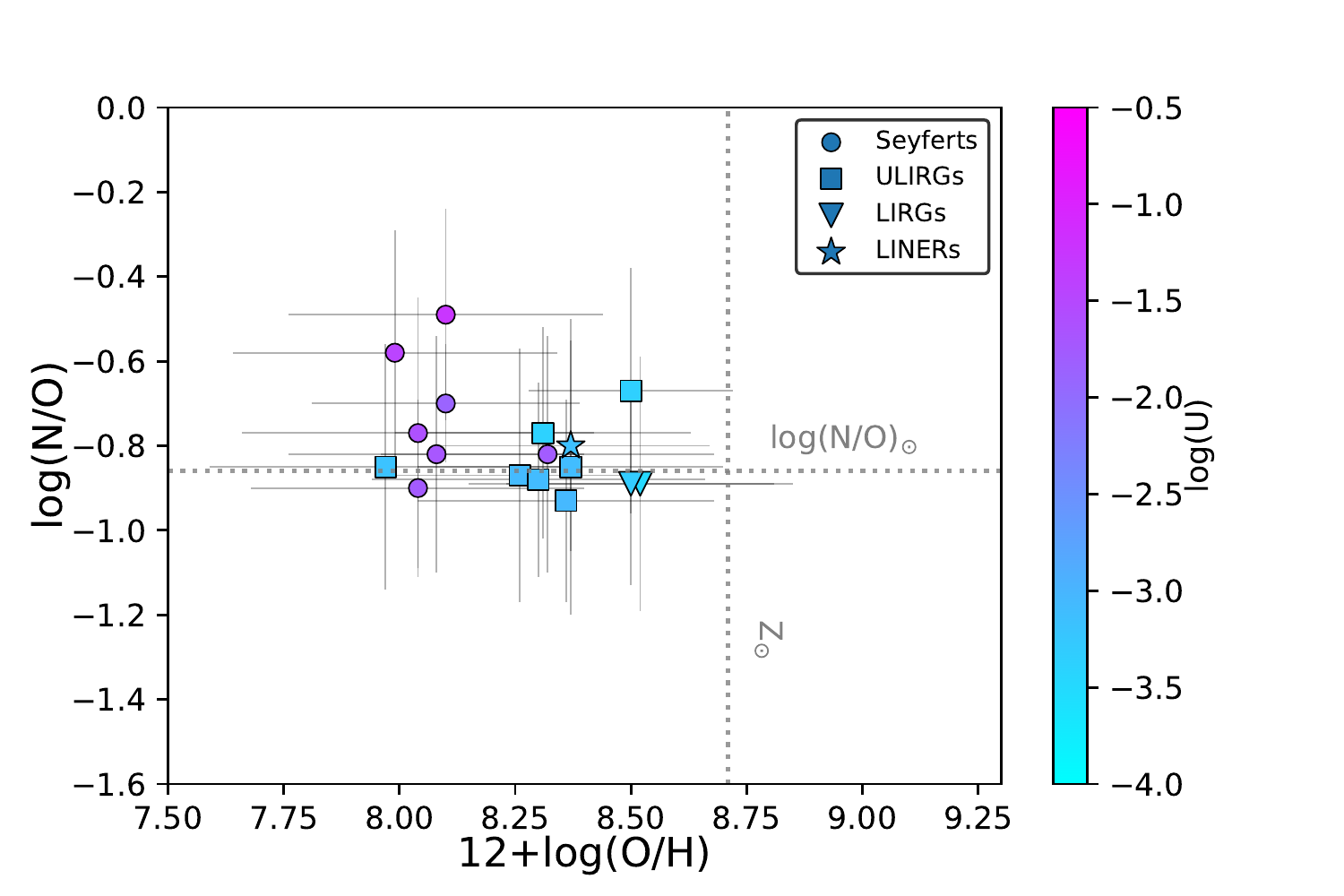} 
	\caption{N/O vs O/H from IR estimations. The values for log($U$) are given by the color bar.}
	\label{Fig7}
\end{figure}

We also present in Fig. \ref{Fig7} the well-known N/O-O/H diagram obtained from our IR estimations. Notice that our statistics for this plot is small because we need galaxies with estimations of both log(N/O) and 12+log(O/H). We find approximately a flattened behavior around log(N/O)$_{\odot }$, although there are some galaxies with higher ratios.

\subsection{Optical estimations}
\label{subsec42}
We compiled from the literature optical emission-line fluxes for our sample of AGN, and corrected  all emission-line ratios for reddening (see Tab. \ref{TabA3}) referred to the Balmer line H$_{\beta }$, following Howarth's extinction curve \citeyearpar{Howarth_1983}, assuming $R_{V} = 3.1$ and a theoretical ratio between H$_{\alpha }$ and H$_{\beta }$ of 3.1, characteristic of the Recombination Case B for the physical conditions of the NLR in AGN.

Although we have additional information from  IR observations, which can be necessary to account for some ionic species  whose emission lines cannot be retrieved from optical emission lines, such as O$^{3+}$, what can in turn lead  to underestimations in the oxygen abundance \citep{Dors_2015, Maiolino_2019, Flury_2020}, we cannot apply the direct method since only 2 galaxies in our sample (namely Mrk 478 and NGC 4151) present measurements of auroral line [\ion{O}{iii}]$\lambda$4363\r{A}, key to determine the electronic temperature T$_{e}$ of the ISM. Thus, we estimated chemical abundances from optical emission lines for our sample of AGN using the optical version of \textsc{HCm} for AGNs \citepalias{Perez-Montero_2019}. The code  takes as input the following reddening-corrected optical emission lines: [\ion{O}{ii}]$\lambda$3727\r{A}, [\ion{Ne}{iii}]$\lambda$3868\r{A}, [\ion{O}{iii}]$\lambda$4363\r{A}, [\ion{O}{iii}]$\lambda$5007\r{A}, [\ion{N}{ii}]$\lambda$6584\r{A} and [\ion{S}{ii}]$\lambda\lambda $6717,6731\r{A}; all of them referred to the Balmer line $H_{\beta }$.

To check our optical estimations, we also considered the calibration proposed by \citet{Flury_2020} 
based on [\ion{O}{iii}]$\lambda$5007\r{A} and [\ion{N}{ii}]$\lambda$6584\r{A} given by:
\begin{equation}
\label{FM20}
\begin{aligned}
12 +  \log \left( O/H \right) _{FM+20} = & 7.863 + 1.170u + 0.027v - 0.406uv \\ & - 0.369u^{2} + 0.208v^{2} + 0.354u^{2}v \\ & - 0.333uv^{2} - 0.100u^{3} + 0.323v^{3}
\end{aligned}
\end{equation}
where $u = \log \left( \mathrm{I} \left( \right. \right. $[\ion{N}{ii}]$\lambda$6584\r{A}/H$_{\alpha } \left. \left. \right) \right) $ and $v = \log \left( \mathrm{I} \left( \right. \right. $ [\ion{O}{iii}]$\lambda$5007\r{A}/H$_{\beta } \left. \left. \right) \right) $. This calibration is valid for the range 7.5 $\leq$ 12+log(O/H) $\leq$ 9.0.

We also considered the calibration based on the N line [\ion{N}{ii}]$\lambda$6584\r{A} given by \citet{Carvalho_2020}:
\begin{equation}
\notag Z/Z_{\odot } = a^{N2} + b
\end{equation}
where $N2 =  \log \left( I \left( \right. \right. $([\ion{N}{ii}]$\lambda$6584\r{A}/H$_{\beta } \left. \left. \right) \right) $ , $a = 4.01 \pm 0.08$ and $b = -0.07 \pm 0.01$. In terms of the oxygen abundance, the calibration is given by\footnote{We assume here the solar abundances from \citet{Asplund_2009}}:
\begin{equation}
\label{Ca20} 12 + \log \left( O/H \right) _{Ca+20} = 8.69 + \log \left( 4.01^{N2} - 0.07 \right)
\end{equation}
which was defined in the range 8.17 $\leq $ 12+log(O/H) $\leq$ 9.0.

\begin{figure*}
	\begin{tabular}{cccc}
		\begin{minipage}{0.05\hsize}\begin{flushright}\textbf{(a)} \end{flushright}\end{minipage}  &  \begin{minipage}{0.43\hsize}\centering{\includegraphics[width=1\textwidth]{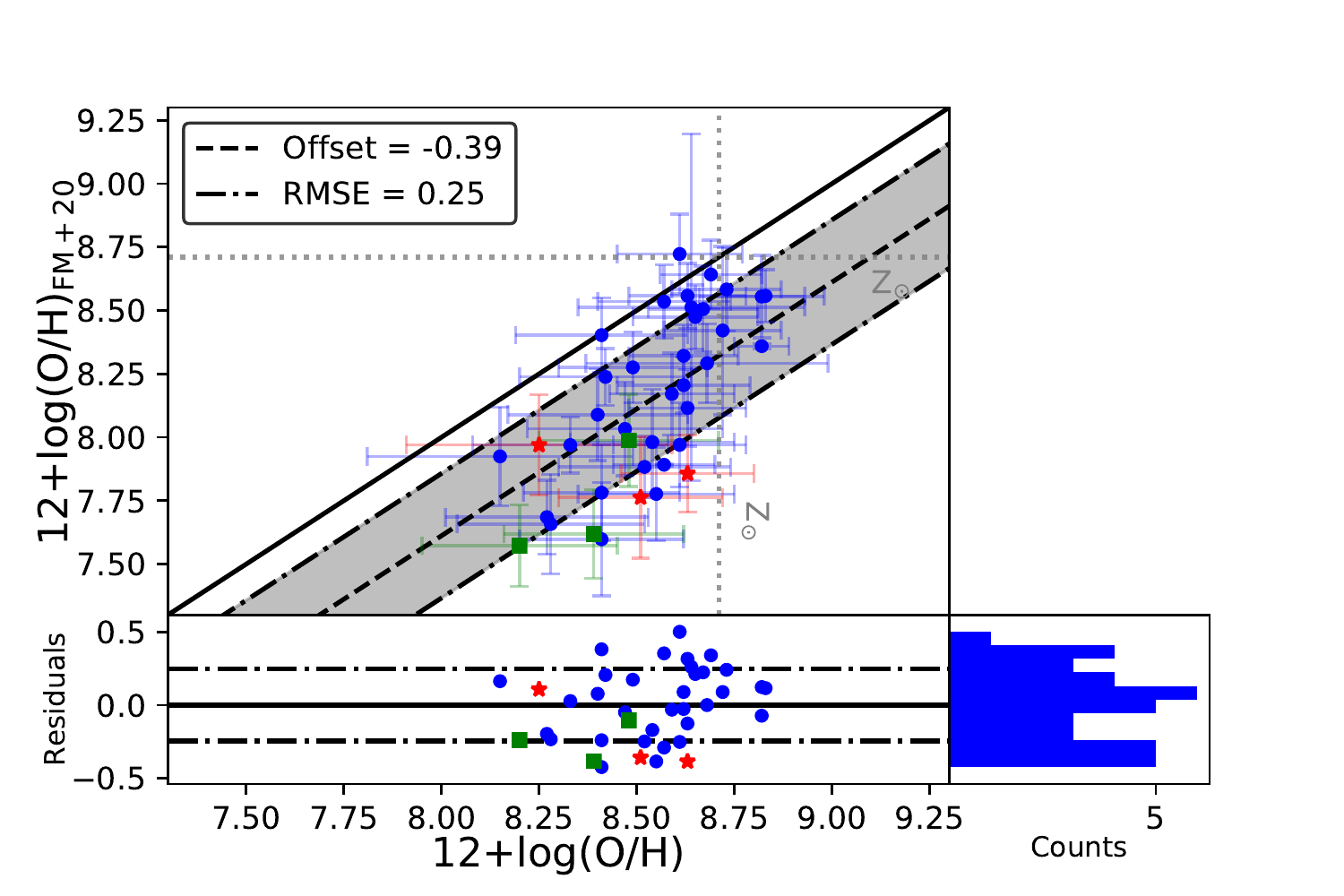}} \vspace{-0.2in} \end{minipage} & \begin{minipage}{0.05\hsize}\begin{flushright}\textbf{(b)} \end{flushright}\end{minipage}  &  \begin{minipage}{0.43\hsize}\centering{\includegraphics[width=1\textwidth]{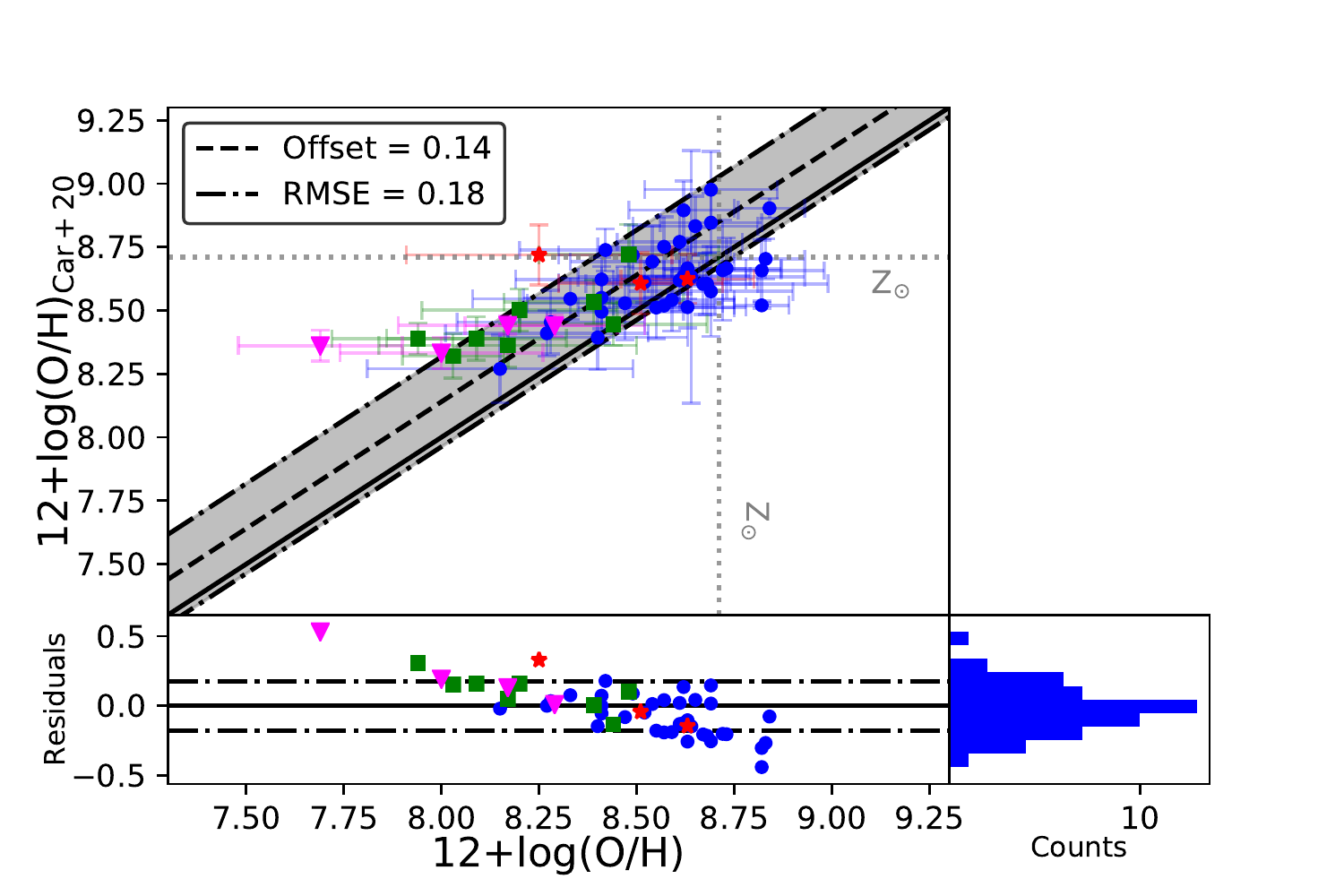}} \vspace{-0.2in} \end{minipage} 
	\end{tabular}
	\caption{Chemical abundances derived from optical emission lines in our sample of AGN. (a) Comparison between the chemical abundances derived with the calibration from \citet{Flury_2020} (y-axis), denoted as FM20, and \textsc{HCm} (x-axis). (b) Comparison between the chemical abundances derived from the calibration from \citet{Carvalho_2020} (y-axis), denoted as FM20, with \textsc{HCm} (x-axis). For all plots we present: Seyferts as blue circles, ULIRGs as green squares, LIRGs as magenta triangles and LINERs as red stars. The offsets are given using the median value (dashed line) and RMSE (dot-dashed lines). Bottom plots show the residuals from the offset and their distribution in a histogram (bottom-right plot).}
	\label{Fig8}
\end{figure*}

We compared the resulting  abundances in our sample of AGN (see Tab. \ref{TabA4}) from the calibrations described above with those from \textsc{HCm} in Fig. \ref{Fig6}. The calibration proposed by \citet{Flury_2020} tends to underestimate \textsc{HCm} abundances, with a median offset of -0.39 dex. 
On the other hand, the calibration proposed by \citet{Carvalho_2020} fits better our results, with a median offset of 0.14 dex. As shown in bottom plot of Fig. \ref{Fig6} (b), the discrepancy is higher for ULIRGs, LIRGs and LINERs than for Seyferts. This could be  explained with the fact that \citet{Carvalho_2020} obtained their calibration from a sample of Seyferts 2 from SDSS, {\em i.e.}, it was obtained using only high-ionization AGNs, covering a different range of the ionization parameter than the values reported for low-ionization AGN \citep[e.g.][]{Kewley_2006}. Another possible source of error is that the calibration is based only on a nitrogen line so it assumes a relation between 12+log(O/H) and log(N/O), although  it has been reported that both quantities might not be related in low ionization AGNs \citep{Perez-Diaz_2021}.

\begin{table*}
	\caption{Same as Tab. \ref{ir_results} but for optical results.}
	\label{opt_results}     
	\centering          
	\begin{tabular}{ll|lll|lll|lll}
		\multicolumn{2}{l}{} &
		\multicolumn{3}{|l}{\boldmath$12+\log_{10} \left( O/H \right) $} & \multicolumn{3}{|l}{\boldmath$\log_{10} \left( N/O \right) $} & \multicolumn{3}{|l}{\boldmath$\log_{10} \left( U \right) $}  \\ \hline \textbf{Sample} & \textbf{N\boldmath$_{tot}^{\circ}$} & \textbf{N\boldmath$^{\circ}$} & \textbf{Median} & \textbf{Std. Dev.} & \textbf{N\boldmath$^{\circ}$} & \textbf{Median} & \textbf{Std. Dev.} & \textbf{N\boldmath$^{\circ}$} & \textbf{Median} & \textbf{Std. Dev.} \\
		All galaxies & 58 & 58 &  8.49 &  0.30 & 51 & -0.84 &  0.24 & 58 & -1.88 &  0.81\\ 
		Seyferts & 43 & 43 &  8.57 &  0.30 & 36 & -0.84 &  0.25 & 43 & -1.81 &  0.23\\ 
		ULIRGs & 8  & 8  &  8.19 &  0.19 & 8  & -0.79 &  0.18 & 8  & -3.51 &  0.25\\ 
		LIRGs & 4  & 4  &  8.09 &  0.23 & 4  & -0.65 &  0.15 & 4  & -3.59 &  0.19\\ 
		LINERs & 3  & 3  &  8.51 &  0.16 & 3  & -0.94 &  0.18 & 3  & -3.73 &  0.18\\  
		
	\end{tabular}      
\end{table*}

We present in Tab. \ref{opt_results} the statistics of the chemical abundances derived from the optical version of the code. Again, we obtain that 12+log(O/H) presents subsolar median values for all types of galaxies. The median N/O values present more variation between different types but, considering the standard deviations, they are still compatible among them and with the solar value obtained from IR estimations.

\begin{figure}
	\centering
	\includegraphics[width=\hsize]{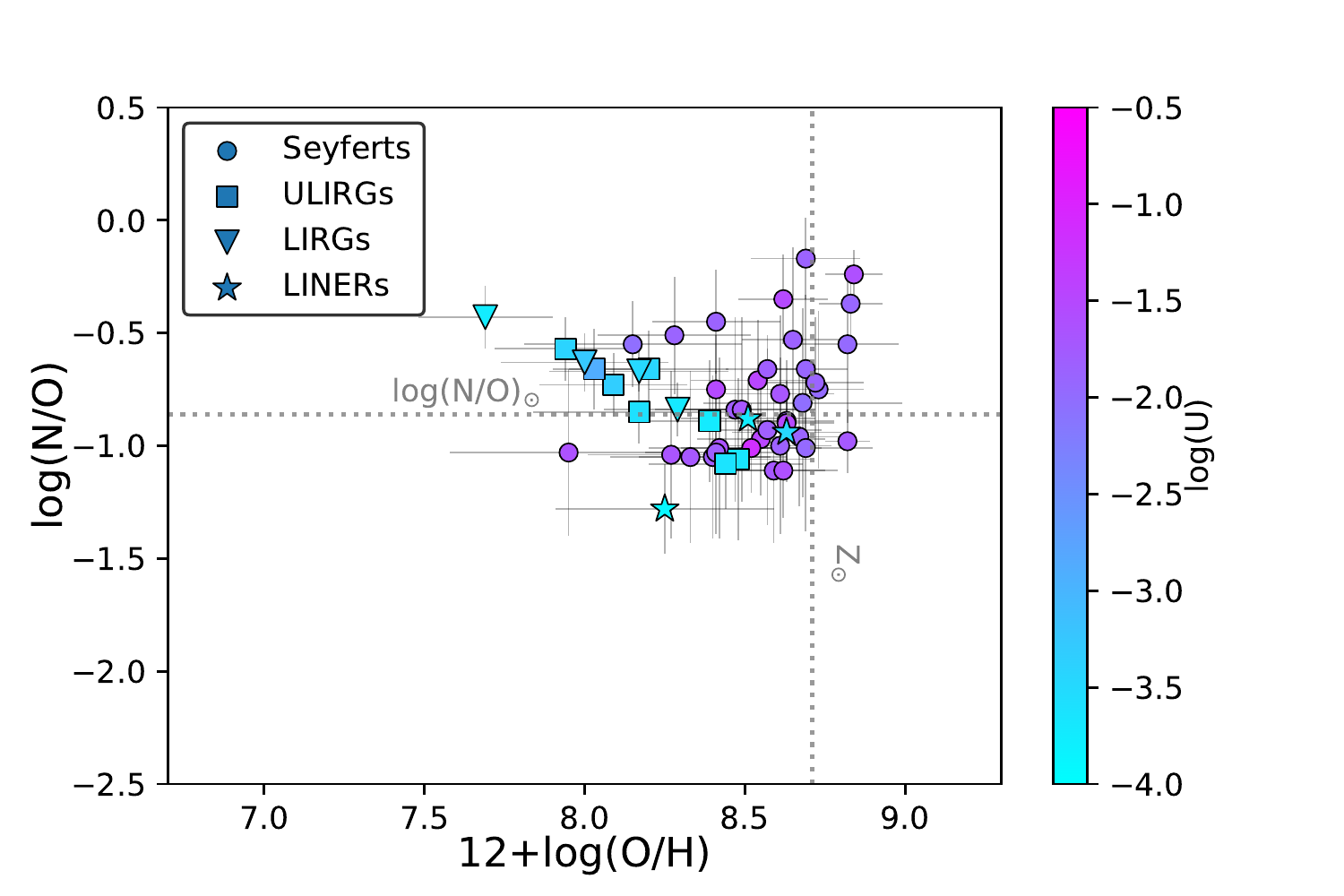} 
	\caption{N/O vs O/H as derived from optical estimations. The log($U$) values are given by the color bar.}
	\label{Fig9}
\end{figure}

We also present in Fig. \ref{Fig9} the N/O-O/H diagram. We can see two different trends based on the two main categories considered through this study. For low-luminosity AGNs (ULIRGs, LIRGs and LINERs) there seems to be an anti-correlation between N/O and O/H (although the corresponding pearsons coefficient correlation is low r$\sim$-0.75). In the case of high-ionization AGNs, both quantities do not seem to be correlated, which was also found by \citet{Perez-Diaz_2021} although with a smaller sample of galaxies. As AGN activity is a rare phenomenon among dwarf galaxies (< 1.8$\%$, \citealt{Latimer_2021}), and these have been challenging targets for previous IR spectroscopic facilities, our N/O vs O/H diagram cannot reproduce with enough statistics the metal-poor regime.

\subsection{Optical vs infrared estimations}
\label{subsec43}

\begin{figure}
	\begin{tabular}{cc}
		\begin{minipage}{0.05\hsize}\begin{flushright}\textbf{(a)} \end{flushright}\end{minipage}  &  \begin{minipage}{0.9\hsize}\centering{\includegraphics[width=1\textwidth]{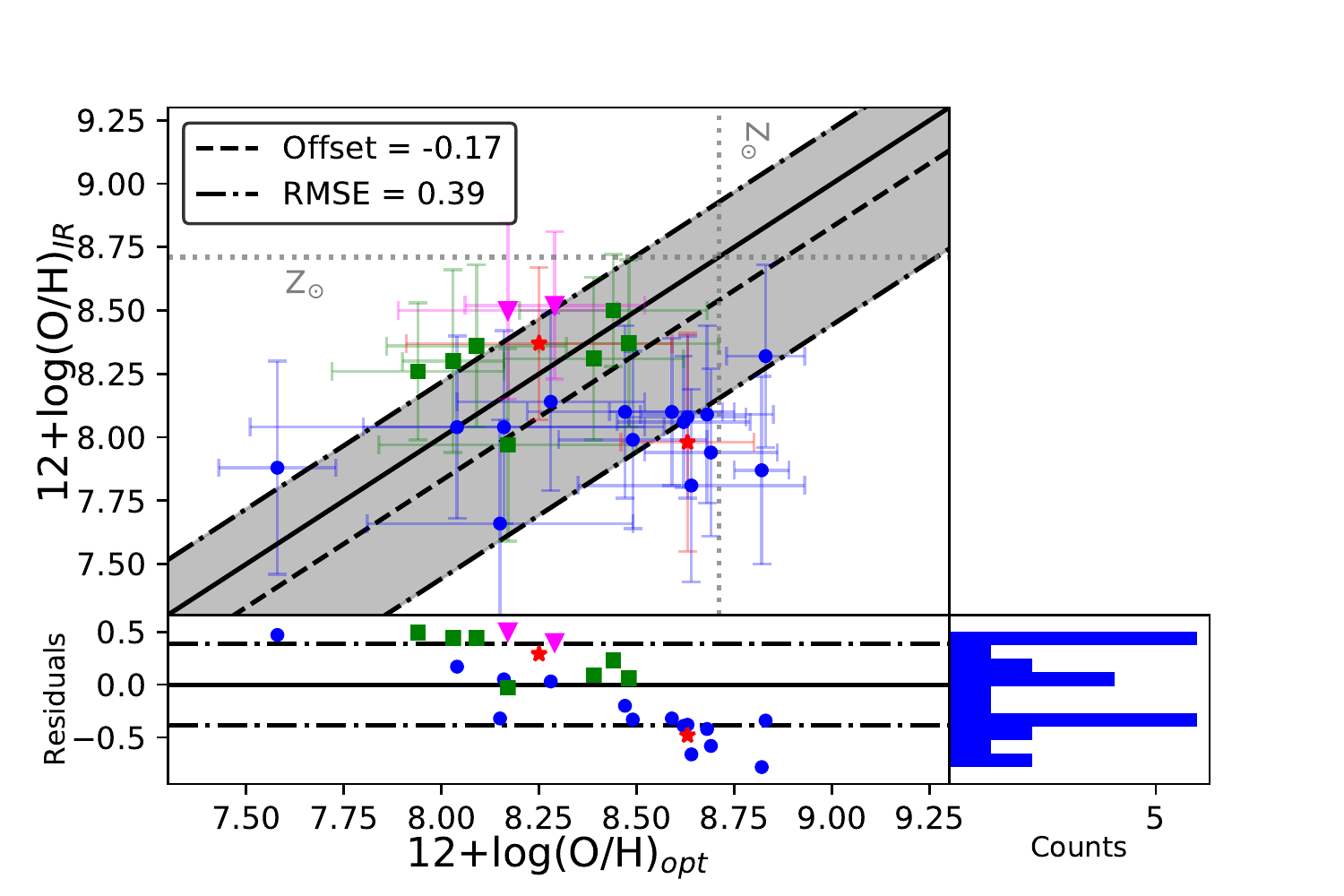}} \vspace{-0.2in} \end{minipage} \\ \begin{minipage}{0.05\hsize}\begin{flushright}\textbf{(b)} \end{flushright}\end{minipage}  &  \begin{minipage}{0.9\hsize}\centering{\includegraphics[width=1\textwidth]{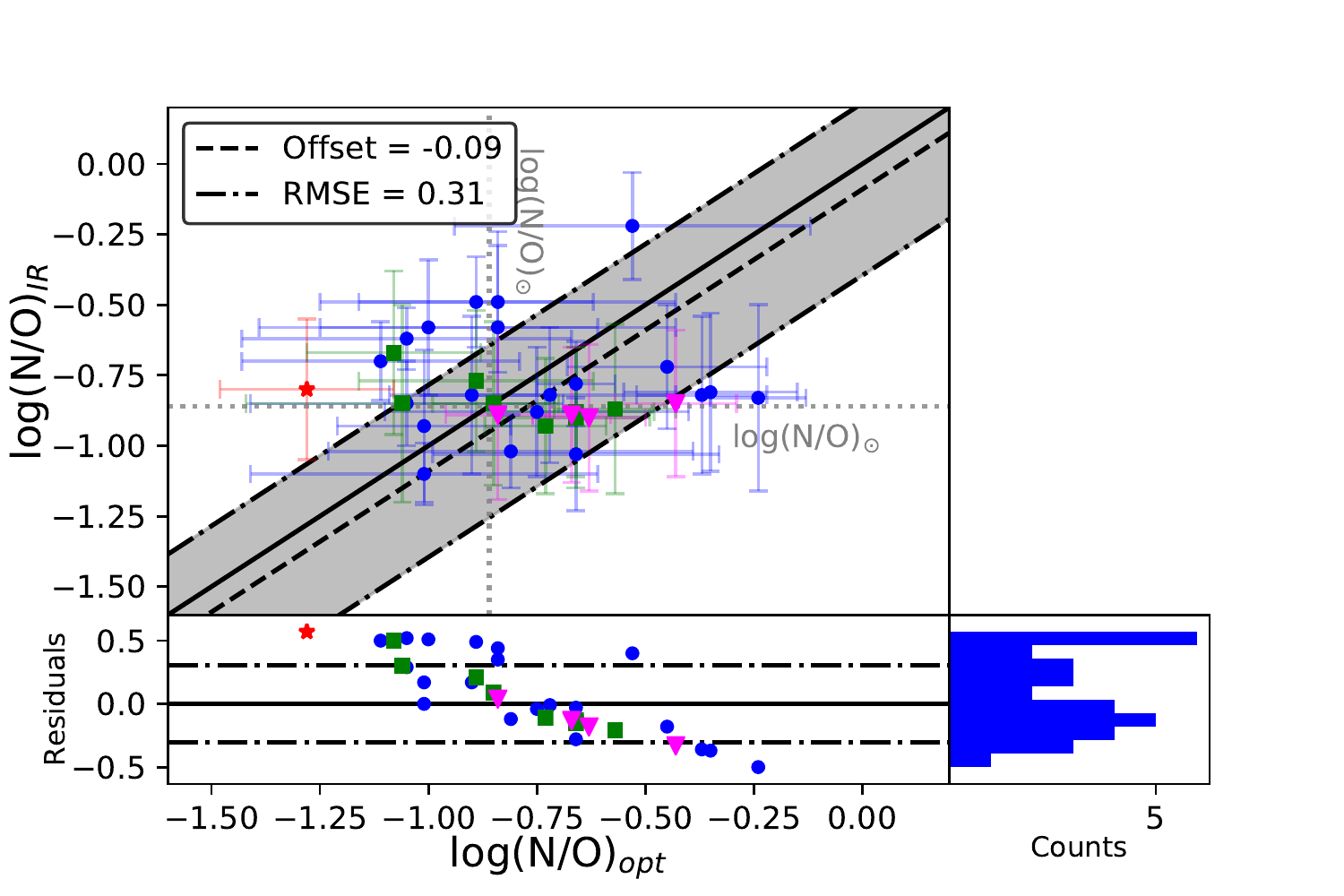}} \vspace{-0.2in} \end{minipage} \\
		\begin{minipage}{0.05\hsize}\begin{flushright}\textbf{(c)} \end{flushright}\end{minipage}  &  \begin{minipage}{0.9\hsize}\begin{flushleft}\includegraphics[width=1\textwidth]{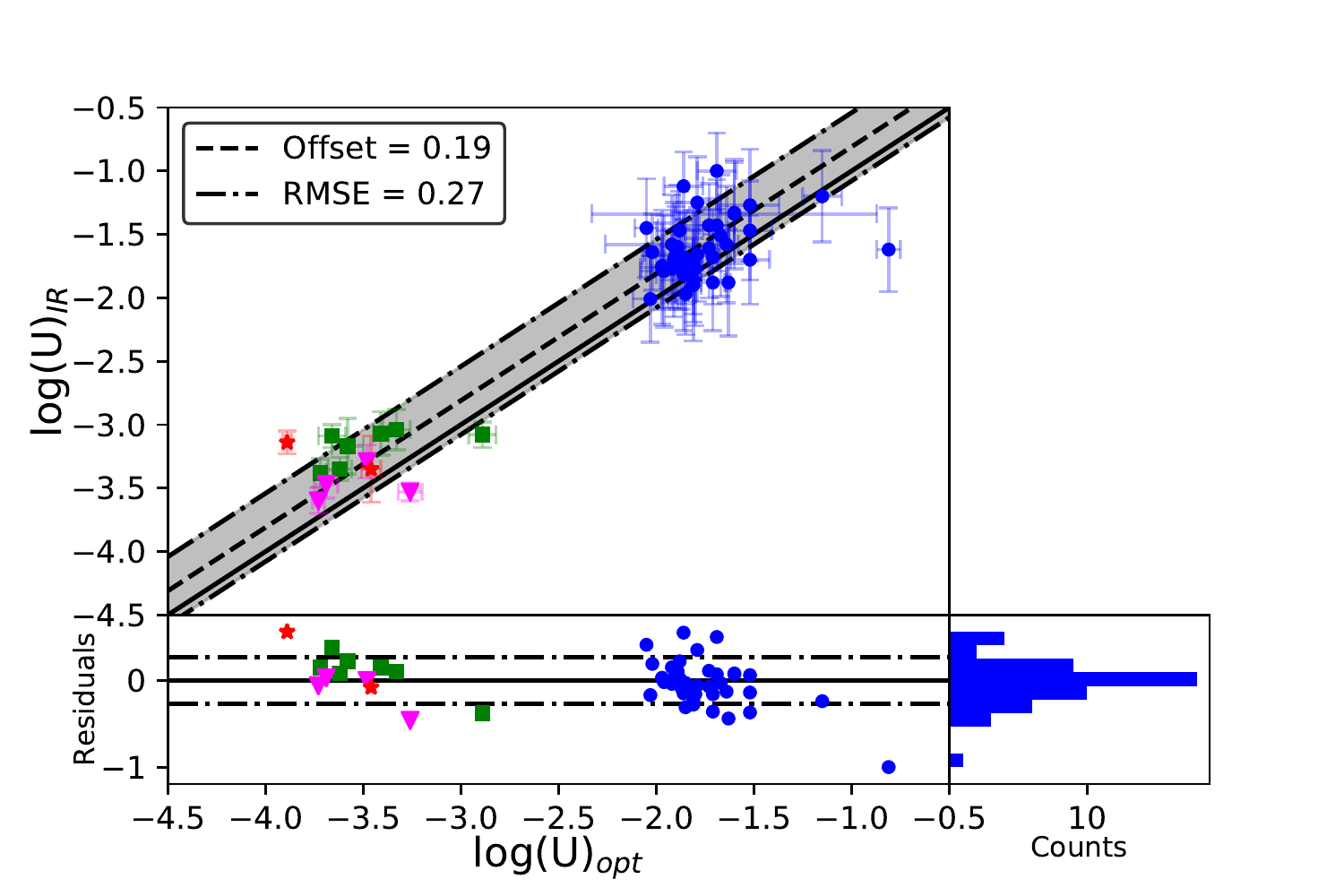}\end{flushleft} \vspace{-0.2in} \end{minipage} \\		 
	\end{tabular}
	\caption{Comparison between the chemical abundances and log($U$) values obtained from optical emission lines using \textsc{HCm} (x-axis) and the corresponding estimations from IR lines using \textsc{HCm-IR} (y-axis). For all plots we present: Seyferts as blue circles, ULIRGs as green squares, LIRGs as magenta triangles and LINERs as red stars. The offsets are given using the median value (dashed line) and RMSE (dot-dashed lines). Bottom plots show the residuals from the offset and their distribution in a histogram (bottom-right plot).}
	\label{Fig10}
\end{figure}

Comparing the results listed in Tab. \ref{ir_results} and Tab. \ref{opt_results}, we can see that Seyferts present lower median oxygen abundances from IR estimations than from their optical counterpart, being the average offset for high-ionization AGNs higher than $0.5$ dex. Although lower, we also found an average offset of $0.3$ dex between optical and IR estimations for LINERs. While ULIRGs present similar oxygen abundances from both methods, in the case of LIRGs we obtain lower abundances from optical observations. However this result must be revisited using  larger samples of galaxies given our low statistic for LIRGs.

From Fig. \ref{Fig10} (a) we can see that 12+log(O/H) values from IR emission lines are systematically lower than the abundances derived using optical lines, which agrees with our previous statement for Seyferts. In the case of N/O, we obtained IR values clustering around the solar abundance, while optical estimations present a wider range of values, as shown in Fig. \ref{Fig10} (b). Finally, in Fig. \ref{Fig10} (c) we compare the resulting log($U$) values, obtaining in overall slightly higher values from IR estimations. However, since the step of the grids is 0.25 dex in log($U$), and both the median offset and RSME are close to this value, we can conclude that little difference is found.

\begin{figure*}
	\begin{tabular}{cccc}
		\begin{minipage}{0.05\hsize}\begin{flushright}\textbf{(a)} \end{flushright}\end{minipage}  &  \begin{minipage}{0.43\hsize}\centering{\includegraphics[width=1\textwidth]{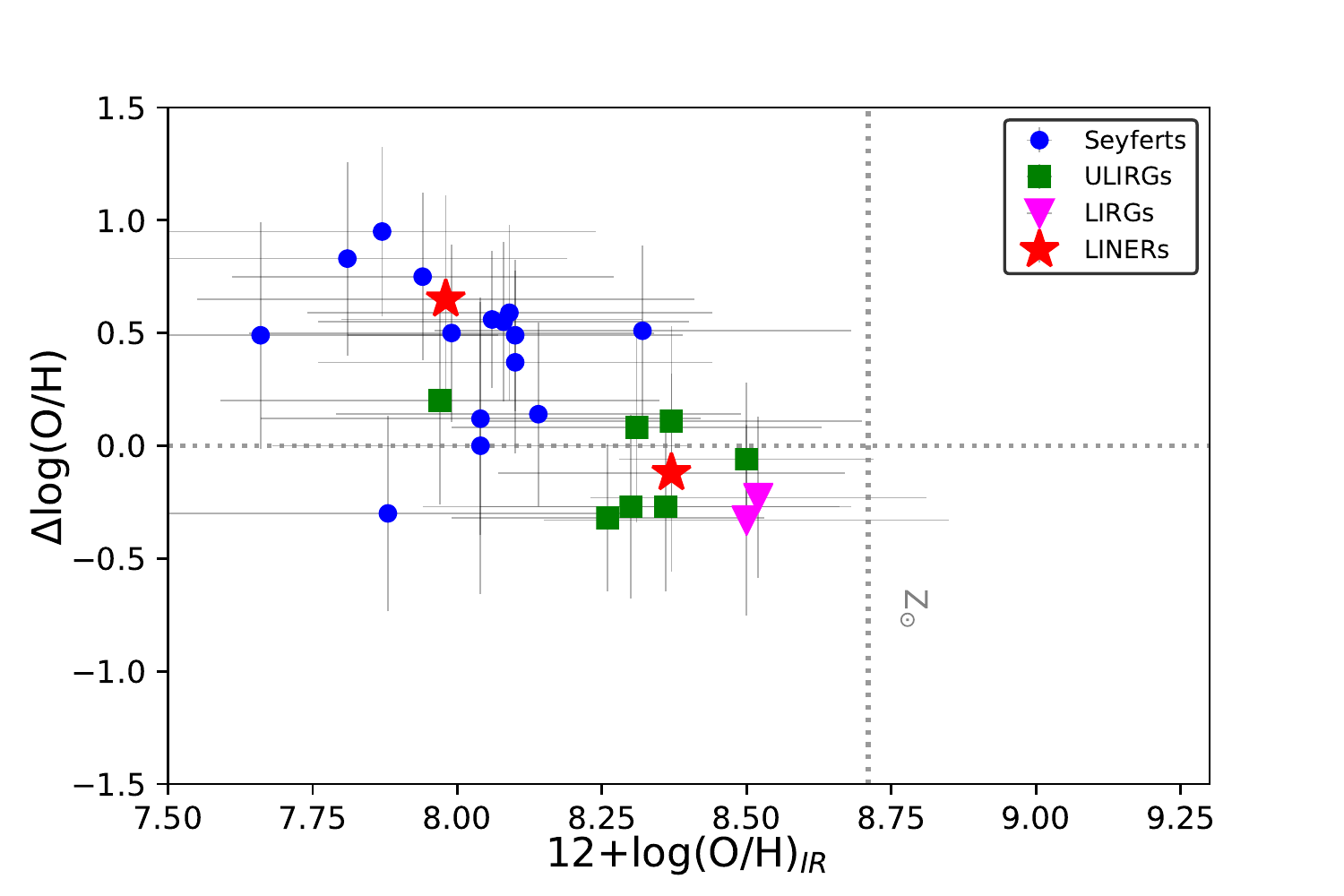}} \vspace{-0.2in} \end{minipage} & \begin{minipage}{0.05\hsize}\begin{flushright}\textbf{(b)} \end{flushright}\end{minipage}  &  \begin{minipage}{0.43\hsize}\centering{\includegraphics[width=1\textwidth]{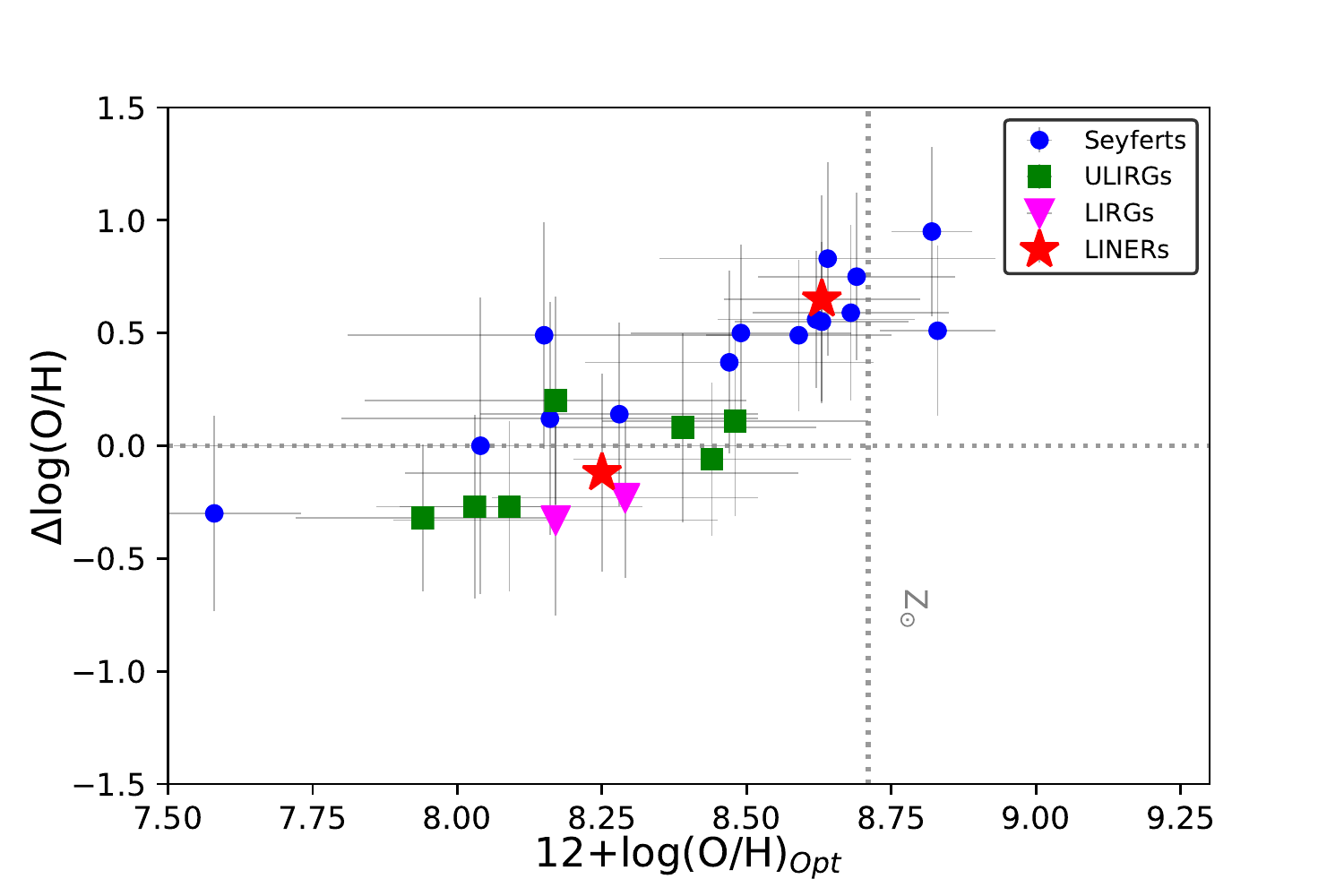}}  \vspace{-0.2in} \end{minipage} \\
		\begin{minipage}{0.05\hsize}\begin{flushright}\textbf{(c)} \end{flushright}\end{minipage}  &  \begin{minipage}{0.43\hsize}\centering{\includegraphics[width=1\textwidth]{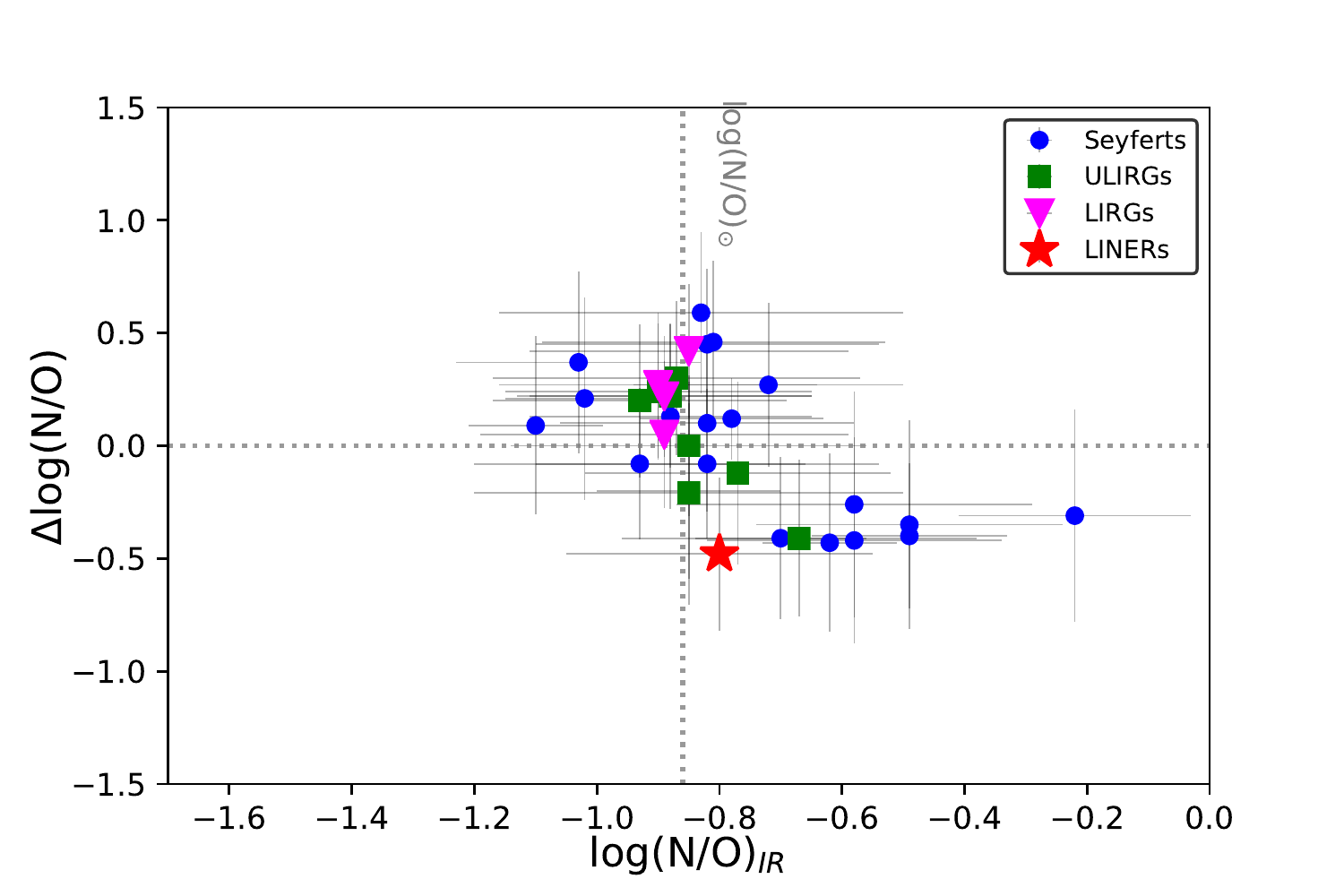}} \vspace{-0.2in} \end{minipage} & \begin{minipage}{0.05\hsize}\begin{flushright}\textbf{(d)} \end{flushright}\end{minipage}  &  \begin{minipage}{0.43\hsize}\centering{\includegraphics[width=1\textwidth]{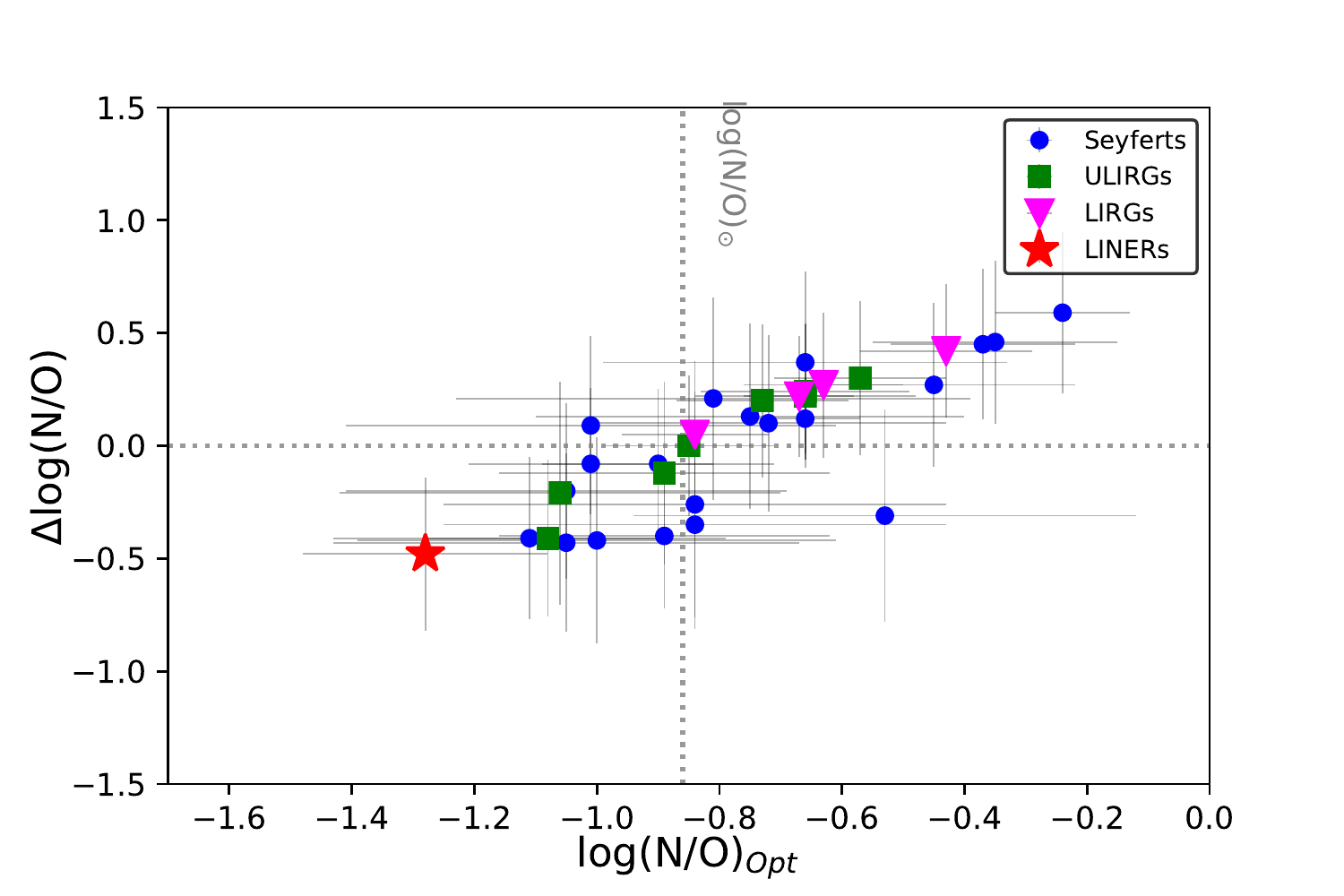}} \\ \vspace{-0.2in} \end{minipage}
	\end{tabular}
	\caption{Discrepancies between the chemical abundance ratios ($\Delta X = X_{opt} - X_{IR}$) as a function of their ratios: (a) 12+log(O/H) and (c) log(N/O) derived both from IR emission lines with \textsc{HCm-IR}, (b) 12+log(O/H) and (d) log(N/O) derived from optical emission lines with \textsc{HCm}. For all plots we present: Seyferts as blue circles, ULIRGs as green squares, LIRGs as magenta triangles and LINERs as red stars.}
	\label{Fig11}
\end{figure*}

\subsection{Dependency of the discrepancies}
\label{subsec44}

As pointed by the previous section, there is a significant difference between optical and infrared estimations of chemical abundances.
Hereinafter, we define the discrepancy for a given quantity $X$ as $\Delta X = X_{opt} - X_{ir}$.

We present in Fig. \ref{Fig11} the discrepancy as a function of the two chemical abundance ratios (12+log(O/H) and log(N/O)) for IR (left column) and optical estimations (right column). Fig. \ref{Fig11} (a) and (c) shows that little correlation is found between the discrepancies and their corresponding abundances from IR emission lines. This is not the case in the optical range as previously discussed: $\Delta$ log(O/H) increases with the oxygen abundance (see Fig. \ref{Fig11} (b)). A similar result is also found for $\Delta $ log(N/O) (see Fig. \ref{Fig11} (d)).

\begin{figure*}
	\begin{tabular}{cccc}
		\begin{minipage}{0.05\hsize}\begin{flushright}\textbf{(a)} \end{flushright}\end{minipage}  &  \begin{minipage}{0.43\hsize}\centering{\includegraphics[width=1\textwidth]{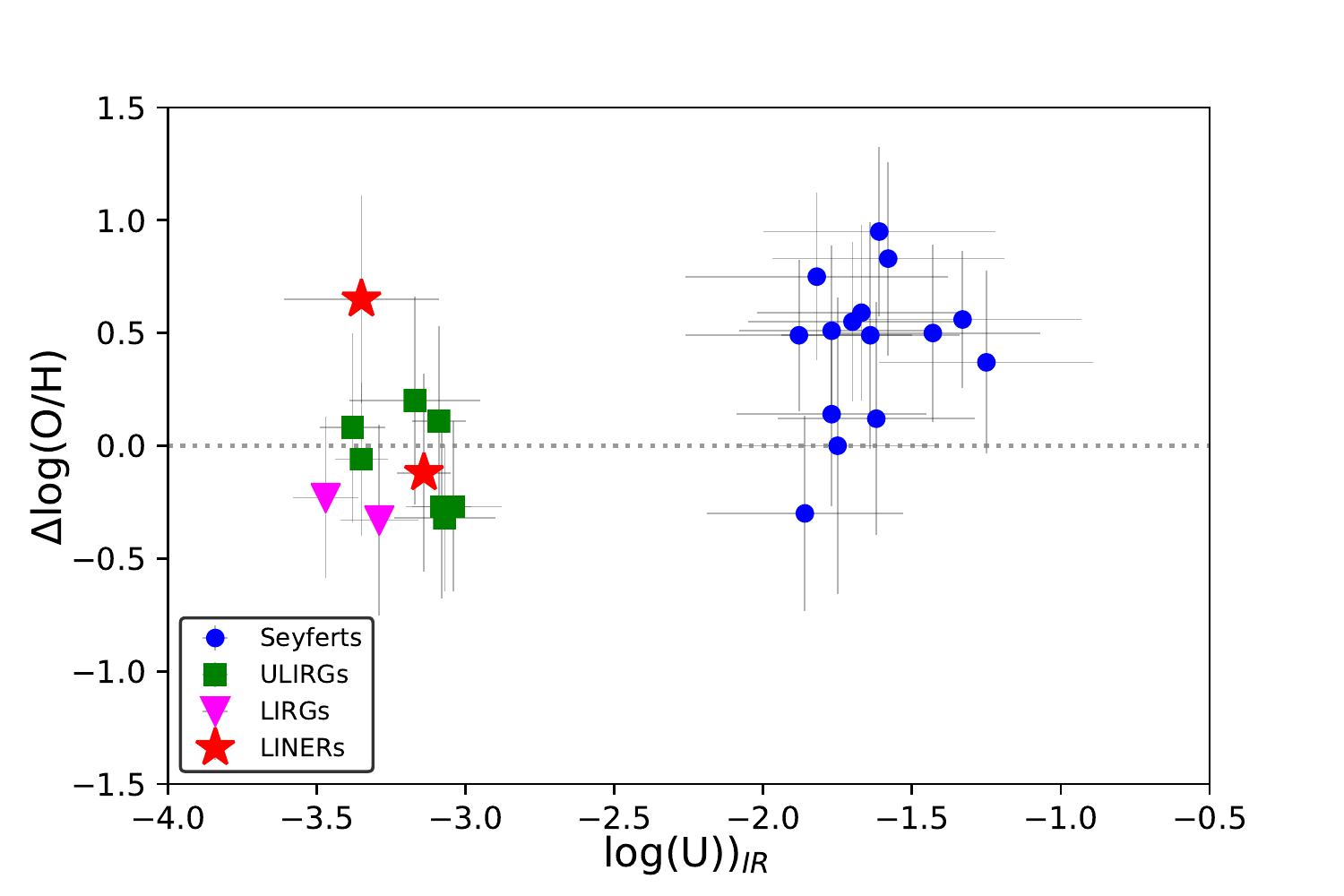}} \vspace{-0.1in} \end{minipage} & \begin{minipage}{0.05\hsize}\begin{flushright}\textbf{(b)} \end{flushright}\end{minipage}  &  \begin{minipage}{0.43\hsize}\centering{\includegraphics[width=1\textwidth]{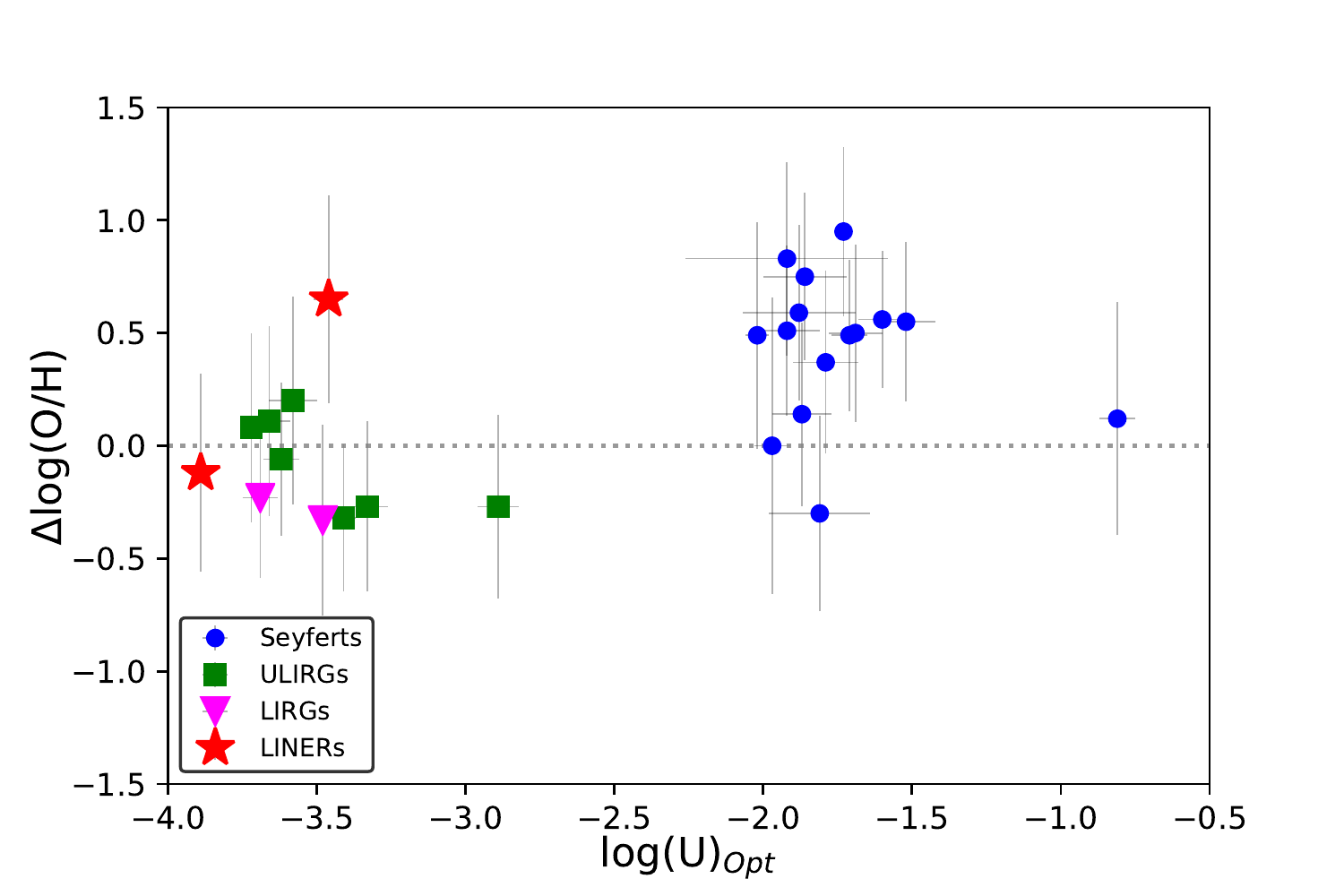}}  \vspace{-0.1in} \end{minipage} \\
		\begin{minipage}{0.05\hsize}\begin{flushright}\textbf{(c)} \end{flushright}\end{minipage}  &  \begin{minipage}{0.43\hsize}\centering{\includegraphics[width=1\textwidth]{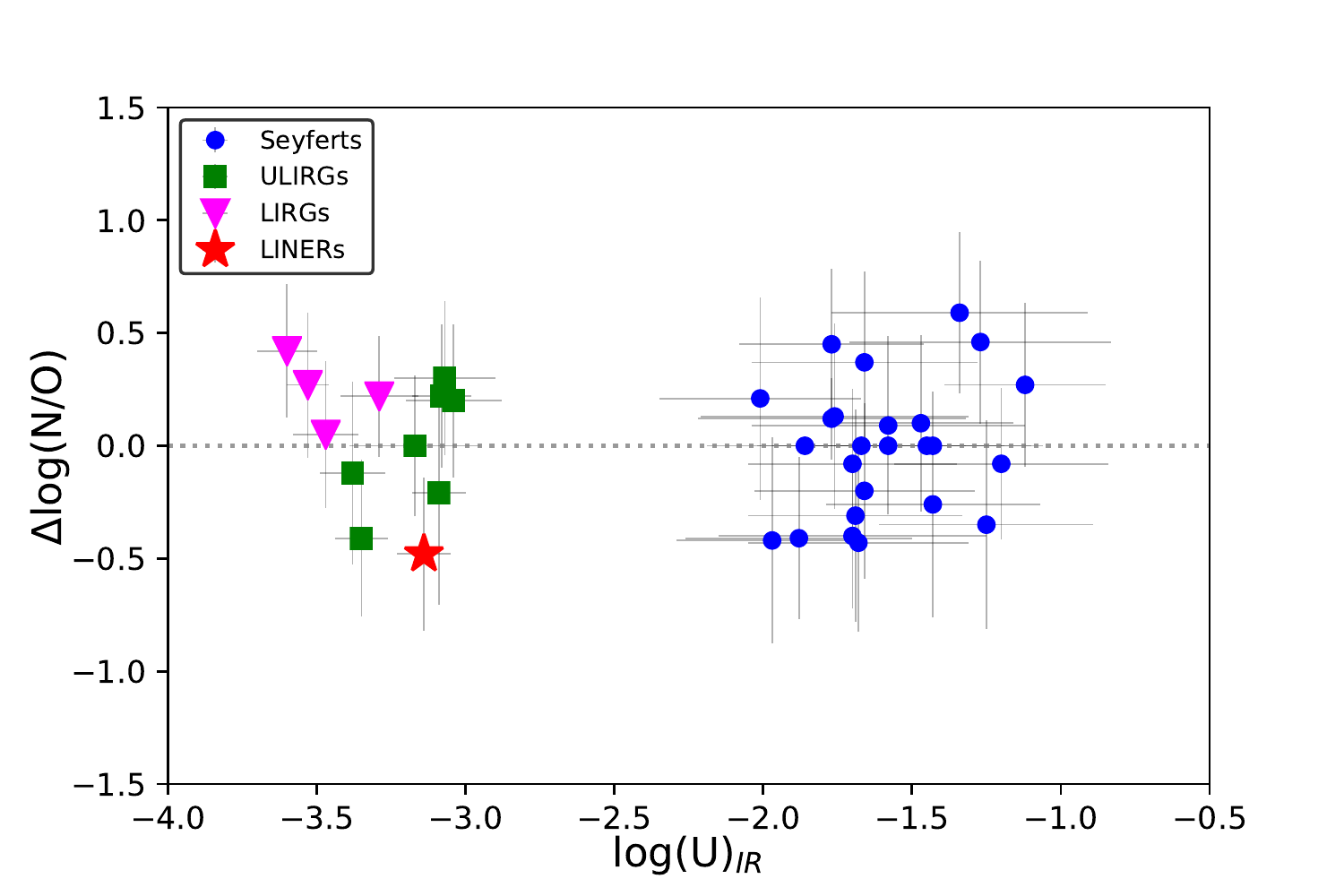}} \vspace{-0.2in} \end{minipage} & \begin{minipage}{0.05\hsize}\begin{flushright}\textbf{(d)} \end{flushright}\end{minipage}  &  \begin{minipage}{0.43\hsize}\centering{\includegraphics[width=1\textwidth]{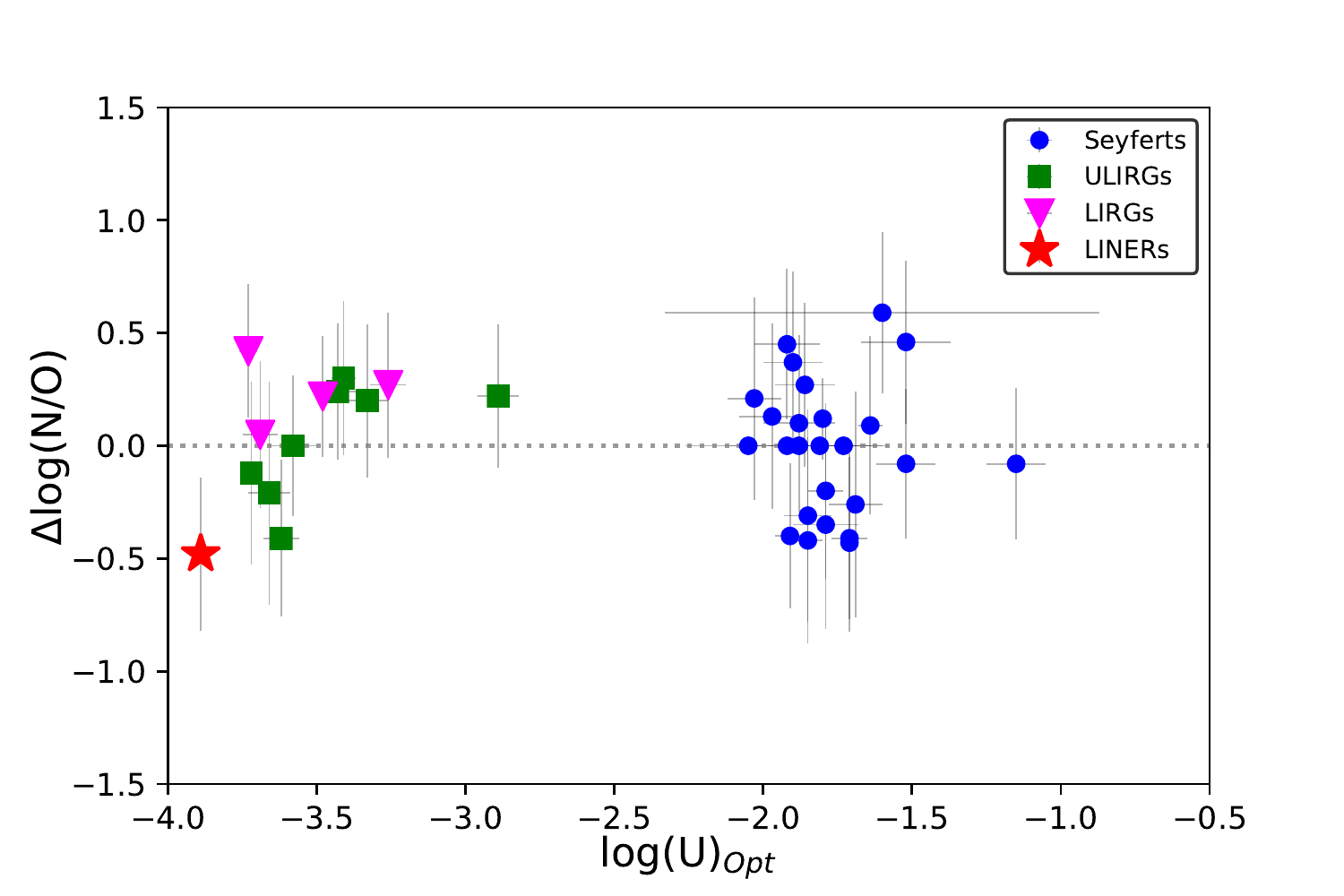}} \\ \vspace{-0.2in} \end{minipage}
	\end{tabular}
	\caption{Discrepancies between the chemical abundance ratios ($\Delta X = X_{opt} - X_{IR}$) as a function of the ionization parameter: (a) and (c) estimated from IR lines, (b) and (d) from optical lines.}
	\label{Fig12}
\end{figure*}

We replicate the same study of the discrepancies as a function of the ionization parameter $U$. Fig. \ref{Fig12} shows that $U$ (either derived from optical or IR emission lines) does not drive the discrepancies found for both O/H and N/O.

In Sec. \ref{subsec31} we explained the importance of electronic density for IR emission lines since for wavelengths in far-IR (above 80 $\mu$m) the corresponding critical densities $n_{c}$ of the lines are closer to the expected $n_{e} \sim 500$ cm$^{-3}$ for the NLR of AGNs \citep{Alloin_2006, Vaona_2012, Netzer_2015}. This is not the case for optical emission lines whose critical densities are in the range [10$^{3.5}$, 10$^{6}$] cm$^{-3}$.

We estimated electronic densities from both, optical and IR emission lines, using \textsc{Pyneb} \citep{Luridiana_2015} and assuming an electronic temperature $T_{e} \sim 2\cdot 10^{4} $ K, which is the average electronic temperature of the different ionic species in the models. We used the sulfur doublet [\ion{S}{ii}]$\lambda\lambda $6717,6731\r{A} for our optical determination and the sulfur lines [\ion{S}{iii}]$\lambda$ 18$\mu $m and [\ion{S}{iii}]$\lambda$ 33$\mu $m to estimate $n_{e}$ from IR lines.

\begin{figure*}
	\begin{tabular}{cccc}
		\begin{minipage}{0.05\hsize}\begin{flushright}\textbf{(a)} \end{flushright}\end{minipage}  &  \begin{minipage}{0.43\hsize}\centering{\includegraphics[width=1\textwidth]{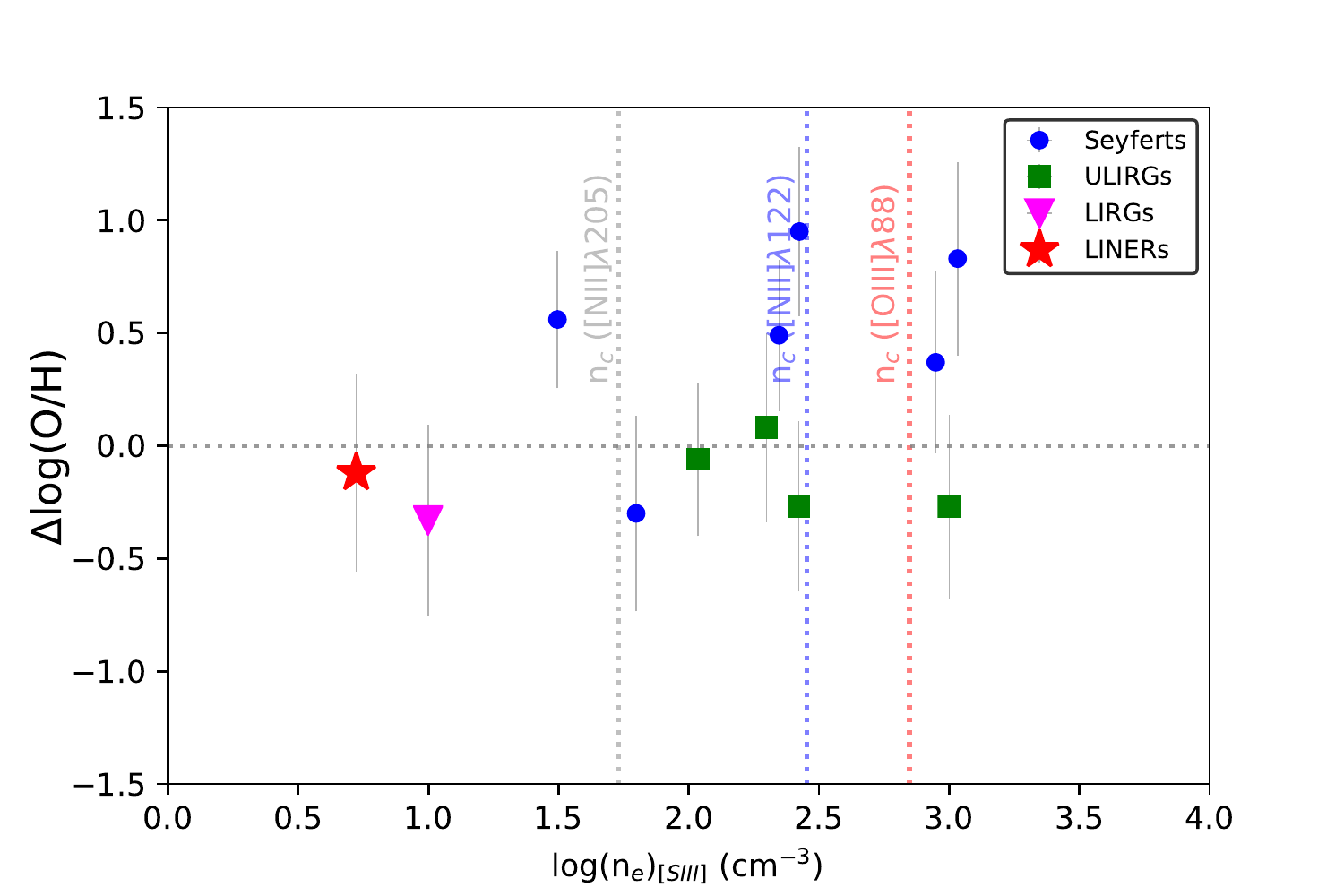}} \vspace{-0.1in} \end{minipage} & \begin{minipage}{0.05\hsize}\begin{flushright}\textbf{(b)} \end{flushright}\end{minipage}  &  \begin{minipage}{0.43\hsize}\centering{\includegraphics[width=1\textwidth]{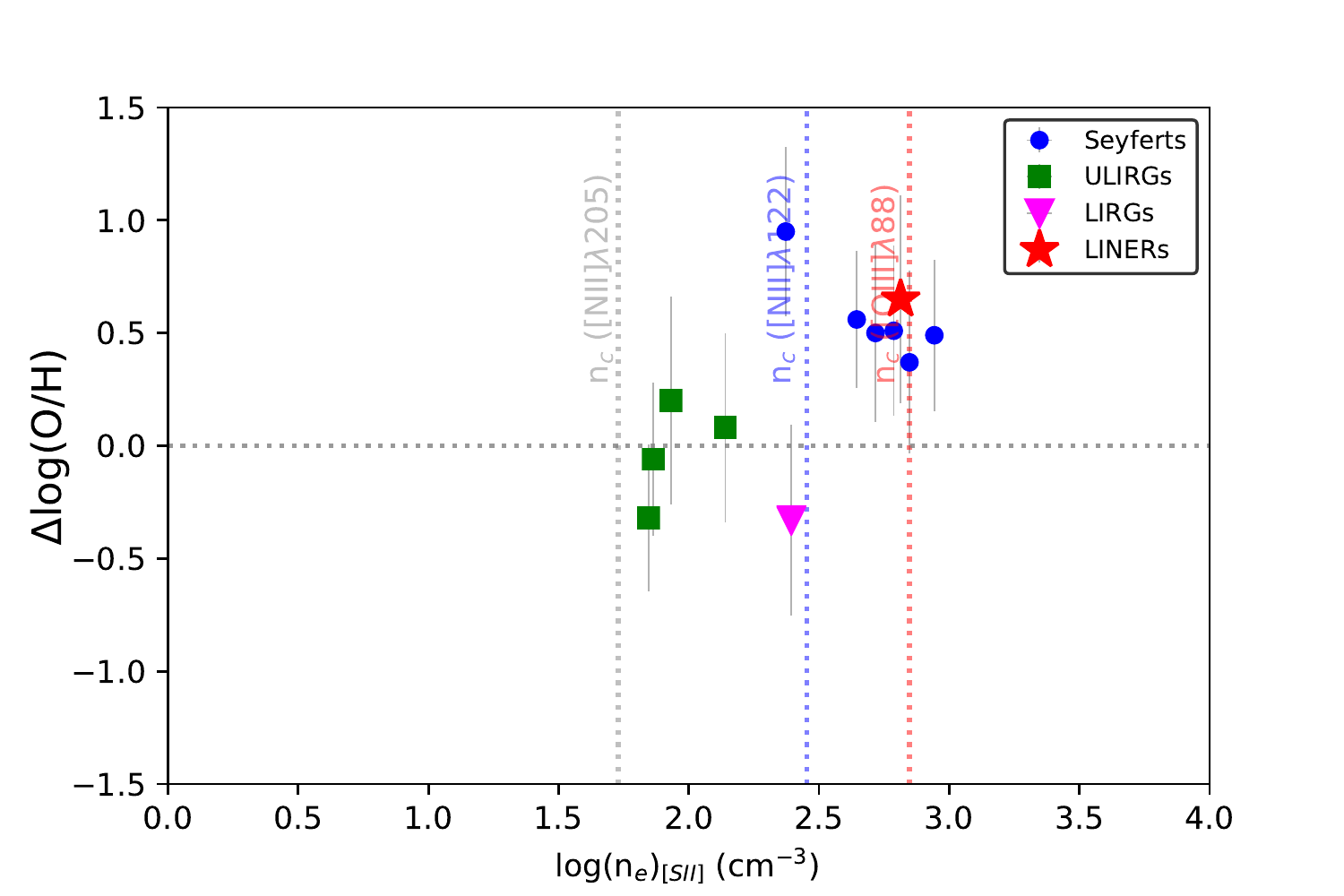}}  \vspace{-0.1in} \end{minipage} \\
		\begin{minipage}{0.05\hsize}\begin{flushright}\textbf{(c)} \end{flushright}\end{minipage}  &  \begin{minipage}{0.43\hsize}\centering{\includegraphics[width=1\textwidth]{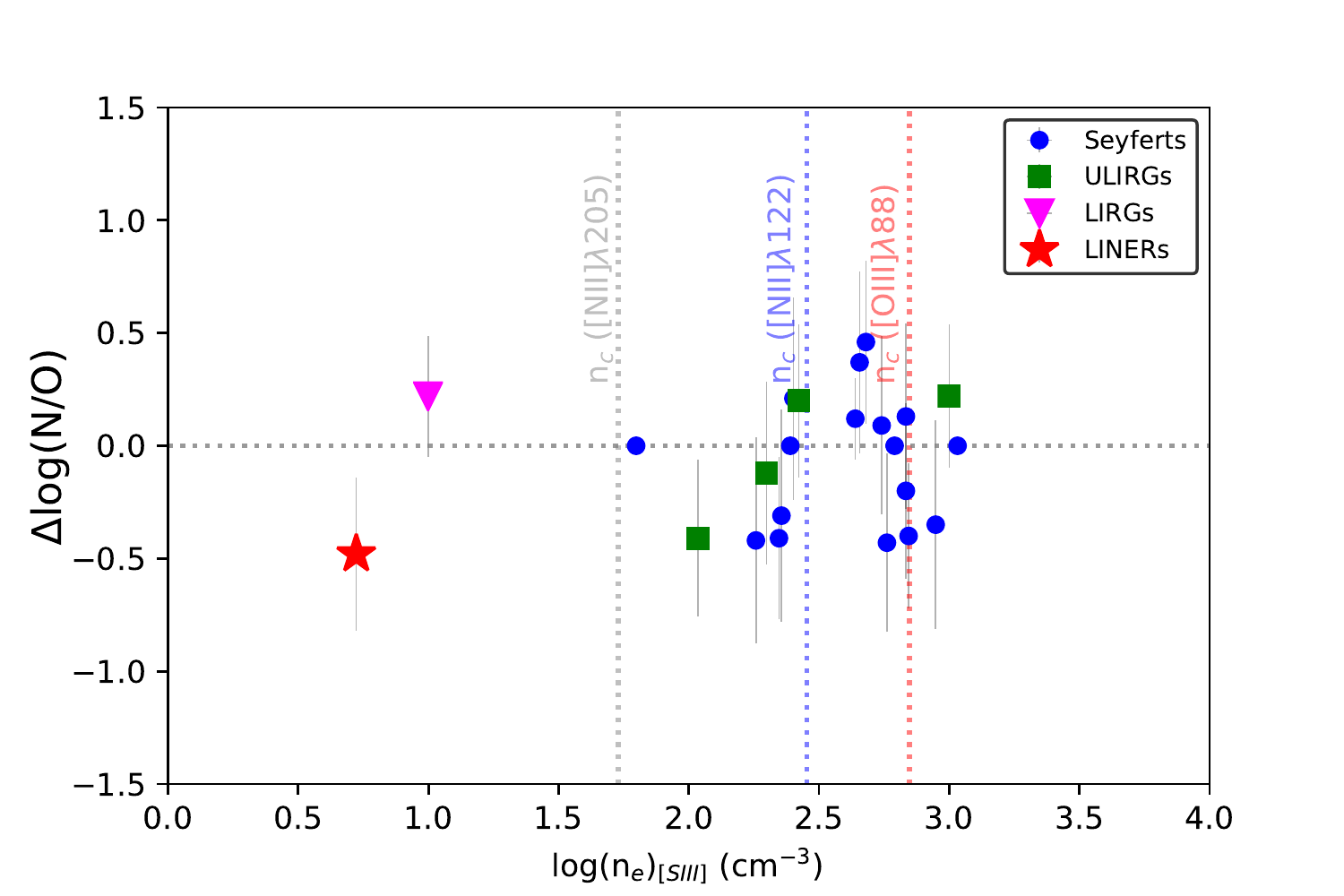}} \vspace{-0.2in} \end{minipage} & \begin{minipage}{0.05\hsize}\begin{flushright}\textbf{(d)} \end{flushright}\end{minipage}  &  \begin{minipage}{0.43\hsize}\centering{\includegraphics[width=1\textwidth]{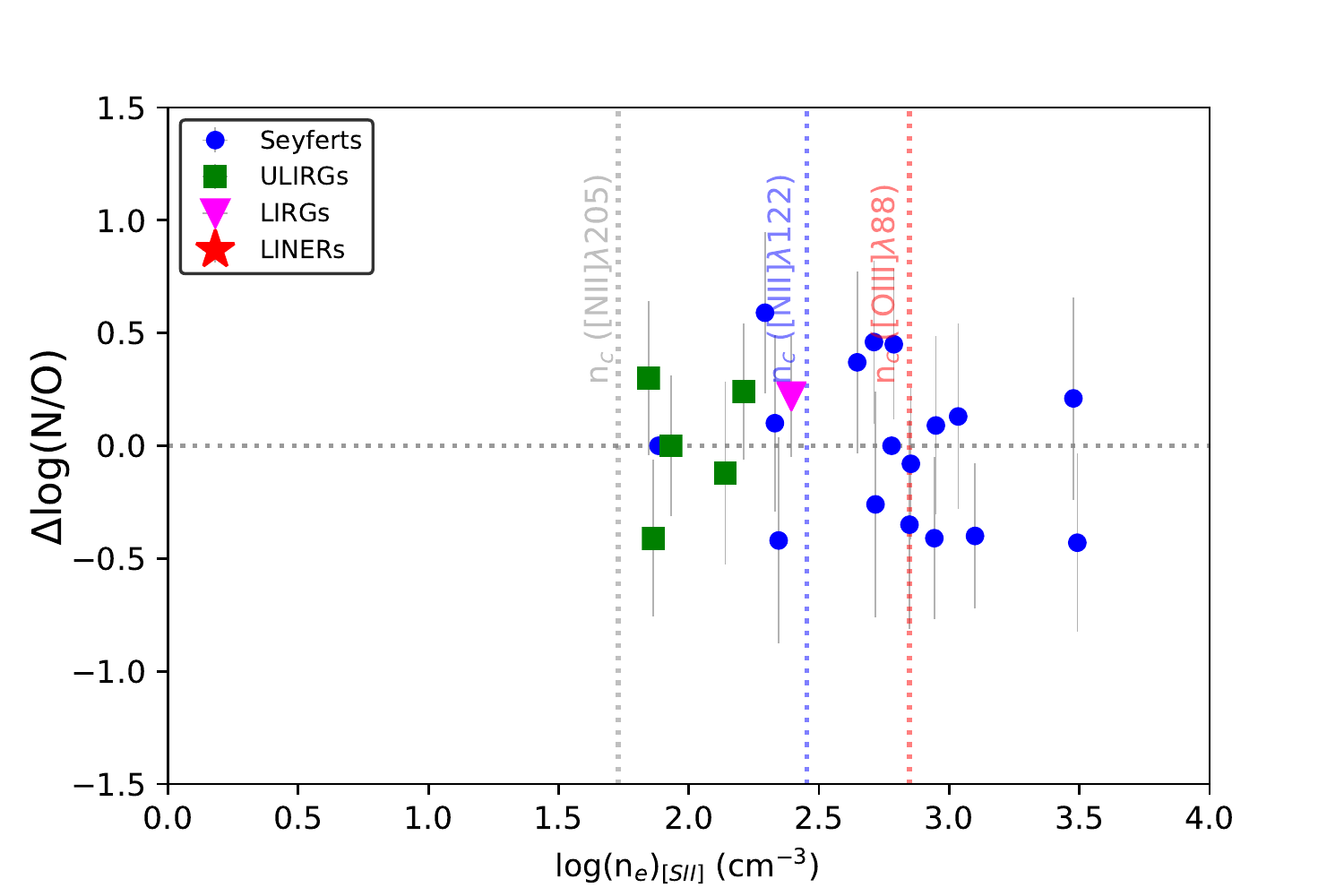}} \\ \vspace{-0.2in} \end{minipage}
	\end{tabular}
	\caption{Discrepancies between the chemical abundance ratios ($\Delta X = X_{opt} - X_{IR}$) as a function of the electronic density: (a) and (c) present electronic densities derived from [S\textsc{iii}] lines, (b) and (d) densities derived from [S\textsc{ii}] lines.}
	\label{Fig13}
\end{figure*}

As shown in Fig. \ref{Fig13} neither $\Delta $log(O/H) nor $\Delta $log(N/O) correlate with electronic density. This result was also found by \citep{Spinoglio_2021}, although they only analyzed nitrogen-to-oxygen abundances in a sample of AGNs from SOFIA due to the spectral coverage. Our results extend this behavior to the oxygen abundances.

\section{Discussion}
\label{sec5}

\subsection{Abundances from IR lines}
\label{subsec51}

IR emission lines are key to analyze chemical abundances in both dusty-embedded regions and from the cold component ($\sim $ 1000 K) of the ionized gas, which is barely accessible  for optical observations. 
However, in general, we warn about the reduced statistics in our sample of galaxies with a reliable derivation of O/H (below $50\% $ of our sample), due to the lack of measurements of hydrogen lines. On the contrary, slightly better statistics are found in N/O ($\sim 60\% $), but again the measurement of [\ion{N}{iii}]$\lambda $57$\mu $m is critical for that estimation. Overall, the estimation of $U$ is almost assured  when running the code ($\sim 90\%$). Nevertheless, these results must be corroborated in larger samples of AGN.

Our estimations of chemical abundances in the NLR of AGN show that the infrared emission is tracing a region characterized by subsolar oxygen abundances (12+log(O/H) < 8.69). Since the measurement of at least one hydrogen line is necessary to provide an estimation of O/H, these subsolar values might be explained by an intrinsic bias: galaxies with measurements of hydrogen emission lines may be characterized by low metallicities. Moreover, as estimations of oxygen abundances required the measurement of faint emission lines as hydrogen recombination lines, these measurements are always obtained with higher uncertainties (see Tab. \ref{Offsets_theoretical}). Unfortunately, the lack of alternative methodologies to directly estimate IR oxygen abundances does not allow us to test this hypothesis.

The nitrogen-to-oxygen ratio seems to be constant for this sample, clustering around the solar value log(N/O)$_{\odot }$ = -1.06. In fact, N/O abundances are well constrained in the range [-1.1, -0.4], which is the same range reported by \citet{Spinoglio_2021} for their sample of AGN.

The lack of an apparent correlation between N/O and O/H (see Fig. \ref{Fig8}), contrary to other assumed relations in the same metallicity regime, evidences that using nitrogen emission lines to estimate oxygen abundances must rely on an independent measurement of N/O, what can also be done by \textsc{HCm-IR}.
The assumption of different relations between N/O and O/H could thus produce non-negligible deviations in the estimated O/H value as derived using N lines. For instance, contrary to our results, \citet{Chartab_2022} reported oxygen abundances above the solar value by assuming a N/O-O/H relation and a fixed ionization parameter $U$. This discrepancy, also observed by \citetalias{Fernandez-Ontiveros_2021} for SFG (showing little offset between IR and optical estimations), might be explained by the different assumption of an N/O-O/H relation for IR estimations.

\subsection{Discrepancies between IR and optical estimations}
\label{subsec52}
While the estimations of the ionization parameter $U$ derived from IR emission-lines  are consistent with those derived from optical lines for low-ionization AGN,  although with slightly more scatter for high-ionization objects (see Fig. \ref{Fig10} (c)), we report an offset between the estimated chemical abundances from IR and from optical lines. From Fig. \ref{Fig10} (a) we obtain that the $\Delta$log(O/H) discrepancy is higher for the more metallic AGN (see also Fig. \ref{Fig11} (d)): using IR emission lines values above solar oxygen abundances cannot be reached, although there are galaxies in our sample whose optical estimations point towards oversolar abundances. This result is found for Seyferts, but also for ULIRGs and LINERs, which contrast with the results from \citet{Chartab_2022} that points to  chemical abundances estimated from IR lines in a sample of ULIRGs higher than those obtained from optical lines.

Regarding nitrogen-to-oxygen abundance ratios, these  follow the same trend: their estimations from optical emission lines are higher than those from IR observations (see Fig. \ref{Fig10} (b)). This result was also found for both SFG \citep{Peng_2021} and AGN \citep{Spinoglio_2021}, although they only use the N3O3 estimator to derive N/O (see Eq. \ref{N3O3}) while we use both N3O3 and N3S34 (see Eq. \ref{N3S34}). Furthermore, we obtained N/O abundances clustering around N/O$_{\odot }$, in agreement with the results by \citet{Spinoglio_2021}.

These discrepancies between N/O and O/H from optical and IR observations also translates into the N/O-O/H diagram (see Fig. \ref{Fig7} and \ref{Fig9}). While there is a trend for decreasing N/O for increasing  O/H for low-ionization AGN, this is not found when IR estimations are considered. In the case of Seyferts, the range of values for O/H and N/O is more limited from IR estimations than from optical. Thus, we warn about using any N/O-O/H relation to estimate oxygen abundances from IR nitrogen lines, specially if this relation was obtained from optical observations. 

As evidenced by Fig. \ref{Fig13}, these discrepancies cannot be explained by a difference in the electronic density in the observed region. On the contrary, as proposed by \citet{Peng_2021},
such a difference could   ultimately indicate a large contribution from the diffuse ionized gas (DIG)  to the estimated chemical abundances. 

Another proposed scenario, based on the results for N/O \citep{Peng_2021, Spinoglio_2021}, is related to the ionization structure of the gas : IR lines are tracing high ionization gas (O$^{++}$, N$^{++}$, S$^{++}$, S$^{3+}$ ) while optical lines are tracing low ionization gas (O$^{+}$, N$^{+}$, S$^{+}$). If the ionization structure plays a role in these discrepancies found for both O/H and N/O, a trend must appear when this variations are analyzed as a function of $U$. As shown by Fig. \ref{Fig12}, the ionization parameter, either obtained from optical lines ((a) and (c)) or from IR lines ((b) and (d)) shows no correlation with the discrepancies in both O/H and N/O. Thus, the different ionization structure cannot explained the differences obtained between IR and optical estimations of chemical abundances.

In any case, since we are analyzing the NLR in AGN, which is not obscured by the dusty torus, it seems unlikely that dusty-embedded regions of the AGN are contributing to these discrepancies, although dust content within NLR might be underestimated. 
However, an alternative  possible explanation for these discrepancies could arise from the contribution of colder parts in the NLR, whose emission is detected in the IR range. According to our results, these zones could then be characterized by solar values of N/O and subsolar oxygen abundances, what could be consistent  with our result that the differences arise above all in the most metallic galaxies.

In the case of AGNs, another possible explanation could rely on the spectral resolution of the IR observations. Due to the emission of the Broad Line Region (BLR), hydrogen recombination lines might present an additional contribution to their fluxes from these broad components, which cannot be spectrally resolved with the current IR data. However, this is not the case for the N/O abundance ratios, whose values are estimated independently of the hydrogen recombination lines, thus an additional contribution to the chemical discrepancies may be present.

\subsection{The importance of N/O}
\label{subsec53}

Overall, we emphasize that determining nitrogen-to-oxygen abundance is fundamental in order to understand the chemical composition and evolution of the ISM. First of all, as shown in Fig. \ref{Fig1}, this ratio does not show a high discrepancy between SFG and AGN models, implying that no bias is introduced if a galaxy is wrongly classified. Although the difference arises for N3S34, N3O3 has probed to be a robust N/O estimator for both AGN and SFG.

Secondly, the estimations of N/O involved close IR emission lines, such as [\ion{O}{iii}]$\lambda $52$\mu$m, [\ion{N}{iii}]$\lambda $57$\mu$m or [\ion{O}{iii}]$\lambda $88$\mu$m, which are more likely to be accessible in the same observational set. Thanks to the ongoing mission SOFIA, some of these emission lines are detected for galaxies in the local Universe, as well as current and future ground-based sub-mm telescopes (e.g. ALMA) will retrieved these lines for the rest-frame IR spectrum of high-redshift galaxies, allowing a redshift-dependent study of N/O.

Thirdly, this chemical abundance ratio is necessary to provide a non-biased estimation of oxygen abundances from nitrogen emission lines. As pointed by several authors \citep[e.g.][]{Perez-Montero_2009, Perez-Diaz_2021, Fernandez-Ontiveros_2021, Spinoglio_2021}, assuming an arbitrary law for N/O-O/H, which is not always followed, can lead to uncertainties in the oxygen content of the gas-phase, and this can be avoidable when data allows an independent previous determination of N/O.

Finally, since N/O involves the abundance of a metal with primary origin (O) and the abundance of another metal with a possible secondary origin (N), its determination also provides key information on the chemical evolution of the metals in the ISM.

\section{Conclusions}
\label{sec6}
We have presented \textsc{HII-CHI-Mistry-IR} for AGN, an updated version of the code proposed for SFG. Thanks to this new method, chemical abundances in the NLR of AGN can be estimated from IR nebular emission lines, which are less affected by extinction and show little dependence on physical conditions of the ISM as the electronic density or temperature. This new tool allows, whenever possible, an independent estimation of N/O, O/H and $U$.

The analysis of a sample of AGN with available IR emission-line fluxes compiled from the literature  shows that their  oxygen abundances tend to be solar and subsolar (12+log(O/H) $\leq$ 8.69), while nitrogen-to-oxygen abundance ratios cluster around solar values (log(N/O) $\sim $ -1). Since both O/H and N/O are calculated independently, these new estimations shows that a relation between N/O-O/H is not found for our sample of AGN.

We also estimated chemical abundances from optical observations of the same sample of AGN. In general, higher oxygen abundances are obtained from these estimations than from IR observations. An analogous result is also found for nitrogen-to-oxygen ratios. We explored if these discrepancies between optical and IR observations may arise from the contribution of diffuse ionized gas, but we concluded that they are not related with electronic density. We also obtained that these discrepancies do not correlate with the ionizing field. As these differences are found for most metallic AGN, IR emission could be tracing zones of the AGN characterized by subsolar oxygen abundances and solar nitrogen-to-oxygen ratios.

In the following years, thanks to JWST and METIS for the local Universe and ALMA, APEX and CSO for high-redshift galaxies, the amount of galaxies (including AGN), whose IR spectral information will be retrieved with high precision, will notably increase, leading to a higher volume of AGN with IR hydrogen recombination lines measured and with many other fine-structure IR lines. With this upcoming data, further constraints can be established for the IR N/O-O/H relation and for the systematic offset between IR and optical estimations.

\begin{acknowledgements}
We acknowledge support from the Spanish MINECO grants AYA2016-76682C3-1-P, AYA2016-79724-C4 and PID2019-106027GB-C41. We also acknowledge  financial  support  from  the  State Agency for Research of the Spanish MCIU through the "Center of Excellence Severo Ochoa" award to the Instituto de Astrof\'{\i }sica de Andaluc\'{\i }a (SEV-2017-0709). JAFO acknowledges the financial support from the Spanish Ministry of Science and Innovation and the European Union -- NextGenerationEU through the Recovery and Resilience Facility project ICTS-MRR-2021-03-CEFCA. We acknowledge the fruitful discussions with our research team. E.P.M. acknowledges the assistance from his guide dog Rocko without whose daily help this work would have been much more difficult.
\end{acknowledgements}

\bibliographystyle{bibtex/aa} 
\bibliography{hcm}

\begin{appendix}
\section{Data}
\label{a1}
We present in this appendix the full dataset of mid- to far-IR spectroscopy of our sample of 58 AGN (Tab. \ref{TabA1}) and optical spectroscopic information retrieved from the literature (Tab. \ref{TabA3}). Tab. \ref{TabA2} and Tab. \ref{TabA4} show our estimations (infrared and optical respectively) of chemical abundances and ionization parameters for our sample.

\begin{landscape}
	\begin{table}
	\caption{List of IR fluxes for our sample of AGN.}\label{TabA1}
	\centering
	\begin{tabular}{llllllllllll}
		\textbf{Name} & \textbf{RA} & \textbf{De} & \textbf{z} & \textbf{Type} & \boldmath$\mathrm{Br}_{\alpha }$ & \boldmath$\mathrm{Pf}_{\alpha }$ & \textbf{[\ion{S}{iv}]10.5$\mu$ m} & ... &  \textbf{[\ion{N}{ii}]122$\mu$ m} & \textbf{[\ion{N}{ii}]205$\mu$ m} & \textbf{Ref.}\\  
		\textbf{(1)} & \textbf{(2)} & \textbf{(3)} & \textbf{(4)} & \textbf{(5)} & \textbf{(6)} & \textbf{(7)} & \textbf{(8)} & ... &   \textbf{(20)} & \textbf{(21)} & \textbf{(22)} \\ \hline 
		IRAS00198-7926 & 00h21m53.6141s &  -79d10m07.9572s & 0.07 & S2 & -  & -  & 8.1$\pm$0.4 & ... & -  & - & FO16,TW  \\
		NGC185 & 00h38m57.8837s & +48d20m14.6616s & -0.0007 & S2 & - & -  & -  & ... & - & - & FO16,TW    \\
		MCG-01-24-012 & 09h20m46.2653s & -08d03m21.9564s & 0.020 & S2 & - & - & 2.33$\pm$0.39 &  ... & - & - & FO16,TW    \\
		NGC4593 & 12h39m39.4550s & -05d20m39.0156s & 0.009 & S1.0 & - & - & 3.9$\pm$0.6 & ... & 2.1$\pm$0.3 & - &  FO16,TW   \\
		NGC5506 & 14h13m14.8757s & -03d12m27.6984s & 0.006 & S1h & - & - & 73.5$\pm$1.6 & ... & 14.1$\pm$1.2 & - & FO16,TW    \\
		IRAS08572+3915 & 09h00m25.3829s & +39d03m54.2988s & 0.058 & ULIRG & 2.53$\pm$0.25 & - & - &  ... & 0.74$\pm$0.15 & - & ARM07,VEI09, \\
		& & & & & & & & & & & HC18,YAN21  \\
		Arp299A & 11h28m33.7s & +58d33m49s & 0.010 & LIRG & - & - & 5.52$\pm $1.11 &  ... & 10.05$\pm$0.78 & - & ALO00,INA13, \\  &  &  &  &  & - & - &  &  &  &  & PEN21,SPI21 \\
		NGC6240 & 16h52m58.8862s & +02d24m0.36288s & 0.024 & LIN & 6.6$\pm$1.8 & - & 2.6$\pm$0.27 &  ... & 23.15$\pm$2.22 & 18.46$\pm$0.39 & INA18,FO16 \\
	\end{tabular}
	\tablefoot{Column (1): name of the galaxy. Columns (2) and (3): coordinates. Column (4): redshift. Column (5): spectral type. Columns (6)-(21): IR emission line fluxes and their errors in 1e$^{-14}$ erg/s/cm$^{2}$. Column (22): references for hydrogen recombination line fluxes. The complete version of this table is available at the CDS.}
	\tablebib{ALO00 \citep{Alonso-Herrero_2000}, ARM07 \citep{Armus_2007}, B-S09 \citep{Bernard-Salas_2009}, BEL03 \citep{Bellamy_2003}, BEL04 \citep{Bellamy_2004}, BEN04 \citep{Bendo_2004}, BRA08 \citep{Brahuer_2008}, DAN05 \citep{Dannerbauer_2005}, FO16 \citep{Fernandez-Ontiveros_2016}, GOL95 \citep{Goldader_1995}, HC18 \citep{Herrera-Camus_2018}, IMA04 \citep{Imanishi_2004}, IMA10 \citep{Imanishi_2010}, INA13 \citep{Inami_2013}, INA18 \citep{Inami_2018}, LAM17 \citep{Lamperti_2017}, LAN96 \citep{Lancon_1996}, LUT02 \citep{Lutz_2002}, MAR10 \citep{Martins_2010}, MUE11 \citep{Muller_2011}, MUR01 \citep{Murphy_2001}, PEN21 \citep{Peng_2021}, PIQ12 \citep{Piqueras_2012}, PS17 \citep{Pereira-Santaella_2017}, REU02 \citep{Reunanen_2002}, REU03 \citep{Reunanen_2003}, RIF06 \citep{Riffel_2006}, SEV01 \citep{Severgnini_2001}, SMA12 \citep{Smajic_2012}, SPI21 \citep{Spinoglio_2021}, VEI97 \citep{Veilleux_1997}, VEI09 \citep{Veilleux_2009}, YAN21 \citep{Yano_2021}, TW (This work).
	}
\end{table}

\begin{table}
	\caption{Chemical abundances estimated from \textsc{HCm-IR}, using the grid of AGN models for $\alpha_{OX} = 0.8$ and the stopping criteria of $2\%$ of free electrons.}\label{TabA2}
	\centering
	\begin{tabular}{llll}
		\textbf{Name} & \boldmath$12+\log(O/H)$ &  \boldmath$\log(N/O)$ & \boldmath$\log(U)$\\ \textbf{(1)} & \textbf{(2)} & \textbf{(3)} & \textbf{(4)} \\ \hline 
		IRAS00198-7926 & 8.16$\pm$0.34 & - & -1.82$\pm$0.32 \\
		NGC185 & - & - & -   \\
		MCG-01-24-012 & 7.87$\pm$0.37 & - & -1.59$\pm $0.39   \\
		NGC4593 & 7.85$\pm$0.39 & -0.9$\pm$0.27 & -1.73$\pm$0.35  \\
		NGC5506 & 7.94$\pm$0.33 & -0.71$\pm$0.14 & -1.77$\pm$0.39    \\
		IRAS08572+3915 & 7.97$\pm$0.38 & -0.83$\pm$0.3 & -3.17$\pm$0.22 \\
		Arp299A & - & -0.88$\pm$0.27 & -3.53$\pm$0.07\\
		NGC6240 & 8.34$\pm$0.31 & -0.8$\pm$0.31 & -3.14$\pm$0.1 \\
		
	\end{tabular}
    \tablefoot{Column (1): name of the galaxy. Columns (2)-(4): chemical abundances and ionization parameters with their corresponding uncertainties. The complete version of this table is available at the CDS.}
\end{table}

\end{landscape}
\newpage
\begin{landscape}

\begin{table}
	\caption{List of optical fluxes for our sample of AGN.}\label{TabA3}
	\centering
	\begin{tabular}{lllllllllll}
		\textbf{Name} & \textbf{RA} & \textbf{De} & \textbf{z} & \textbf{Type} & \textbf{[O\textsc{ii}]3727$\AA$} & \textbf{[Ne\textsc{iii}]3868$\AA$} & ... &  \textbf{[S\textsc{ii}]6717,31$\AA$} & \boldmath$F(H_{\beta})$ & \textbf{Ref.}\\  
		\textbf{(1)} & \textbf{(2)} & \textbf{(3)} & \textbf{(4)} & \textbf{(5)} & \textbf{(6)} & \textbf{(7)} & ... &   \textbf{(11)} & \textbf{(12)} & \textbf{(13)} \\ \hline 
		IRAS00198-7926 & 00h21m53.6141s &  -79d10m07.9572s & 0.07 & S2 & -  & -  & ... & 0.96$\pm$0.30  & 4.42$\pm$1.83 & LUM01  \\
		NGC185 & 00h38m57.8837s & +48d20m14.6616s & -0.0007 & S2 & - & -  & ... & 4.5$\pm$1.0 & 0.09$\pm$0.24 & HO97    \\
		MCG-01-24-012 & 09h20m46.2653s & -08d03m21.9564s & 0.020 & S2 & 4.56$\pm$0.08 & 1.27$\pm$0.03 &  ... & 1.86$\pm$0.03 & 1.04$\pm$0.21 & KOS17    \\
		NGC4593 & 12h39m39.4550s & -05d20m39.0156s & 0.009 & S1.0 & 0.10$\pm$0.03 & 0.12$\pm$0.03 &  ... & 0.023$\pm$0.007 & 367$\pm$57 & MOR88,MAL17  \\
		NGC5506 & 14h13m14.8757s & -03d12m27.6984s & 0.006 & S1h & 8.0$\pm$2.2 & 1.83$\pm$0.48  & ... & 1.92$\pm$0.42 & 139$\pm$23 & MAL17,SHU80,    \\ & & & & & & & & & & DUR88\\
		IRAS08572+3915 & 09h00m25.3829s & +39d03m54.2988s & 0.058 & ULIRG & 2.707$\pm$0.49 & - &  ... & 1.22$\pm$0.18 & 1.11$\pm$0.47 & RUP08 \\
		Arp299A & 11h28m33.7s & +58d33m49s & 0.010 & LIRG & - & - & ... & 0.69$\pm$0.10 & 21.2$\pm$2.1 & GAR06  \\
		NGC6240 & 16h52m58.8862s & +02d24m0.36288s & 0.024 & LIN & 45.9$\pm$12.5 & 0.99$\pm$0.26 & ... & 3.43$\pm$1.06 & 89.99$\pm$15.55 & MAL17,CON12 \\
		
	\end{tabular}
	\tablefoot{Column (1): name of the galaxy. Columns (2) and (3): coordinates. Column (4): redshift. Column (5): spectral type. Columns (6)-(11): optical emission line fluxes and their errors referred to H$_{\beta }$ emission and reddening-corrected. Column (12): sum of emission lines [\ion{S}{ii}]6717$\AA$ and [\ion{S}{ii}]6731$\AA$ referred to H$_{\beta }$ emission and reddening-corrected. Column (13): Flux of hydrogen line H$_{\beta }$ and its error in 1e$^{-14}$ erg/s/cm$^{2}$ and reddening-corrected. Column (13): References for optical emission lines. The complete version of this table is available at the CDS.}
	\tablebib{
	BOK75 \citep{Boksenberg_1975}, BUC06 \citep{Buchanan_2006}, BUT09 \citep{Buttiglione_2009}, CON12 \citep{Contini_2012}, COS77 \citep{Costero_1977}, DUC97 \citep{Duc_1997}, DUR88 \citep{Durret_1988}, ERK97 \citep{Erkens_1997}, GAR06 \citep{Garcia-Marin_2006}, GOO83 \citep{Goodrich_1983}, GU06 \citep{Gu_2006}, HO93 \citep{Ho_1993}, HO97 \citep{Ho_1997}, KIM95 \citep{Kim_1995}, KIM98 \citep{Kim_1998}, KOS78 \citep{Koski_1978}, KOS17 \citep{Koss_2017}, KRA94 \citep{Kraemer_1994}, LUM01 \citep{Lumsden_2001}, MAL86 \citep{Malkan_1986}, MAL17 \citep{Malkan_2017}, MOR88 \citep{Morris_1988}, MOO96 \citep{Moorwood_1996}, MOU06 \citep{Moustakas_2006}, OLI94 \citep{Oliva_1994}, OST75 \citep{Osterbrock_1975}, OST76 \citep{Osterbrock_1976}, OST93 \citep{Osterbrock_1993}, PAS79 \citep{Pastoriza_1979}, PHI78 \citep{Phillips_1978}, PHI83 \citep{Phillips_1983}, RUP08 \citep{Rupke_2008}, SHA07 \citep{Shang_2007}, SHU80 \citep{Shuder_1980}, SIM98 \citep{Simpson_1998}, STO97 \citep{Storchi-Bergmann_1997}, VAC97 \citep{Vaceli_1997}, VEI99 \citep{Veilleux_1999}, WES85 \citep{Westin_1985}, WIL79 \citep{Wilson_1979}, WIN92 \citep{Winkler_1992}, WU98 \citep{Wu_1998}	
	}
\end{table}

\begin{table}
	\caption{Chemical abundances estimated from optical emission lines.}\label{TabA4}
	\centering
	\begin{tabular}{llllll}
		\textbf{Name} & \boldmath$12+\log(O/H)$ &  \boldmath$\log(N/O)$ & \boldmath$\log(U)$ & \boldmath$12+\log(O/H)_{FM+20}$ & \boldmath$12+\log(O/H)_{Ca+20}$ \\ \textbf{(1)} & \textbf{(2)} & \textbf{(3)} & \textbf{(4)}& \textbf{(5)} & \textbf{(6)} \\ \hline 
		IRAS00198-7926 & 8.28$\pm$0.24 & -0.51$\pm$0.26 & -1.87$\pm$0.10 & 7.65$\pm$0.20 & 8.45$\pm$0.13 \\
		NGC185 & 8.55$\pm$0.2 & -0.97$\pm$0.25 & -1.47$\pm$0.1 & 7.78$\pm$0.18 & 8.51$\pm$0.12 \\
		MCG-01-24-012 & 8.82$\pm$0.07 & -0.98$\pm$0.14 & -1.73$\pm$0.01 & 8.36$\pm$0.01 & 8.52$\pm$0.1 \\
		NGC4593 & 8.04$\pm$0.53 & - & -1.97$\pm$0.06 & - & - \\
		NGC5506 & 8.59$\pm$0.16 & -1.11$\pm$0.32 & -1.71$\pm$0.06 & 8.17$\pm$0.16 & 8.54$\pm$0.12    \\
		IRAS08572+3915 & 8.17$\pm$0.33 & -0.85$\pm$0.14 & -3.58$\pm$0.08 & - & 8.36$\pm$0.09 \\
		Arp299A & 8.00$\pm$0.26 & -0.63$\pm$0.13 & -3.26$\pm$0.06 & - & 8.33$\pm$0.06\\
		NGC6240 & 8.25$\pm$0.34 & -1.28$\pm$0.20 & -3.89$\pm$0.02 & 7.96$\pm$0.20 & 8.72$\pm$0.11 \\
		
	\end{tabular}
	\tablefoot{Column (1): name of the galaxy. Columns (2)-(4): chemical abundances and ionization parameters with their corresponding uncertainties derived from \textsc{HCm} using the grid of AGN models for $\alpha_{OX} = 0.8$ and the stopping criteria of $2\%$ of free electrons. Column (5): oxygen abundances and their uncertainties derived with the calibration proposed by \citet{Flury_2020} (Eq. \ref{FM20}). Column (6): oxygen abundances and their uncertainties derived with the calibration proposed by \citet{Carvalho_2020} (Eq. \ref{Ca20}). The complete version of this table is available at the CDS.}    
\end{table}
\end{landscape}
\end{appendix}

\end{document}